%% file: mcprimer.tex
\documentclass{SciPost}
\input{incl_settings.tex} 
\input{incl_shortcuts.tex}

\graphicspath{{./figs/}}

\usepackage{soul}
\usepackage{todonotes}

\newcommand{\responsible}[1]{}
\makeatletter
\if@todonotes@disabled

\else

\fi
\makeatother

\begin{document}
\pagestyle{SPstyle}

\begin{center}{
\Large \textbf{\color{scipostdeepblue}{
The Monte Carlo Ecosystem in High-Energy Physics: A Primer}}}
\end{center}

\begin{center}
Melissa van Beekveld\textsuperscript{1},
Enrico Bothmann\textsuperscript{2},
Andy Buckley\textsuperscript{3},\\
Christian Gütschow\textsuperscript{4},
Peter Skands\textsuperscript{5},
Ramon Winterhalder\textsuperscript{6}
\end{center}

\begin{center}
\textbf{1} Nikhef, Amsterdam, the Netherlands\\
\textbf{2} CERN, Geneva, Switzerland\\
\textbf{3} School of Physics \& Astronomy, University of Glasgow, UK\\
\textbf{4} Centre for Advanced Research Computing, University College London, UK\\
\textbf{5} School of Physics and Astronomy, Monash University, Melbourne, Australia\\
\textbf{6} TIFLab, Universit\`a degli Studi di Milano \& INFN Sezione di Milano, Italy
\end{center}

\begin{center}
\today
\end{center}


\section*{Abstract}
\textbf{%
Monte Carlo event generators are the central interface between theoretical calculations and experimental measurements in collider physics. Over several decades, a comprehensive and highly modular ecosystem of tools has developed around them, encompassing matrix-element calculations, parton showers, hadronisation models, and their integration with detector simulation, event-level analysis and statistical inference. While these tools are ubiquitous in modern research, the conceptual scope and technical structure of the full simulation chain can be challenging to navigate, particularly for researchers entering the field.
In this primer, we provide a structured and up-to-date overview of the high-energy physics Monte Carlo ecosystem, focusing primarily on event-generator methodologies and their role within the broader collider workflow. We discuss the conceptual foundations of modern generators, the computational and organisational challenges of large-scale simulations, and the principles that enable interoperability and reproducibility across theory and experiment. We also examine the evolving computing landscape and sustainability considerations that will shape the future development of these tools.
Aimed primarily at early-stage doctoral researchers while serving as a reference for the broader community, this article seeks to clarify architecture, methodology, and long-term trajectory of Monte Carlo event generation in collider physics.}

\clearpage
\vspace{10pt}
\noindent\rule{\textwidth}{1pt}
\tableofcontents\thispagestyle{fancy}
\noindent\rule{\textwidth}{1pt}
\vspace{10pt}


\section{Introduction}
\label{sec:introduction}

\responsible{Chris}

Monte Carlo (MC) simulations are a cornerstone of modern particle physics, enabling us to connect first-principles theoretical models with the complex signatures observed in detectors. At the energy and precision frontier explored by the Large Hadron Collider (LHC), our ability to interpret experimental results -- and to search for signs of physics beyond the Standard Model (SM) -- relies critically on detailed and accurate simulations of particle collisions. From the calculation of hard scattering processes through the modelling of hadronisation and detector effects, MC tools provide a flexible and systematically improvable bridge between theory and experiment.

The scale of the LHC programme poses not only a conceptual challenge but a computational one. With hundreds of petabytes of data already collected and more on the way in the High-Luminosity LHC (HL-LHC) era, we cannot afford to simulate every possible theory from scratch. Instead, we rely on a rich and interoperable ecosystem of MC tools that allow us to simulate SM processes with high fidelity, explore the consequences of new physics models, and efficiently compare predictions to data. These tools are indispensable for everything from detector design and event reconstruction to theory validation and precision measurement. Yet, for newcomers to the field, this ecosystem can appear overwhelming -- comprising a patchwork of specialised tools, each handling a different part of the simulation chain, often with different interfaces, assumptions, and histories.

In this spirit, the present article is meant as a primer and point of orientation for the Monte Carlo community. It aims to provide a coherent map of the current tool landscape, clarify the roles and interfaces of the main components, and place recent developments into a broader context of long-term sustainability and interoperability. Such an overview is also motivated by community-driven efforts around MC support and coordination, which seek to lower the entry barrier for new users, facilitate the exchange between tool developers and practitioners, and ensure that the rapidly evolving simulation chain remains transparent, reliable, and systematically improvable.

\begin{figure}[htbp!]
    \centering
    \includegraphics[width=\textwidth]{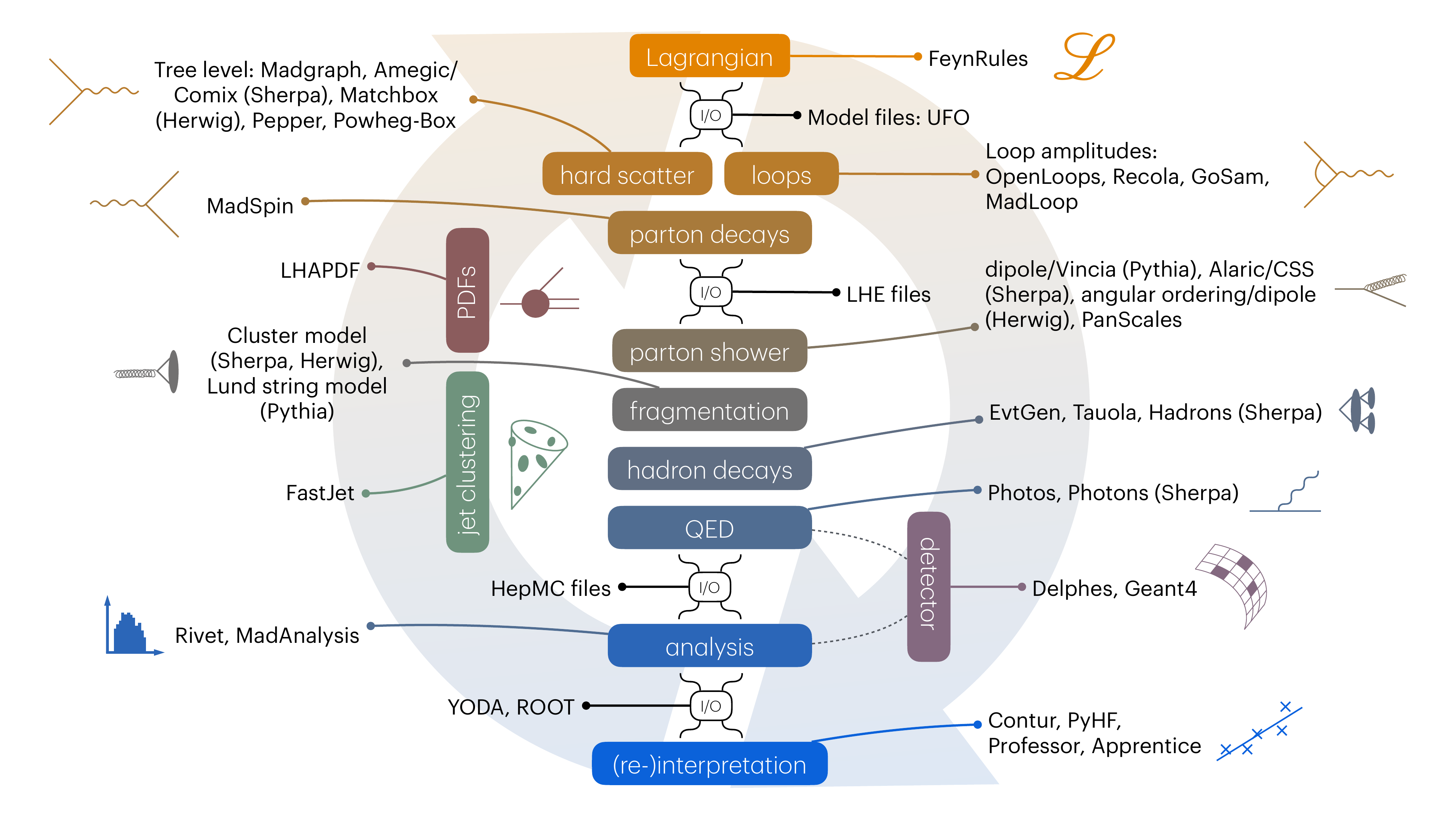}
    \caption{Overview of the most commonly used Monte Carlo tools and their roles within the full simulation and analysis chain. The data flow proceeds from the underlying theory at the top to the statistical interpretation at the bottom, which in turn feeds back into model building, completing the cycle.}
    \label{fig:ecosystem_overview}
\end{figure}

\subsection{The Monte Carlo event generation chain}
\label{sec:overview}

\responsible{Andy}

Before turning to the structure and components of event simulation, it is a good idea to be clear about our overall aim and motivation. In an ideal world, the output of a full event-generation chain would be a large number of final-state particles, distributed in position and kinematics exactly corresponding to what would be seen by an ideal detector. Passed through a detector-simulation toolchain (see later), these simulated events would be indistinguishable from real events.

In practice, such perfect faithfulness is a pipe-dream, but it is still instructive to indulge fantasy and ask how we would achieve it. 
In the generation step, using quantum field theory (QFT) as our most complete mathematical framework, we would construct the (squared) scattering amplitude for the transition from the initial-state beams to the large number of final-state particles that ultimately propagate into our detector.
Unfortunately, several factors mean that we must immediately curtail the fantasy, and return to the land of the possible:
\begin{description}
    \item[Non-perturbative effects:] this desired output is at ``particle level'', i.e.~including not just the perturbative core of parton interactions, but also their transitions from and to hadrons. The detailed operation of these transitions is not fully understood, being beyond the current capabilities of e.g.~lattice quantum chromodynamics (QCD), and hence a complete, bottom-up QFT modelling of the particle-level event is not yet possible. (Beyond hadronisation, much multiplicity is generated by the decays (and chains of decays) of primary hadrons; these are hadronic effects, but generally better understood or constrained than hadronisation itself, and hence not a major focus of this discussion.)

    \item[Perturbative tractability:] even within the more limited partonic event, where perturbative QFT calculations are generally valid\footnote{More to come later on specifics such as orders of loop correction and all-orders ``resummation'' treatments of photon and gluon radiation.}, our fantasy is not viable as the number of terms in the perturbation series grows factorially with the number of final-state partons. Current practical limits for QCD calculations involve fewer than 10 final-state partons; extension to the hundreds that would be required -- there must be more final-state partons than primary hadrons -- is orders of magnitude beyond computational feasibility.

    \item[Phase-space dimensionality:] and finally, even if somehow a tractable expression were obtained for the scattering amplitude with these enormous numbers of final-state partons, $N_\mathrm{fs}$, this (squared) amplitude would be a function of all the momentum components of all those particles: an on-shell phase-space of dimensionality $\sim 3N_\mathrm{fs} = \mathcal{O}(1000)$. The structure of field theories implies that the dominant regions of probability density are tightly clustered islands (at mostly-unknown locations) in this enormous space, again requiring unrealistic resources to obtain representative event samples.
\end{description}

And so we have the challenge: the object we want to simulate or calculate is not remotely computable by a direct approach. But, thankfully, good approximations are not so far away. Much of the partonic multiplicity can be emulated via a Markov process (the ``parton shower''), based on iterated individual parton splittings in place of a fully correlated high-dimensional correlator: for small or unimportant partonic correlations, this approach is the key to extrapolating from achievable, low-multiplicity matrix elements to exclusive partonic final states at the hadronisation scale. And the non-perturbative effects, both in initial and final states, are also open to attack via approximations: factorisation of beam-hadron structure and modelling of beam-remnants and multiple interactions can be applied in the initial state, and hadronisation models based on improved asymptotic behaviours of QCD in the final state. By piecing these elements together, event generators truly can produce realistic-looking simulated events ready to be injected into detector simulation and analysis. 
However, uncertainties enter at several stages: the hard-scattering calculation and parton-shower kernels are limited by finite perturbative accuracy, while the parton shower, multiple scattering, and hadronisation rely on modelling approximations. Some of these effects are constrained by ``tuning'' to data, whereas others must be propagated to the final predictions.

This leads naturally to a modular, multi-stage structure. Each component addresses a specific aspect of the physics modelling or computational workflow, and over time the community has developed specialised tools that are now widely adopted. \Cref{fig:ecosystem_overview} provides a schematic overview of this tool chain, spanning the stages from the underlying theory to analysis-level observables. In the following, we briefly introduce its main components.

\section{From theory to partons}
\label{sec:partons}

We now turn to the first stages of the simulation chain, in which the underlying theory is used to generate parton-level scattering events. This involves combining information about the incoming hadrons, encoded in parton distribution functions, with perturbative calculations of the hard scattering process, followed by the modelling of additional QCD radiation through parton showers. The methods used to combine these ingredients in a consistent and systematically improvable way are central to modern event generation.

\subsection{Parton densities}
\label{sec:pdfs}

\responsible{Andy}

We present the structure of MC event modelling in the order followed by a fully exclusive generator chain: starting from the most perturbative, largest-momentum-transfer part of the process, the hard scattering, and then moving step by step toward lower scales and higher particle multiplicities, eventually entering the non-perturbative regime of hadronisation.

In doing so, however, it is necessary -- especially for hadron collisions, but also in more detailed treatments of lepton or photon beams -- to specify the structure of the incoming beam particles.\footnote{Here ``parton'' is used in a broad sense to denote any field participating in the initial state of the hard scattering.} Clearly, the types of particles available to initiate the hard interaction strongly affect the probabilities of the possible subprocesses. These effects are encoded in parton distribution functions (PDFs), which combine perturbative and non-perturbative physics. While PDFs are not exactly part of the Monte Carlo simulation chain, they are essential \emph{inputs}. Other essential inputs include the strong coupling \alphaS and the electroweak parameters; in particular, \alphaS also enters PDF determinations.

At leading order,\footnote{Beyond leading order, PDFs no longer admit a simple probabilistic interpretation. Rather, they are scheme-dependent objects which, when convolved with the hard-scattering matrix elements, yield physical cross-sections.} a longitudinal PDF $f_{i/h}(x,\muF^2)$ can be interpreted as the probability density for finding a parton of flavour $i$ inside a hadron $h$, carrying a fraction $x$ of the hadron's longitudinal momentum. 
The scale \muF\ -- also often denoted by $Q$ and conventionally given squared -- is the \emph{factorisation scale}, 
an auxiliary scale introduced when separating long-distance dynamics absorbed into the PDFs from the short-distance hard scattering.\footnote{``Sensible'' choices for \muF are by no means unique or even well-defined. In the absence of a ``true'' value, such magic numbers must be treated as uncertainties; indeed, the factorisation scale and its cousin, the renormalisation scale, are major sources of uncertainty.}

A PDF set is obtained by parametrising the distributions at a low input scale $Q_0 \sim 1~\GeV$ and evolving them perturbatively to higher scales using the DGLAP equations~\cite{Gribov:1972ri,Altarelli:1977zs,Dokshitzer:1977sg}.
A standard parametrisation at fixed input scale $Q_0$ for each flavour takes the form
\begin{equation}
x f_i(x,Q_0^2) = A_i x^{\lambda_i} (1-x)^{\beta_i} F_i(x),
\label{eq:pdf_param}
\end{equation}
where $A_i$ is a normalisation parameter, fixed or constrained such that the PDFs satisfy the appropriate valence-number and momentum sum rules, the exponents $\lambda_i$ and $\beta_i$ control the small-$x$ and large-$x$ behaviour, respectively, and a flexible residual function $F_i(x)$ -- for example a polynomial or a neural network -- captures additional structure~\cite{Thorne:2007fe}.
The scale dependence away from $Q_0$ is not parametrised independently, but is generated perturbatively by DGLAP evolution.
The parameters entering this form are \emph{not} predicted perturbatively, but are determined from data in global fits. Denoting them collectively by $\vec a$, one adjusts them such that the evolved PDFs provide the best possible description of a wide range of measurements, typically quantified through a goodness-of-fit function of the form
\begin{equation}
\chi^2(\vec a)
=\sum_{m,n} \bigl(D_m - T_m(\vec a)\bigr)\,
C^{-1}_{mn}\,
\bigl(D_n - T_n(\vec a)\bigr) ,
\end{equation}
where $D_m$ are the experimental data points, $C_{mn}$ is their covariance matrix, and $T_m(\vec a)$ are the corresponding theory predictions obtained from the evolved PDFs convolved with perturbative partonic cross-sections.\footnote{The perturbative partonic cross-section will be discussed in \cref{sec:hard_scatter}.}
Since DGLAP evolution couples different flavours through parton splittings, individual PDFs cannot be evolved independently. We do not discuss the details of the evolution equations here, but note that they are conceptually closely related to the parton-shower evolution discussed in \cref{sec:showers}. The origin of the name ``factorisation scale'' is hopefully now clearer: \muF represents the scale at which radiative evolution within the hadron is (hopefully smoothly) exchanged for explicitly radiated partons via the matrix element and parton showers.

Using this strategy, many global PDF fits have been constructed over the past decades. Widely used modern PDF sets are provided by the CT, MSHT, and NNPDF collaborations~\cite{Dulat:2015mca,Bailey:2020ooq,NNPDF:2017mvq,NNPDF:2021njg}, as well as the PDF4LHC ``meta PDFs'' which aim to compactly provide behaviours and uncertainty coverage representative of all these families~\cite{PDF4LHCWorkingGroup:2022cjn}. In practice, PDFs are usually accessed through the \mc{LHAPDF} library~\cite{Buckley:2014ana}.

When we refer to a PDF ``set'', we usually mean not only the collection of parton flavours in a single fit, but also an ensemble of PDFs that encodes the uncertainties of the fit. This structure makes it possible to propagate PDF uncertainties through event generation and into physical observables, as discussed later in \cref{sec:uncertainties}. It is therefore useful to briefly summarise the main ways in which PDF uncertainties are represented.
\begin{description}
    \item[Replica sets:]
    In the replica approach, the experimental input data are resampled many times according to their quoted uncertainties and correlations, typically assuming an approximately Gaussian uncertainty model. An independent PDF fit is then performed for each pseudo-dataset, yielding an ensemble of $N_\mathrm{rep}$ replica PDFs whose distribution represents the propagated fit uncertainty. Predictions are obtained by evaluating the observable with each replica in turn, so that the mean, variance, and uncertainty intervals can be estimated directly from the ensemble. Typical replica sets contain $N_\mathrm{rep}=100$ members, although larger sets with $\mathcal{O}(1000)$ replicas are sometimes provided for precision studies. Depending on the PDF-set convention, member~0 may correspond either to the central fit (with no data-smearing) or to the average over the replica ensemble.

    \item[Hessian sets:]
    The main alternative to the replica approach is the Hessian representation, which provides a more compact encoding of the fit uncertainty. Rather than sampling the PDF probability distribution explicitly, one approximates the goodness-of-fit function $\chi^2(\vec a)$ in the vicinity of its minimum. 
    Expanding $\chi^2(\vec a)$ around the best-fit parameters $\vec a_0$, one can write
    \begin{equation}
    \chi^2(\vec a)
    =
    \chi^2(\vec a_0)
    +
    \cancel{
    \sum_i
    \left.
    \frac{\partial \chi^2}{\partial a_i}
    \right|_{\vec a_0}
    (a_i-a_{0,i})
    }
    +
    \frac{1}{2}
    \sum_{i,j}
    (a_i-a_{0,i})
    \left.
    \frac{\partial^2 \chi^2}{\partial a_i \partial a_j}
    \right|_{\vec a_0}
    (a_j-a_{0,j})
    +\cdots ,
    \end{equation}
    where the linear term vanishes because $\vec a_0$ corresponds to the best fit.
    Hence, the leading non-trivial dependence is governed by the quadratic term motivating the definition of the Hessian matrix,
    \begin{equation}
    H_{ij}
    =
    \frac{1}{2}
    \left.
    \frac{\partial^2 \chi^2}{\partial a_i\,\partial a_j}
    \right|_{\vec a=\vec a_0},
    \end{equation}
    which, under the usual assumption that the fit is approximately Gaussian near its minimum, determines the local covariance structure of the PDF parameters, with the covariance matrix approximately given by $C_{ij} = H_{ij}^{-1}$.\footnote{These assumptions are generally not exactly true, leading to a ``dynamic tolerance'' approach to $\Delta \chi^2$ construction to recover desirable statistical coverage in PDF fits.}
    As usual in linear algebra, the most natural representation is obtained in the principal basis in which the covariance matrix $C_{ij}$ (and its inverse) are diagonal. Each eigenvector then defines an independent uncertainty direction in parameter space. In practice, Hessian PDF sets are typically provided either as pairs of positive and negative variations along each eigenvector direction, giving $2N_\mathrm{eig}$ error members for $N_\mathrm{eig}$ eigenvectors, or in a more compressed symmetric form with a single effective variation per eigenvector. Observable uncertainties are reconstructed by evaluating the observable for the full set of error members and combining the deviations according to the appropriate Hessian prescription. Because this construction relies on an explicit parameter-space representation of the fit, early NNPDF sets based on neural-network parametrisations were provided primarily in replica form, with Hessian versions only introduced later through dedicated post-processing techniques~\cite{Carrazza:2015aoa}.
\end{description}
Further sources of uncertainty may also be provided in addition to the experimental fit uncertainty encoded by replica and Hessian sets, for example dedicated variations corresponding to different values of \alphaS or heavy-quark masses. In practice, these uncertainty components can be propagated consistently using the \mc{LHAPDF} library functions.

\subsection{Physics model and hard scattering}
\label{sec:hard_scatter}

\responsible{Ramon}

The simulation chain begins with a choice of the underlying physics model: the particle content, parameters, and interaction vertices that define the hard scattering.
In other words, before generating any events, we must decide \emph{what theory we are simulating} and \emph{which interactions are allowed}.
In practice, this information may come from a built-in model implementation, or be derived from a user-provided Lagrangian using tools such as \mc{FeynRules}~\cite{Alloul:2013bka}, \mc{SARAH}~\cite{Staub:2008uz}, or \mc{LanHEP}~\cite{Semenov:2014rea}, which export models in the \mc{UFO}~\cite{Degrande:2011ua,Darme:2023jdn} format for use within matrix-element generators.
Once the model is fixed, the hard-scattering step specifies the partonic process and provides the corresponding matrix elements. From these, parton-level events are generated according to the fully differential partonic cross-section.

At leading order (LO)\footnote{Leading-order results are also often referred to as \emph{Born-level} predictions, especially when distinguishing them from higher-order corrections.}, the task reduces to a familiar problem from quantum field theory: compute a scattering amplitude, square it, and integrate it over the allowed phase space to obtain a cross-section.
For a partonic scattering process with incoming momenta $p_a$ and $p_b$ and $n$ final-state particles with momenta $p_i$, the LO differential cross-section can be written as
\begin{equation}
\d \hat\sigma_{n}(\hat s)
=
\frac{1}{2\hat s}\;
\overline{\big|\mathcal{M}_{n}(\Phi_n)\big|^2}\;
\frac{1}{S_{\{n\}}}\d\Phi_n(\hat s;\,p_1,\ldots,p_n)\,,
\label{eq:partonic_xsec}
\end{equation}
where $\hat s=(p_a+p_b)^2$ is the squared partonic centre-of-mass energy, the bar indicates the average (sum) over initial-state (final-state) spin and colour degrees of freedom, $S_{\{n\}}$ is the symmetry factor for identical particles, and $\d\Phi_n$ is the Lorentz-invariant $n$-body phase-space measure.

\Cref{eq:partonic_xsec} makes explicit that a fixed-order partonic calculation consists of two conceptually distinct pieces: (i) the squared matrix element, which encodes the dynamics of the underlying theory, and (ii) the phase-space measure, which specifies the allowed kinematics.
It is often convenient to combine all process-dependent factors into a single differential weight
\begin{equation}
w_n(\Phi_n)
= \frac{1}{2\hat s}\;
\overline{\big|\mathcal M(\Phi_n)\big|^2}\;
\frac{1}{S_{\{n\}}}\,.
\label{eq:partonic_weight}
\end{equation}
Any physical observable $\mathcal O$ is then obtained by integrating this weight over phase space against the corresponding measurement function,
\begin{equation}
\langle {\mathcal O} \rangle
= \int \d\Phi_n \; w_n(\Phi_n)\,
\mathcal O(\Phi_n)\,.
\end{equation}
For collisions involving composite initial states, this partonic prediction is further convolved with PDFs, as discussed in \cref{sec:pdfs}, thereby accounting for the non-perturbative structure of the incoming beams.
Similar effective parton densities can also arise in other contexts, such as photon radiation from leptonic beams.

\subsubsection{Matrix elements and perturbative expansion}

\begin{figure}[t]
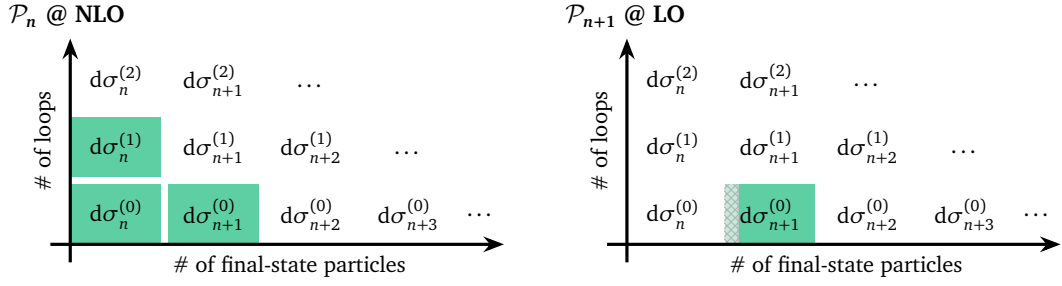

\centering
\scalebox{0.80}{
\arrayrulecolor{white}
\begin{tabular}{l}
\large\bf $\mathcal P_n$ @ NLO\\
\begin{loopsnlegsFO}
  \emptyBox[2]{n} & \emptyBox[2]{n+1} & \ldots
\\[2mm]\hline
  \exactBox[1]{n} & \emptyBox[1]{n+1}
  & \emptyBox[1]{n+2} & \ldots
\\[2mm]\hline
   \exactBox[0]{n} & \exactBox[0]{n+1}
   & \emptyBox[0]{n+2} & \emptyBox[0]{n+3}
   & \parbox{0.6\boxlen}{\centering\ldots}
\end{loopsnlegsFO}
\end{tabular}
\begin{tabular}{l}
\large\bf $\mathcal P_{n+1}$ @ LO\\
\begin{loopsnlegsFO}
  \emptyBox[2]{n} & \emptyBox[2]{n+1} & \ldots
  &
\\[2mm]\hline
  \emptyBox[1]{n} & \emptyBox[1]{n+1}
   & \emptyBox[1]{n+2} & \ldots
\\[2mm]\hline
  \emptyBox[0]{n} & \exactBox[0]{n+1}\addCut{0.22}
   & \emptyBox[0]{n+2}
   & \emptyBox[0]{n+3}
   & \parbox{0.6\boxlen}{\centering\ldots}
\end{loopsnlegsFO}
\end{tabular}}
\caption{Schematic ordering of perturbative coefficients $\d\sigma^{(\ell)}_{\!n+j}$ in loop order $\ell$ (vertical axis) and final-state multiplicity $n+j$ (horizontal axis). {\sl Left:} $\mathcal P_n$ at NLO, combining the Born contribution ($\ell=0$, multiplicity $n$), the virtual correction ($\ell=1$, multiplicity $n$), and the real-emission term ($\ell=0$, multiplicity $n+1$). {\sl Right:} $\mathcal P_{n+1}$ at LO, corresponding exactly to the tree-level real-emission contribution entering $\mathcal P_n$ at NLO. The hatched region indicates the need for kinematic cuts to ensure that the additional radiation remains resolved.}
\label{fig:NLO_vs_LO}
\end{figure}

In perturbative quantum field theory, scattering amplitudes are expanded in powers of the relevant coupling constants, such as the strong coupling \alphaS or the electroweak coupling $\alpha$.
For fixed multiplicity $n$, the amplitude can be written schematically as
\begin{equation}
\mathcal M_n
= \sum_{\ell=0}^{\infty} \mathcal M_n^{(\ell)} \,,
\end{equation}
where $\ell$ denotes the number of loops.
Physical observables depend on the \emph{squared} amplitude. The perturbative expansion of the cross-section is therefore obtained only after squaring:
\begin{align}
|\mathcal M_n|^2
&=
\left|\mathcal M_n^{(0)}\right|^2
+ 2\,\mathrm{Re}\!\left(\mathcal M_n^{(0)} \mathcal M_n^{(1)\,*}\right)
+ \left|\mathcal M_n^{(1)}\right|^2
+ \cdots \,.
\end{align}
A fixed-order calculation consists of truncating this expansion at a given power of the coupling constants.
To make the discussion concrete, we consider a generic hard scattering process, denoted by $\mathcal P_n$, where the subscript $n$ indicates the number of resolved final-state partons at Born level. The corresponding tree-level amplitude $\mathcal M_n^{(0)}$ defines the Born contribution,
\begin{equation}
\d\sigma_n^{\mathrm{LO}}
=
\d\sigma_n^{(0)}
\sim
\left|\mathcal M_n^{(0)}\right|^2 \,.
\end{equation}
At this order, the final state contains exactly $n$ resolved partons.
At next-to-leading order (NLO), both one-loop corrections at multiplicity $n$ and tree-level amplitudes at multiplicity $n+1$ contribute,
\begin{equation}
\label{eq:nlo-contribs}
\d\sigma_n^{\mathrm{NLO}}
= \d\sigma_n^{(0)} + \d\sigma_n^{(1)} + \d\sigma_{n+1}^{(0)}\,,
\end{equation}
where
\begin{equation}
\d\sigma_n^{(1)}
\sim
2\,\mathrm{Re}\!\left(\mathcal M_n^{(0)} \mathcal M_n^{(1)\,*}\right) 
\qquad \mand \qquad
\d\sigma_{n+1}^{(0)}
\sim
\left|\mathcal M_{n+1}^{(0)}\right|^2 \,.
\end{equation}
The additional tree-level term at multiplicity $n+1$ corresponds to real radiation.
Importantly, the LO prediction for the process $\mathcal P_{n+1}$ is exactly the real-emission contribution entering the NLO correction to $\mathcal P_n$. This hierarchical structure naturally links predictions of different multiplicities:
\begin{itemize}
\item LO for $\mathcal P_n$ involves tree-level amplitudes with $n$ partons;
\item LO for $\mathcal P_{n+1}$ involves tree-level amplitudes with $n+1$ partons;
\item NLO for $\mathcal P_n$ combines one-loop amplitudes at multiplicity $n$ with tree-level amplitudes at multiplicity $n+1$.
\end{itemize}
This structure is illustrated schematically in \Cref{fig:NLO_vs_LO}, where the coefficients $\d\sigma^\ell_{\!j}$ are organised according to loop order $\ell$ and final-state multiplicity $n+j$.
At LO for $\mathcal P_n$, only the Born configuration with $n$ resolved partons contributes.
In contrast, the LO contribution for $\mathcal P_{n+1}$ involves an additional parton, and kinematic cuts (indicated by the hatched region) are required to ensure that this extra radiation remains resolved.

Unlike the Born contribution, the real-emission term contains regions of phase space in which the additional parton becomes soft or collinear.
These unresolved configurations give rise to infrared (IR) divergences, which cancel only after combining the real-emission contribution with the virtual correction $\d\sigma_n^{(1)}$ for an infrared-safe observable.
We will discuss this cancellation mechanism in more detail when considering the phase-space integration.

In addition to these IR singularities, loop amplitudes also contain ultraviolet (UV) divergences arising from integrations over large internal momenta.
These UV divergences are typically regularised using dimensional regularisation, and absorbed into a redefinition of the parameters of the theory through renormalisation. After renormalisation and IR cancellation, physical observables become finite.
However, at any fixed perturbative order they retain a dependence on the renormalisation scale and, more generally, on the used renormalisation scheme.
This residual dependence formally appears at higher orders and thus provides a practical estimate of missing perturbative corrections.

In practice, most users do not implement these amplitudes by hand. Instead, they rely on a broad ecosystem of computational tools. Besides symbolic and semi-automated frameworks such as \mc{QGraf}~\cite{Nogueira:1991ex,Nogueira:2021wfp}, \mc{FeynArts}/\mc{FormCalc}~\cite{Hahn:2000kx,Hahn:1998yk}, and \mc{FeynCalc}~\cite{Mertig:1990an,Shtabovenko:2023idz}, widely used tree-level matrix-element generators include
\mc{HELAS}~\cite{Murayama:1992gi,Stelzer:1994ta} within \mc{MadGraph5\_aMC@NLO}~\cite{Alwall:2014hca,Frederix:2018nkq} (shortened to \mc{MG5aMC} hereafter),
\mc{Amegic}~\cite{Krauss:2001iv} and
\mc{Comix}~\cite{Gleisberg:2008fv} within
\Sherpa~\cite{Sherpa:2019gpd,Sherpa:2024mfk},
\mc{Matchbox}~\cite{Platzer:2011bc} within
\mc{Herwig}~\cite{Bellm:2025pcw},
\mc{CalcHEP}~\cite{Belyaev:2012qa},
and \mc{O'Mega}~\cite{Moretti:2001zz} within
\mc{WHIZARD}~\cite{Kilian:2007gr}.
These programs automatically generate Feynman diagrams or recursion-based representations, perform colour and helicity summations, and provide numerical implementations of the corresponding tree-level amplitudes.

At NLO, one-loop amplitudes are supplied by dedicated providers such as \mc{MadLoop}~\cite{Hirschi:2011pa},
\mc{OpenLoops}~\cite{Cascioli:2011va,Buccioni:2019sur}, \mc{Recola}~\cite{Actis:2016mpe,Denner:2017wsf}, and \mc{GoSam}~\cite{Cullen:2011ac,GoSam:2014iqq},
which are typically interfaced to general-purpose event generators. 
Other frameworks such as \mc{BlackHat}~\cite{Berger:2008sj,Berger:2008ag} played an important role in establishing the automated computation of multi-leg one-loop amplitudes. 
Modern providers build on these developments and implement advanced reduction techniques, together with dedicated numerical stability checks~\cite{Passarino:1978jh,Denner:2005nn,Ossola:2006us,Ossola:2007ax}.

Beyond NLO, the level of automation decreases significantly. At next-to-next-to-leading order (NNLO), the computation of two-loop amplitudes remains a highly non-trivial task. While substantial progress has been made, fully general and process-independent automation is not yet available at the same level as at NLO.
Two- and multi-loop virtual amplitudes are often obtained either analytically, using integration-by-parts reduction and differential-equation methods implemented in tools such as \mc{FIRE}~\cite{Smirnov:2019qkx}, \mc{Reduze}~\cite{vonManteuffel:2012np},
\mc{LiteRed}~\cite{Lee:2013mka}, and
\mc{Kira}~\cite{Klappert:2020nbg,Lange:2025fba}, or numerically via sector-decomposition techniques as implemented in \mc{SecDec}~\cite{Borowka:2012yc}, \mc{PySecDec}~\cite{Heinrich:2023til}, and \mc{FIESTA}~\cite{Smirnov:2021rhf}.
For multi-leg processes, integrand-level approaches based on numerical unitarity, for example as implemented in
\mc{Caravel}~\cite{Abreu:2020xvt}, are increasingly being explored.
Dedicated parton-level generators providing fully differential NNLO predictions for selected classes of processes have also been developed, such as \mc{MCFM}~\cite{Campbell:2011bn,Campbell:2019dru} and \mc{NNLOJET}~\cite{NNLOJET:2025rno}. 
These frameworks implement sophisticated subtraction schemes and process-specific ingredients, enabling precision phenomenology for a wide range of observables, albeit without the level of general automation available at NLO.
As a result, NNLO and, more generally, N$^{k}$LO calculations typically require more specialised workflows and remain process-dependent to a much larger extent than their NLO counterparts.

The coexistence of several, partly overlapping implementations reflects both the intrinsic complexity of high-multiplicity and higher-order
perturbative calculations and the need for independent cross-checks.\footnote{ Many physicists also implement streamlined versions of such calculations themselves -- partly as a rite of passage, partly for pedagogical insight, and occasionally in the hope of improving upon existing methods. This tradition of reimplementation has historically played an important role in both training and innovation within the field.}
Automated perturbative calculations involve a large number of algorithmic ingredients -- diagram generation or recursion relations, colour algebra, helicity summation, ultraviolet renormalisation, and numerical stability control -- each of which is highly non-trivial.
Independent implementations based on different algorithms and design philosophies therefore play a crucial role in validating results and ensuring the long-term reliability of precision predictions.

\subsubsection{Phase space and Monte-Carlo integration}

Even once the squared matrix element is known, computing a cross-section requires integrating it over the allowed phase space. For $n$ final-state particles, the Lorentz-invariant phase space has dimension
\begin{equation}
\dim(\Phi_n) = 3n - 4 ,
\end{equation}
after imposing on-shell constraints and overall momentum conservation.
Analytic evaluation of such integrals is rarely possible. 
Event generators therefore rely on Monte Carlo integration~\cite{Weinzierl:2000wd}, which estimates integrals by sampling random phase-space points and averaging the integrand\footnote{The name ``Monte Carlo'' was popularised in the 1940s by Stanislaw Ulam and John von Neumann at Los Alamos, inspired by the Monaco casino and reflecting the central role of randomness in the method. In contrast to roulette, however, increasing the number of events systematically improves the result.}.
Its statistical uncertainty scales as
\begin{equation}
\Delta I \sim \frac{1}{\sqrt{N}} ,
\end{equation}
where $N$ denotes the number of sampled phase-space points.
In contrast, deterministic quadrature rules can converge as $\Delta I \sim N^{-4}$ in one dimension, but in $d$ dimensions their scaling typically degrades to $\Delta I \sim N^{-4/d}$ due to the tensor-product construction of multidimensional grids. While superior at low dimensionality, such methods become inefficient for the high-dimensional phase-space integrals encountered in collider physics, where the dimension-independent $1/\sqrt{N}$ scaling of Monte Carlo integration is superior.
While the statistical uncertainty decreases as $1/\sqrt{N}$, its prefactor is set by the variance of the sampled integrand. 
Rather than relying solely on increasing the number of phase-space points, one can therefore improve convergence by reducing this variance. To this end, modern integrators employ importance sampling. Writing
\begin{equation}
\int \mathrm{d}\Phi\,w(\Phi)
= \int \mathrm{d}\Phi\,\frac{w(\Phi)}{g(\Phi)}\, g(\Phi)
= \int \mathrm{d}G(\Phi)\,\frac{w(\Phi)}{g(\Phi)}
\qquad \mwith \qquad g(\Phi) = \det\left|\frac{\partial G(\Phi)}{\partial \Phi}\right|,
\end{equation}
one samples phase-space points according to a probability density $g(\Phi)$ that approximates the weight $w(\Phi)$.
The Monte Carlo estimator then scales as
\begin{equation}
\Delta I \sim \frac{\sqrt{\mathrm{Var}[w/g]}}{\sqrt{N}} ,
\end{equation}
so that an appropriate choice of $g$ can substantially reduce the variance while preserving the characteristic $1/\sqrt{N}$ behaviour. In the ideal case $g(\Phi) \propto |w(\Phi)|$, the ratio $w/g$ becomes constant and the variance formally vanishes. 
In practice, constructing an optimal density $g(\Phi)$ is highly non-trivial, since scattering amplitudes exhibit multiple resonances, threshold structures, and soft or collinear enhancements. A common strategy is therefore to approximate 
the target density through a multi-channel decomposition,
\begin{equation}
g(\Phi) = \sum_i \alpha_i \, g_i(\Phi) \qquad \mwith \qquad 0\le\alpha_i\le1\;, \quad \mand \quad  \sum_i \alpha_i =1\;,
\end{equation}
where each density $g_i$ is an analytic mapping designed to capture a specific integrand structure. In the original Kleiss--Pittau formulation~\cite{Kleiss:1994qy}, the coefficients $\alpha_i$ are global weights that are tuned iteratively to 
minimise the overall variance.
In the single-diagram enhanced (SDE) approach~\cite{Maltoni:2002qb,Mattelaer:2021xdr}, one instead introduces a phase-space dependent partition of unity,
\begin{equation}
\sum_i \beta_i(\Phi) = 1 \quad\longrightarrow \quad w(\Phi)= \sum_i \beta_i(\Phi)\,w(\Phi) = \sum_i w_i(\Phi)\;,
\end{equation}
which decomposes the integrand accordingly. Each term is then associated with a dedicated mapping adapted to the structure of the channel integrand $w_i$. Typically, the weights $\beta_i(\Phi)$ are constructed from matrix--element-level information.
Beyond such analytic constructions, adaptive algorithms such as \mc{VEGAS}~\cite{Lepage:1977sw,Lepage:1980dq,Lepage:2020tgj} iteratively refine the sampling grid based on previously evaluated phase-space points. In modern approaches, these adaptive refinements can themselves be enhanced using machine-learning techniques, as discussed in \cref{sec:acceleration}.
At LO, event generation then ultimately amounts to drawing phase-space points distributed according to \cref{eq:partonic_weight}, evaluating the matrix element, and storing the resulting kinematics together with the associated event weight $w(\Phi)$. For many downstream applications -- in particular detector simulation and experimental analyses -- it is advantageous to work with (approximately) unit-weight events. This avoids large statistical fluctuations from a few large-weight events and simplifies histogramming and uncertainty propagation. Therefore, an additional \emph{unweighting} step is usually performed to convert weighted events into approximately equal-weight events. 
The standard procedure is the \emph{hit-and-miss} (or accept--reject) method. Given a set of weighted events with weights $w(\Phi)$ and a known upper bound $w_{\max}$, each event is accepted with probability
\begin{align}
P_{\text{acc}}(\Phi) = \frac{w(\Phi)}{w_{\max}} \, .
\end{align}
Accepted events are stored with unit weight, while rejected events are discarded. The unweighting efficiency is therefore given by $\langle w \rangle / w_{\max}$ and depends directly on how well the sampling distribution matches the integrand. A perfectly matched importance-sampling density would yield $w(\Phi)=\text{const.}$ and hence perfect efficiency.

In practice, however, $w_{\max}$ is not known \emph{a priori}.
It must be estimated from a finite sample of already generated events, and one can never guarantee that the true global maximum has been found. To avoid large statistical fluctuations from rare outliers, generators often determine $w_{\max}$ from a pre-sampling phase or from robust estimators (e.g.\ based on quantiles or median-based measures) rather than the single largest observed weight. 
If during event generation a phase-space point with $w(\Phi) > w_{\max}$ is encountered\footnote{Every Monte-Carlo practitioner eventually encounters the event that exceeds the supposedly known $w_{\max}$. This often happens shortly after the first large production run has been launched. It serves as a reminder that rare regions of phase space matter -- a lesson that becomes even more pronounced once negative weights enter the story at NLO.}, several strategies exist.
A common choice is to keep such events with an \emph{overweight}, i.e.\ assign them a weight $w(\Phi)/w_{\max} > 1$, thereby slightly spoiling strict unit-weighting but preserving correctness. Alternatively, the maximum can be dynamically increased, which in turn reduces the efficiency of subsequent unweighting.

\subsubsection{Infrared structure and subtraction at NLO}
\label{sec:nlo-subtraction}

To improve perturbative accuracy and reduce theoretical uncertainties, we often must go beyond LO. At NLO, the cross-section schematically takes the form
\begin{equation}
\label{eq:xsnlo}
\sigma^{\mathrm{NLO}}
= \int \d\Phi_n\,\left[ B(\Phi_n) + V(\Phi_n) \right] 
+\int \d\Phi_{n+1}\;R(\Phi_{n+1})\,,
\end{equation}
where $B$ denotes the Born contribution, $V$ the virtual correction, and $R$ the real-emission contribution. Individually, $V$ and $R$ contain infrared (soft and/or collinear) divergences.
Only after combining them for infrared-safe observables do these divergences cancel. 
Since this cancellation occurs only after integration over unresolved regions of phase space, numerical evaluation requires rearranging the integrand. Modern event generators achieve this either with \emph{subtraction} or with \emph{slicing} methods.

In subtraction approaches~\cite{Catani:1996vz,Nagy:1998bb,Catani:2002hc,Frederix:2009yq}, one introduces a local counter-term $S(\Phi_{n+1})$ that reproduces the infrared limits of the real-emission contribution locally in phase space. Moreover, upon integration over the one-parton emission phase space, the counter-term generates the same infrared singularities as the virtual correction,
\begin{equation}
\label{eq:sigma_NLO}
\sigma^{\mathrm{NLO}}
=
\int \d\Phi_n\,\left[ B(\Phi_n) + V(\Phi_n) + \int\d\Phi_1\, S(\Phi_{n+1})\right] 
+\int \d\Phi_{n+1}\,\left[R(\Phi_{n+1}) - S(\Phi_{n+1})\right]\,,
\end{equation}
so that the poles in $V$ are cancelled analytically by the integrated counterterm $\int\d\Phi_1\, S$.
As a result, both contributions are separately infrared finite and suitable for Monte Carlo integration.

In slicing approaches~\cite{Giele:1991vf,Harris:2001sx}, one instead introduces a resolution parameter $\tau_{\rm cut}$ that partitions the phase space into unresolved and resolved regions.
In the unresolved region, the real-emission matrix element factorises and can be integrated analytically using its universal infrared limits. The resolved region is finite and evaluated numerically. The final result becomes independent of $\tau_{\rm cut}$ in the limit $\tau_{\rm cut}\to 0$, up to power-suppressed corrections.

A practical consequence of fixed-order calculations beyond leading order is that the resulting weighted parton-level samples often contain \emph{negative weights}.
While this may appear counter\-intuitive, such weights do not represent physical negative probabilities.
Rather, they arise from the structure of the perturbative expansion and from the procedures used to render it finite.
First, fixed-order perturbation theory truncates the squared amplitude at a given order in the coupling.
For instance, at NLO one retains
\begin{equation}
|{\cal M}_n|^2
= |{\cal M}_n^{(0)}|^2
+ 2\,\mathrm{Re}\!\left({\cal M}_n^{(0)}{\cal M}_n^{(1)\,*}\right)
+\mathcal O(\alpha_s^2)\,,
\end{equation}
while the positive-definite contribution $|{\cal M}_n^{(1)}|^2$ is formally of NNLO and therefore omitted.
The interference term can be locally negative in phase space, so the truncated expression is not positive definite.
Second, subtraction or slicing procedures introduce compensating contributions that implement the cancellation of infrared divergences between real and virtual parts. In practice, this rearrangement is often the dominant source of locally negative weights. Negative weights therefore serve as a bookkeeping device that ensures the correct perturbative accuracy at the level of \emph{weighted event sums}.

Extensions of these techniques to NNLO accuracy exist, based on more sophisticated subtraction or slicing formalisms and forming the basis of modern NNLO and NNLO+PS predictions.
Prominent examples include antenna subtraction~\cite{Gehrmann-DeRidder:2005btv}, sector-improved residue subtraction (\mc{STRIPPER})~\cite{Czakon:2010td}, $q_T$ subtraction~\cite{Catani:2007vq}, and $N$-jettiness slicing~\cite{Boughezal:2015dva}.
However, the increased infrared complexity means that NNLO calculations are, in general, less automated than at NLO and currently only available for a more restricted set of process classes.

Up to this point, our discussion has focused on LO event generation with unweighted parton-level events and on fixed-order NLO calculations producing weighted samples. In practice, however, one ultimately aims to generate fully exclusive NLO-accurate events suitable for detector simulation. At fixed order, this is not straightforward: NLO event weights arise from cancellations between real and virtual contributions, and although the total cross-section is finite, the local event weight is not guaranteed to be bounded in infrared-sensitive regions. A naive hit-and-miss unweighting procedure, which requires a finite global maximum weight, is therefore not strictly well defined. Achieving fully exclusive event generation instead requires matching fixed-order calculations to parton showers and, when combining different multiplicities, performing consistent merging procedures. Parton showers reorganise the perturbative expansion into a Sudakov-resummed form with a probabilistic interpretation, thereby restoring a bounded event-generation framework in which fully exclusive events can be generated. These techniques will be discussed in the following sections.

\subsection{Parton showers}
\label{sec:showers}

Massless gauge theories share a common feature: they really like to emit particles with very low energy $E$, or particles with very small angle $\theta$ with respect to their emitters.%
\footnote{Massive quantum field theories will still have the soft divergence, but the collinear divergence is shielded by the mass of the emitting/emitted particle. 
This physical feature leads to an observable region of phase space where radiation is very sparse: the \emph{dead cone}.}
The probability, which here we take to be double-differential, for such an emission to occur is proportional to
\begin{equation}
\label{eq:simp-prob}
C \frac{\d E}{E} \frac{\d\theta}{\theta}\,,
\end{equation}
where $C$ is a constant that depends on the couplings of the gauge theory.%
\footnote{
In reality the phase-space for emitting a massless particle has three independent variables (the fourth one is fixed by the on-shell condition $\delta(k^2)$). 
In \cref{eq:simp-prob} we have omitted the azimuthal angle, $\phi$, as \emph{helicity-averaged} splitting function, which govern the splitting probability, do not depend on $\phi$.
}
One readily sees that the quantity in \cref{eq:simp-prob} grows large when $E\to 0$ (the particle becomes soft) and when $\theta \to 0$ (the particle becomes collinear).
The problematic end-points at $E=0$ and $\theta=0$ are removed by including the virtual corrections, 
but away from those the large probabilities remain. 
Sufficiently \emph{inclusive} observables, such as the total cross-section, will not be affected by the presence of additional radiation: such observables are defined to be inclusive over whatever may or may not happen.
These kinds of observables are therefore well described by fixed-order methods, as explained in \cref{sec:hard_scatter}.
However, more exclusive observables, such as the transverse momentum of a heavy boson in $pp$ scattering, are very sensitive to the presence of additional radiation, especially when that additional radiation appears at a very different scale than that of the hard process itself. 
In such situations, any fixed-order computation of these observables is ill-defined, as the perturbative series at any truncated fixed order in the coupling constant does not converge.
The reason for this non-convergence is that the coefficients that appear in the perturbative expansion grow large.
This is easily seen by considering the integral of \cref{eq:simp-prob} over $E \in [0,1]$ and $\theta \in [0,1]$ with a constraint such that the transverse momentum $\pT \sim E \theta > e^{-L}$, where $L$ is a large number. 
We then obtain
\begin{equation}
C \int_0^1 \frac{\d E}{E} \int_0^{1} \frac{\d\theta}{\theta}\,\Theta(E\theta - e^{-L}) = \frac{1}{2}CL^2\,.
\end{equation}
One sees that even for small $C$ (corresponding to a small coupling, in the perturbative regime), this contribution gets large as $L$ increases. 

Luckily, quantum field theories exhibit another remarkable feature. 
Matrix elements and phase spaces for $(n+1)$-body processes, where the additional ``$+1$'' particle is much softer or more collinear than all other particles in the process, factorise into a matrix element and phase-space describing the $n$-body process, ``times'' a matrix element and phase space describing the radiation of the $+1$ soft or collinear particle.%
\footnote{``Times'' does not denote an exact multiplication. For QCD, which is a non-Abelian gauge theory, the kinematics factorise, but the colour (in the case of soft radiation) and spin (in the case of collinear radiation) quantum correlations do not. For this reason, QCD shower algorithms are usually formulated in the leading-colour and spin-averaged limits, although systematic improvements exist.}
If we then consider a sequence of soft or collinear radiations, where the next emission is always more soft or collinear than the current (i.e.\ the emissions are \emph{strongly ordered}), the matrix element and phase space are proportional to
\begin{align}
    \prod_{i = 0}^n C \frac{\d E_i}{E_i}\frac{\d\theta_i}{\theta_i}\frac{\d\phi}{2\pi}\,.
\end{align}
The phase space for the radiation of one additional particle is four-dimensional (as the momentum of that particle is described by a four-vector), but we have used the mass constraint $\delta(m^2 - p^2)$ to integrate out one of those dimensions.
This leaves behind a three-dimensional phase-space, with one variable describing the energy fraction.
Recall that each factor $C$ contains a power of the coupling constant, which is smaller than $1$ in the perturbative limit.
Written in this form, the lack of convergence of the perturbative series becomes immediately apparent.

The solution to this problem is resummation, which is a technique that takes these large contributions into account for all orders in the perturbative constant. 
Likewise, showers aim to guarantee the perturbative convergence of the series for observables sensitive to disparate scales, making shower algorithms an indispensable ingredient of event simulation. 
However, unlike resummation, the end product of a Monte Carlo shower is a collection of identified particles with definite momenta.
This set of particles then feeds into hadronisation models, see \cref{sec:hadrons}, another crucial component of MC event generation. 

The starting point for the formulation of a shower algorithm is a definition of the \emph{Sudakov form factor}, or no-emission probability. 
This is defined as one minus the probability to emit any particle with, for example, a transverse momentum above a given scale $\pT$. 
If we were to consider only one such emission, the probability reads
\begin{align}
P(\text{no emission above } \pT) = 1 - C\int \frac{\d E}{E}\int \frac{\d\theta}{\theta}\,\Theta(E\theta - \pT)\,.
\end{align}
Accounting for many of such emissions in a strongly-ordered sequence, one arrives at the \emph{Sudakov form factor} 
\begin{align}
\label{eq:no-emsn-prob}
\Delta(\pT,Q) = \exp\left(-C \int^Q \frac{\d E}{E} \int \frac{\d\theta}{\theta}\Theta(E\theta -\pT)\right)\,.
\end{align}
This no-emission probability between the scales $\pT < Q$ can be used to calculate, for example, the distribution of the transverse momenta $\pT$.
Such a distribution can be generated using Monte Carlo methods. We start with $i=1$ and $\pTi[0] = Q$, after which we 
\begin{enumerate}
    \item sample uniformly a random number $r_i$;
    \item find $\pTi[i]$ such that $\Delta(\pTi,\pTi[i-1]) = r_i$, i.e.\ the next emission has a smaller $\pT$ than the previous one;
    \item repeat until $\pTi$ becomes less than some cut-off scale $\Lambda$. 
\end{enumerate}
Repeating this many times yields an ensemble of events.
This algorithm, together with the precise formulation of the no-emission probability, forms the backbone of shower algorithms. 
However, in state-of-the-art showers that can generate ``physical'' events, many additional considerations are required.

Firstly, we have now assumed that there is only one kind of splitting probability, whereas in reality we need to consider $q\to qg$, $g\to gg$, $g\to q\bar{q}$, quantum electrodynamic (QED) or electroweak emissions, etc.
One can use the leading-order expression for such branchings, based on soft/collinear factorisation, include terms beyond strict soft/collinear factorisation (which goes under the hood of matrix-element corrections), or implement also higher-order contributions. 
The coupling will in general also depend on the splitting kinematics (i.e.\ \alphaS is a \emph{running} coupling).
Furthermore, with initial-state branchings, one needs to be careful with the treatment of PDFs.

Secondly, an ensemble of real particles that are supposed to reflect a physical high-energy scattering need to obey momentum conservation, meaning that one cannot simply add an additional particle to a pre-existing ensemble without introducing an appropriate \emph{momentum mapping}.
Other than not violating momentum conservation and being infrared safe, such momentum mapping has many degrees of freedom in its formulation, leading to many different shower algorithms. 

Concerning the ordering of emissions, in the algorithm above we have implicitly assumed an ordering in transverse momentum, but one may choose other ordering conditions, such as an ordering in virtuality, or in angle (the latter of which ties closely together with the original formulations of event-shape resummation, using a principle called \emph{colour coherence}).

Finally, one can even alter the algorithm itself to make it faster in specific contexts. 

\begin{figure}[tb!]
\includegraphics[width=\textwidth]{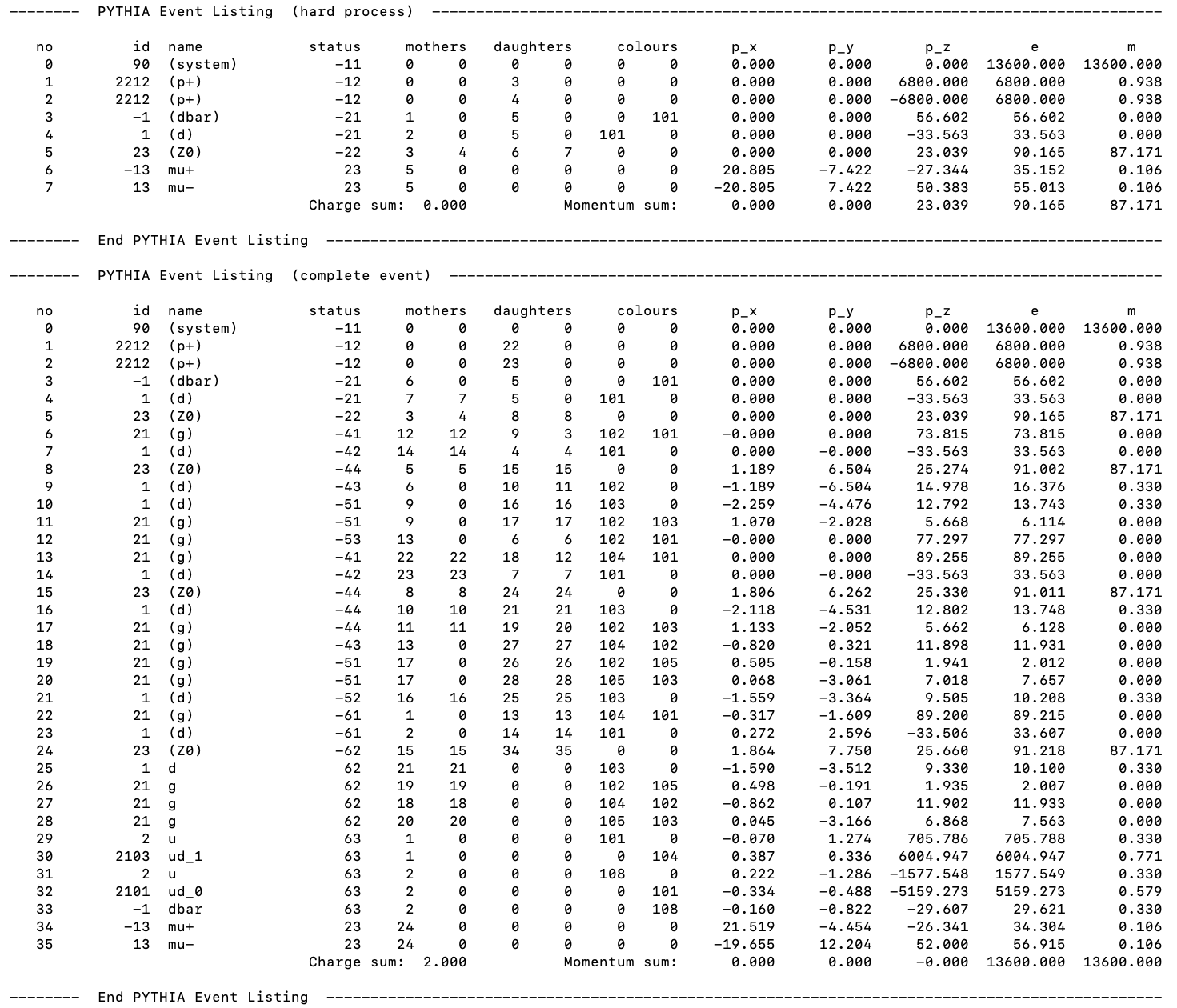}
\caption{An event record for the \Pythia[8] shower. The starting point is a hard process for the creation and decay of a $Z$ boson in a $pp$ collision with a centre-of-mass energy of 13.6~\TeV.}
\label{fig:pythia8-event-record}
\end{figure}

After taking various of such considerations into account, the end result is a showered event record, as shown in \Cref{fig:pythia8-event-record} for the \Pythia[8] shower.%
\footnote{Outputs of other shower programs such as \mc{Herwig} and \mc{Sherpa} will look slightly different.}
Such an event record is hugely informative, and it is instructive to familiar oneself with it. 
At the top of the Event Listing we see the hard process. 
Here, we see that we started off with a scattering of two protons, each carrying an energy of $6.8$~\TeV.
From the proton travelling in the positive $z$-direction one extracts a $\bar{d}$-quark (which in the event record has `mother' label $1$), and from the one travelling in the negative $z$-direction a $d$ (which in the event record has `mother' label $2$).
They then annihilate into a $Z$ boson (with mother labels $3,4$, which are the $\bar{d},d$ quarks), which is unstable and decays into a pair of muons (who carry mother label $5$, which is the $Z$ boson).
One sees that the $p_x$ and $p_y$ momenta of the muons add up to 0: the system starts of with $0$ transverse momentum. 

The next event is the output of the \Pythia[8] default QCD shower after showering this hard event.%
\footnote{Which is called \mc{SimpleShower}, but not because it is simple!}
We see that a lot more particles are present. 
First of all, we can trace back the original hard event by comparing the entries under `no 0-5'. 
We also see that now the mothers of $d$ and $\bar{d}$ are no longer the protons.
Instead, the mother of the hard-scattering particle $\bar{d}$ is now a gluon (`no 6'), and that of $d$ is another $d$ (`no 7'), but with different momentum. 
This is a result of the backwards initial-state evolution of the shower: the original initial-state particle backwards evolves to another initial-state particle by emitting a final state particle. 
The final-state particles are listed without a set of parenthesis around them. 
Firstly, note that the momenta of these final state particles add up to the original centre-of-mass energy: total momentum is conserved.
Secondly, one also sees that now the $p_x$ and $p_y$ components of the two muons no longer add up to $0$: the $Z$ boson acquired a transverse momentum while QCD radiation was emitted (also see the particle in the event record with `no 24'). 
This is precisely why we need the shower!

\begin{figure}
\includegraphics[width=\textwidth]{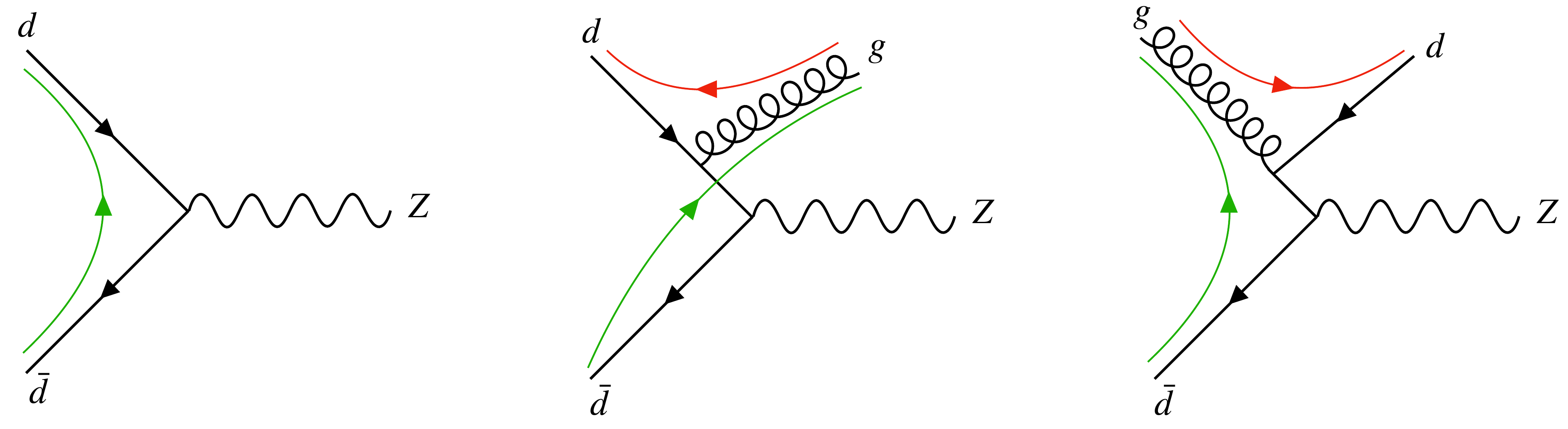}
\caption{Colour labels for a Born $d\bar{d}\to Z$ event (left), a $d\bar{d} \to Zg$ (middle) and a $\bar{d}g\to Z\bar{d}$ event (right). The black arrows indicate the particle--anti-particle connections, whereas the curved, coloured arrows indicate the colour flow.}
\label{fig:colour-labels}
\end{figure}

\Pythia[8] is a so-called \emph{dipole} shower.
The construction of such a shower relies on the fact that in a particular representation of the QCD Lagrangian, namely in the colour-flow basis, each particle can be represented by its colour flow (see e.g.\ Refs.~\cite{Maltoni:2002mq,Kilian:2012pz}). 
That means that quarks get a colour flow number, anti-quarks an anti-colour flow number, and gluons are represented with a (different) colour and anti-colour flow number at the same time. 
In this representation, it is straightforward to neglect terms of $1/\Nc$ (at the amplitude level), where \Nc is the number of colours.%
\footnote{
If we were to include those $1/\Nc$ terms in full consistency, we would need to develop a shower that radiates with \emph{QCD multipoles} instead of dipoles.
In addition, since also a virtual exchange of a particle can change the colour state, we would need to keep track of the action of vetoed emissions on the colour state as well. 
This is a highly ambitious goal, which is (in the very least) computationally hard to achieve. 
Nonetheless, \emph{amplitude-level} shower evolution is a field that is under active development. 
}
The contribution that gets ignored in the $N_c \to \infty$ limit is the one where a gluon does not propagate a colour, but instead forms a colour flow at each connecting end. 
By then imagining that we can draw a line between each colour and anti-colour flow number, we create leading-colour dipoles or antennas, from which radiation is generated.
An example of this is seen in \Cref{fig:colour-labels}. 

One of the main motivations for formulating a shower in terms of dipoles is that it correctly incorporates colour coherence effects. 
Showers that are based on independent emissions off individual partons correctly describe the collinear limit, but do not reproduce the interference pattern of soft-gluon radiation (unless a particular ordering is chosen, see below). 
In QCD, soft gluons are emitted coherently by pairs of colour-connected partons, rather than by single emitters, and this coherence leads to a suppression of radiation outside the opening angle of the colour partners.
An alternative way to account for this effect is provided by angular-ordered showers, in which emissions are ordered in decreasing opening angle, which is the ordering chosen by one of the showers that is present in the \mc{Herwig} framework.

\subsubsection{Overview of shower algorithms in general-purpose MC event generators}
\label{sec:shower-overview}

It should come as no surprise that, given the substantial amount of freedom one has in designing a shower algorithm, 
more than one shower formulation is available ``on the market''. 
The three most widely used general-purpose MC generators are \Pythia, \mc{Herwig} and \mc{Sherpa}, and each have their own set of shower implementations (besides many other generational aspects!). 
Here we will give a brief overview of the shower implementations in these MC generators.
More information can be found in their (very complete) manuals. 

We already came across \Pythia~\cite{Bierlich:2022pfr} in our previous discussion. 
The development of \Pythia traces back to early work on Monte Carlo event generators in the late 1970s and early 1980s. 
Its origins lie in the \mc{JETSET} program, which focused on modelling parton showers and the Lund string fragmentation model (see \cref{sec:string}). 
In parallel, the \Pythia program was created to simulate hard scattering processes in high-energy collisions. 
During the 1980s and 1990s these two codes were gradually merged, forming the combined \mc{PYTHIA/JETSET} package~\cite{Sjostrand:1994kzr}. 
Over time the program evolved substantially, culminating in the modern C++ rewrite \Pythia[8].
While earlier versions used virtuality-ordered showers, which does not automatically guarantee colour coherence in absence of additional constraints, \Pythia[8] only supports transverse-momentum ordered showers. 
In its default shower~\cite{Sjostrand:2004ef}, radiation is generated using a dipole-style picture in which each colour-connected pair of partons acts as the source of radiation, while the evolution variable is an approximated version of transverse momentum of the branching. 
Since its original formulation in Ref.~\cite{Sjostrand:2004ef}, \Pythia[8]'s default shower has undergone significant improvements to e.g.~allow for interleaving shower and multi-parton interactions~\cite{Corke:2010yf} (also see \cref{sec:mpi}), and the addition of weak~\cite{Christiansen:2014kba} and hidden-sector~\cite{Carloni:2010tw,Carloni:2011kk} showers.
In addition to the default shower, \Pythia also supports the \mc{Vincia} shower~\cite{Ritzmann:2012ca,Brooks:2020upa}, which is based on an antenna-shower formalism. In this approach, radiation is generated from colour-connected antennae rather than individual partons. 
The \mc{Vincia} shower also supports QED~\cite{Skands:2020lkd} and electroweak (EW)~\cite{Brooks:2021kji} emissions.

The event generator \Herwig was originally developed in the late 1980s to provide a Monte Carlo framework for simulating high-energy particle collisions with a strong emphasis on QCD coherence~\cite{MARCHESINI19841,Webber:1983if,MARCHESINI1988461}. 
Early versions of the program~\cite{MARCHESINI1992465,Corcella:2000bw}, written in Fortran, included an angular-ordered parton shower, where successive emissions are ordered according to their opening angle. 
As explained above, this ordering directly reflects the coherent radiation pattern expected from colour-charged partons in QCD, and thus ensures that soft gluon interference effects are properly captured. 
During the 1990s and early 2000s, \Herwig evolved into \mc{HERWIG} and later \mc{HERWIG++}~\cite{Bahr:2008pv}, which was a major rewrite in C++.
The latest version, \Herwig[7]~\cite{Bellm:2025pcw} unifies the earlier \mc{HERWIG++} development with additional improvements, including automated matching to NLO matrix elements. 
Besides an angular-ordered shower algorithm (with various recoil prescriptions, see e.g.~Refs.~\cite{Gieseke:2003rz, Reichelt:2017hts,Bewick:2019rbu}), \Herwig also provides a dipole shower algorithm~\cite{Platzer:2009jq}. 
The dipole shower uses a transverse--momentum-like evolution variable.
\Herwig also supports QED~\cite{Bellm:2015jjp}, EW~\cite{Masouminia:2021kne} and dark sector~\cite{Kulkarni:2024okx} showers, colour matrix-element corrections~\cite{Platzer:2018pmd,Platzer:2012np}, and the computation of spin correlations~\cite{Knowles:1988vs,Knowles:1988hu,Collins:1987cp,Richardson:2001df}.%
\footnote{Most showers use spin averaged splitting functions, but hard-collinear radiation can change the helicity of their emitters.}

The event generator \Sherpa~\cite{Sherpa:2024mfk} was developed in the early 2000s to provide a flexible and modern framework for simulating high-energy particle collisions. 
It was designed from the outset to combine matrix-element calculations (see also \cref{sec:matching_and_merging}) with parton showers in a consistent way. 
Originally, \Sherpa only supported the Catani--Seymour dipole shower~\cite{Nagy:2005aa,Nagy:2006kb,Schumann:2007mg}, which was specifically designed to ease parton-shower matching. 
Although \mc{Dire}~\cite{Hoche:2015sya} was later added as a dipole-inspired alternative with modified Catani--Seymour kernels, it is no longer supported. Consequently, the Catani--Seymour shower remains the relevant formulation for current \Sherpa applications.\footnote{%
\mc{Dire} was for long also part of \Pythia[8], but has been removed from there since version 8.315.}
\Sherpa also supports QED radiation in the Catani--Seymour shower.%
\footnote{Like \Herwig, \Sherpa has a separate module that simulates QED, based on the so-called YFS formalism~\cite{Yennie:1961ad}.}

\subsection{Matching, merging, and logarithmic accuracy}
\label{sec:matching_and_merging}

The perturbative accuracy of Monte Carlo generators can be improved along four complementary directions:
\begin{itemize}
    \item consistent combination of fixed-order matrix elements with parton showers;
    \item control of the logarithmic accuracy of parton showers;
    \item going beyond leading colour in the parton shower;
    \item and by going beyond pure-QCD corrections. 
\end{itemize}
%

The first direction involves consistently combining fixed-order perturbative calculations with parton showers. 
Hard, wide-angle emissions are described accurately by exact perturbative matrix elements, while the shower handles only process-independent soft and collinear radiation. 
The combination of these two descriptions allows for a more accurate simulation of events and reliable predictions over all kinematic regions.
A key consideration in these procedures is avoiding double counting of emissions and preserving the accuracy of both the hard scattering and the shower evolution.
Matching and merging schemes are now largely automated and applicable in principle to any process and final-state multiplicity, enabling NLO accuracy for most processes. 
These will be the topics of \cref{sec:matching,sec:merging}, respectively,
with \cref{sec:matching_merging_in_event_generators} providing
additional generator-specific details.
Selected techniques extend these ideas to NNLO for specific cases, as discussed in \cref{sec:matching_and_merging_at_higher_orders}.

The second direction focuses on assessing and improving the logarithmic accuracy of the parton shower.
Parton showers resum large logarithms that appear in soft and collinear emissions to all orders in \alphaS.
The accuracy is defined by the number of logarithmic towers correctly reproduced.
To make this concrete, consider again \cref{eq:no-emsn-prob}.
After performing the double integral analytically, one obtains the \emph{Sudakov form factor}
\begin{align}
\Delta(\pT,Q) = \exp\left(-\frac{1}{2}C\ln^2 \frac{\pT}{Q}\right)\,.
\end{align}
This is an example of a resummed expression, where an infinite series of logarithmic contributions is captured in a compact form, rather than order by order in \alphaS, which instead would be appropriate for a fixed-order prediction. 
The logarithmic accuracy is determined by how powers of the coupling (contained in $C$) are associated with the logarithmic term $L$.
For an N$^{k}$LL resummation of the Sudakov, one would need to correctly include terms up to $C^{k-1} L^{k}$.%
\footnote{
The Sudakov resummation is an example of a resummation that follows \emph{leading-logarithmic} counting.
For other observables, such as fragmentation functions, \emph{single-logarithmic} counting is more relevant. 
In that case, N$^{k-1}$SL means correctly reproducing terms of order $C^kL^k$, because the $C L^2$ term is absent in the calculation. 
For yet other observables, the resummation does not organise itself as an exponent, where people refer to the accuracy in \emph{double-logarithmic} counting, where N$^{k}$DL means getting $C^n L^{2n-k}$ for all $n$.
Note that this precise terminology varies between (sub)fields and research groups, so it is always important to clarify the convention being used.
}
Because the Sudakov factor generated by a parton shower has a resummed form, one might expect that a shower can reproduce resummed predictions for physical observables.
But exactly there lies the difficulty: different observables obey different types of resummation, and some do not obey any resummation at all. 
To then prove that a shower is of a certain logarithmic order means that one needs to prove that it would correctly describe \emph{all} observables at that order for which a sensible formulation of resummation can be formulated. 
This will be the topic of \cref{sec:shower_accuracy}.

The final directions to improve the accuracy of Monte Carlo generators are the inclusion of corrections beyond leading-colour QCD or even beyond pure QCD, such as QED and electroweak (EW) effects.
We will not discuss these types of corrections in further detail, but we refer back to \cref{sec:shower-overview} which describes public shower algorithms that partially include such corrections. 

\subsubsection{Matching}
\label{sec:matching}

Matching schemes define how to consistently generate events at NLO accuracy matched with parton-shower emissions. A central requirment is to avoid double-counting between real emission contributions that appear in both the fixed-order NLO calculation and the parton shower.
Another requirement is to preserve NLO accuracy under \alphaS expansion not only in the total cross-section, but also in differential distributions, maintaining the logarithmic accuracy of the parton shower, and ensuring a smooth transition between the hard emission regime and the infrared (i.e.\ soft and/or collinear) region. The two principal approaches that successfully achieve a matching meeting these requirements are MC@NLO~\cite{Frixione:2002ik} and POWHEG~\cite{Nason:2004rx,Frixione:2007vw}.

They both split the $(n+1)$-particle real-emission correction $R$, see \cref{eq:xsnlo}, into an infrared ($S$) and a hard ($H$) part: $R = S + H$.
MC@NLO defines this split by identifying $S$ with splitting kernels
(these can be identified with the ones used by the parton shower~\cite{Hoeche:2011fd}),
which exactly reproduce $R$ in the infrared-singular limits; we are then left with an infrared-finite remainder $H = R - S$.
POWHEG instead uses a function $F$ to define the split with $S=FR$
and $H=(1-F)R$~\cite{Alioli:2008tz}.
The function $F$ is defined to approach 1 in the infrared-singular limits.
The split allows both schemes to apply corrections to the first parton-shower emission of the infrared events only,
ensuring that the NLO emission spectrum is not spoiled at order \alphaS.
Parton-shower emissions from the hard events are left unmodified.

The main conceptual difference is the definition of the partition as defined above:
While MC@NLO uses a \emph{subtractive} procedure to define $H=R-S$,
POWHEG uses a \emph{multiplicative} procedure to define $H=(1-F)R$.
The advantage of the MC@NLO approach is that the one-step shower exponentiates only those contributions that are truly present at all orders in the perturbative series, while
POWHEG only enforces this partially using a suitably defined $F$.
The differences introduced by this, however, are beyond the claimed NLO accuracy of the schemes.
Numerically, they can still be significant, in particular for processes with large higher-order corrections,
such as Higgs production.
Given the freedom of choosing technical parameters that define $F$
(e.g.\ $h_\mathrm{damp}$, as introduced below),
POWHEG also introduces an ``exponentiation uncertainty''.
The advantage of the POWHEG approach is that by using $0 \leq F \leq 1$
one can guarantee that hard real-emission events generated from the $H$ term
(sometimes labelled ``hard'' $\mathbb H$ events) have strictly positive weights,
while MC@NLO implementations typically generate a significant
fraction of negatively weighted $\mathbb H$ events,
although this can be mitigated somewhat by choosing suitable splitting kernels to define $S$
with $S \leq R$ throughout most of the phase-space.
Both MC@NLO and POWHEG can suffer from negative weights
for events that start from an $n$-particle configuration
(sometimes labelled ``standard'' $\mathbb S$ events\footnote{%
    The $\mathbb S$ and $\mathbb H$ event labels, sometimes also written $\mathcal S$ and $\mathcal H$,
    can be confusing: While hard $(n+1)$-particle $\mathbb H$ events correspond to the hard term $H$ in the matched NLO differential cross section,
    the standard $\mathbb S$ events do \emph{not} correspond to the subtraction term $S$ used in MC@NLO,
    but simply to events that start from an $n$-particle configuration matrix element.}),
because the sampled integrand, i.e.~the differential NLO cross-section
(which now includes the $S$ term integrated over the infrared emission),
can become negative.\footnote{%
In principle, both POWHEG and MC@NLO can use folding techniques
to reduce (but not eliminate) the negative-weight fraction
of the $\mathbb S$ events~\cite{Frixione:2007nw}.
Folding means that the integrand
for the one-emission phase-space of the infrared emission
is sampled not just once,
but twice or even more times for each $\mathbb S$ event.
The sum then typically has a lower probability
of being negative.
The disadvantage of this method is the run-time cost
of the additional integrand evaluations.}

\begin{figure}
\centering
\includegraphics[width=0.9\textwidth]{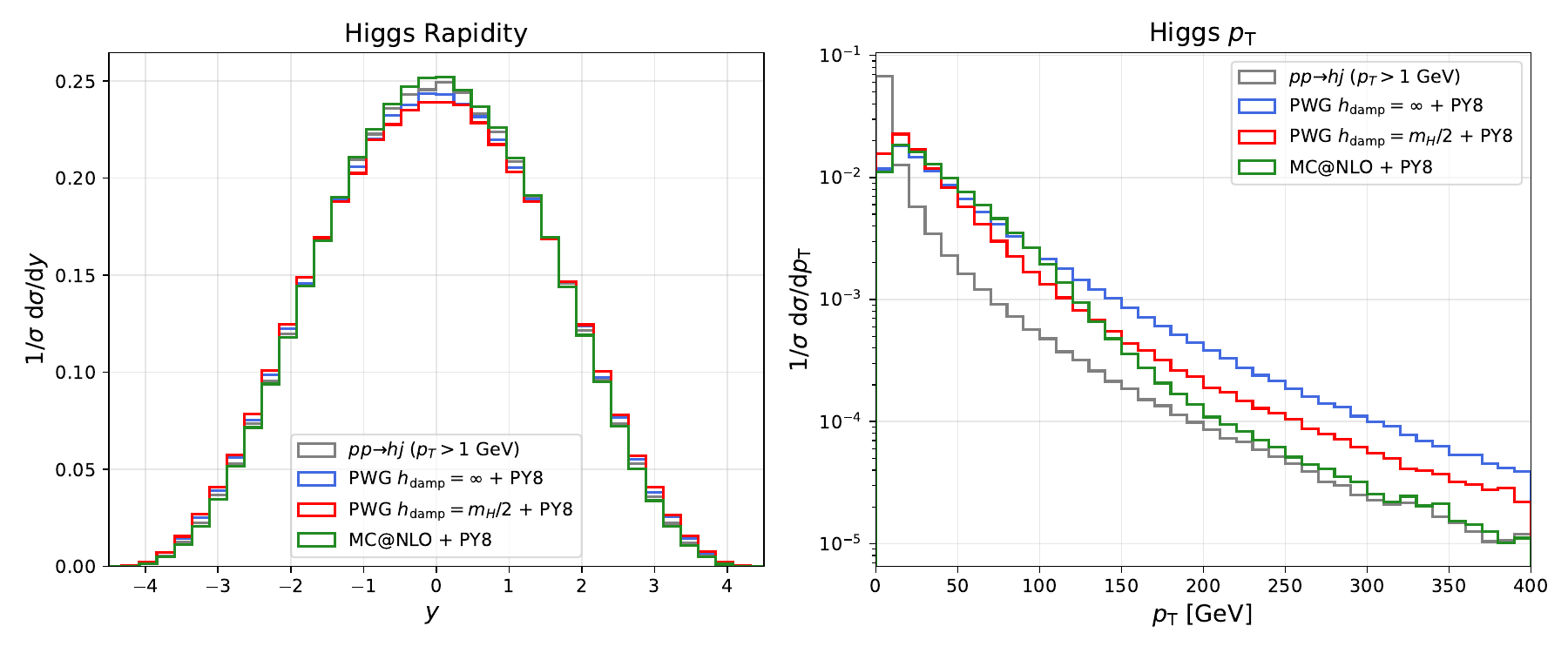}
\caption{%
\label{fig:matching}
A comparison of the MC@NLO and POWHEG (labelled ``PWG'') matching methods
to consistently combine parton shower events
with a calculation at NLO accuracy
for Higgs production events in association with a single jet ($pp \to hj$).
The POWHEG distribution comes in two variants,
$\hdamp = \infty$ and $\hdamp = m_h/2$,
corresponding to two different POWHEG damping parameters
that controls the relative fractions of $\mathbb S$ and $\mathbb H$ events (see main text).
The distributions are compared with a fixed-order calculation
of $pp \to hj$ at leading order.
On the left side, the distribution of the Higgs boson's rapidity, $y$, is shown, while the right plot shows its transverse momentum, \pT.
}
\end{figure}

\Cref{fig:matching} shows a comparison of MC@NLO and POWHEG event samples
for Higgs production $pp \to h$, with a LO calculation
for Higgs production in association with one jet, $pp \to hj$.
The $pp \to hj$ LO calculation does not include any parton showering.
It is equivalent to the real-emission correction $R$ of the NLO calculation
for $pp \to h$.
A cut on the jet $\pT$ is used to remove the singularity at $\pT = 0$.
The left-hand plot shows the normalised distribution of the Higgs rapidity $y$.
The observable is already described well by a LO $pp \to h$ calculation,
and is not sensitive to additional emissions.
We indeed find that all distributions agree with each other.
For the transverse momentum \pT of the Higgs,
shown in the right-hand plot,
the distributions differ.
Note that for a non-zero Higgs \pT, the Higgs needs to recoil against
at least one additional jet, and the observable is maximally sensitive
to the radiation pattern of the hardest few jets.
For lower energies, $\pT < m_h$, it is likely that the recoil is built up
against several jets.
This region cannot be adequately described by the LO $pp \to hj$
calculation, which only includes exactly one jet to recoil against.
In particular at very low energies $\pT \lesssim 30\gev$,
the matched distributions feature the so-called ``Sudakov peak''
(here between 10 and 20\,\gev),
which corresponds to the recoil against many soft emissions
generated by the parton shower.
The LO calculation cannot predict this feature.
The matched distributions, on the other hand,
are in good agreement for $\pT \lesssim m_h$,
since this region is dominated by the emissions of the parton shower
(which is here chosen to be \Pythia[8] for both MC@NLO and POWHEG distributions).
Beyond this, the matched distributions are sensitive to the details
of the matching scheme and parameters.
While the MC@NLO distribution begins to follow the $pp \to hj$ calculation
at this point, the POWHEG distribution's behaviour is sensitive
to the definition of $F$, which is given here by
the damping factor \hdamp:
$$F=\frac{\hdamp^2}{\pT^2+\hdamp^2}.$$
The choice $\hdamp = \infty$ corresponds
to ``standard'' POWHEG, for which all events are $\mathbb S$
events and hence carry the $k$-factor $\bar B / B$,
with $\bar B$ corresponding to the first integrand
of \cref{eq:sigma_NLO}.
With lower $\hdamp$,
more and more events are $\mathbb H$ events,
that do not carry the $k$-factor,
and the distribution becomes more similar to the MC@NLO one.
It should be emphasised again that these differences are formally of higher order;
however, given the particularly large $k$-factor in the case of Higgs production, they can lead to numerically significant effects away from the parton--shower-dominated region.

\subsubsection{Multi-jet merging}
\label{sec:merging}

Multi-jet merging schemes define how to consistently
combine matrix-element based predictions
not just for a fixed final-state multiplicity,
but for a range of final-state multiplicities.
As an example, instead of using the exact matrix element information
for the $pp \to e^+e^-$ process (i.e.~the 0-jet multiplicity),
and using only the parton shower with its logarithmic approximation
for the respective matrix elements for higher jet multiplicities,
one can improve the accuracy of the resulting sample
by evaluating exact matrix elements
for all $pp \to e^+e^- + 0,\ldots,n$\,jets.
These matrix elements can be evaluated at LO or at NLO.
For NLO, the merging scheme must be combined with a matching scheme
such as MC@NLO or POWHEG
to combine the matrix-element calculation and the parton shower
for the given NLO jet multiplicity.
Given the higher computational complexity of NLO calculations,
it is common to compute lower jet multiplicities at NLO accuracy,
while relying on cheaper LO matrix elements for some of the higher multiplicites.
This is often indicated using a notation such as in this example:
``$pp \to e^+e^- + 0,1,2\,\text{jets @ NLO} + 3,4,5\,\text{jets @ LO}$''.
Here,
the first three jet multiplicities are simulated at NLO,
while the higher multiplicities are described at LO.
Even higher multiplicities would be generated by the parton shower approximation,
i.e.~at its given logarithmic accuracy.

However, a parton shower is also run for the lower multiplicities -- otherwise,
its resummation would not be consistently taken into account.
First, the phase space is separated into a ``hard''
and a ``soft'' region.
Then the parton shower is used to populate the soft domain,
while the hard domain is generated by matrix elements.
To define this separation, a criterion is required
to categorise soft and hard emissions,
giving rise to a jet measure $Q$
and a corresponding threshold \Qcut,
which is sometimes called the \emph{merging cut}.
Below this cut, the parton shower is allowed
to freely generate emissions, whereas parton-shower emissions above this cut lead to a veto of the event.

Such vetoes are required to address
a further challenge when
combining matrix elements of different jet multiplicities
is the avoidance of double counting.
Fixed-order calculations
are \emph{inclusive} with respect to additional emissions, which is why they can be consistently supplemented by a unitary parton shower without altering the total cross-section.

When combining samples at different multiplicities,
however, this inclusiveness becomes problematic: each sample can populate overlapping regions of phase space. To prevent double counting, the individual contributions must be made \emph{exclusive},
i.e.~restricted to exactly $n$ resolved emissions,
rather than ``$n$ or more''.

In practice, this can be achieved by applying no-emission probabilities
(Sudakov form factors, cf.~\cref{eq:no-emsn-prob})
to each parton that could produce an additional resolved emission.
Equivalently, one may allow the parton shower to generate trial emissions and veto an event in which such emissions occur.

Identifying the relevant partons requires more than just inspecting the external legs of a given matrix element, since the same final state could also arise from a lower-multiplicity configuration followed by subsequent emissions. To resolve this ambiguity, one reconstructs a plausible branching history by iteratively clustering the external partons according to the inverse splitting probabilities
of the shower~\cite{Andre:1997vh}.
This procedure, often referred to as clustering,
is continued until the underlying core process ($n=0$) is reached.
The resulting sequence of clusterings then defines a consistent parton-shower history for the event.
The no-emission probability is then evaluated for each propagating parton between successive splittings, using the corresponding transverse-momentum scales,
$\pT^{(i)}$ and $\pT^{(i+1)}$,
as input for the Sudakov form factor, $\Delta(\pT^{(i)}, \pT^{(i+1)})$~\cite{Lonnblad:2001iq}.
For external legs, the lower scale is replaced by the parton-shower cut-off scale, $\Lambda$.

\begin{figure}[tb]
\centering
\includegraphics[width=0.6\textwidth]{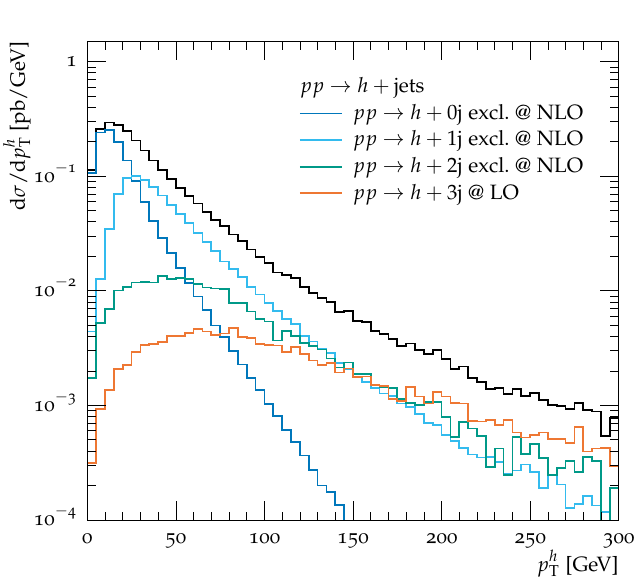}
\caption{%
\label{fig:merging}
Illustration of the different contributions
for different jet multiplicities
when multi-jet merging is used (coloured non-solid lines),
and their combination into a single
inclusive sample (black solid line)
as a prediction for the Higgs transverse momentum $\pT^h$
in Higgs production $pp \to h + \text{jets}$.
For each multiplicity except the highest, a jet veto is used to make the predictions \emph{exclusive} (see main text),
and parton shower emissions are included only below the merging cut; for the highest multiplicity,
$pp \to h + 3j$, is treated normally, i.e.\ it is inclusive and then allowed to shower normally.
The first three jet multiplicities -- 0, 1 and 2 jets -- are described at NLO
within the merged sample, whereas the highest 3-jet multiplicity
is described at LO. Adapted from a figure kindly provided by Frank Siegert.%
}
\end{figure}

\Cref{fig:merging} illustrates the summation of several \emph{exclusive} jet multiplicities
(dressed with the parton shower below the merging cut)
into one inclusive sample for Higgs production via gluon fusion $pp \to h + \text{jets}$.
Note that the highest multiplicity, here $pp \to h + 3\,\text{jets}$ is not rendered
exclusive, and the shower can evolve without being constrained by the merging cut, as there is no more double-counting with higher multiplicites at this point.

The cluster history can also be used to improve the matching
of the soft and the hard region.
The parton shower evaluates the strong coupling
at each splitting scale, $\alphaS(\pT^{(i)})$.
With the cluster history,
we can replace the coupling factor of the hard scattering, $\alphaS(\muR^2)^p$,
where $p$ is the perturbative order in QCD
and $\muR^2$ is the renormalisation scale,
with $\prod_{i=1}^p \alphaS(\pT^{(i)})$.
This mimics the behaviour of the parton shower,
and ensures a smooth matching across the merging cut \Qcut.
This dynamical scale choice addresses a problem
that is not specific to multi-jet merging,
but arises more generally in processes with many final-state jets.
Such configurations often exhibit a hierarchy of splitting scales spanning a wide energy range.
In these situations, assigning a single scale to all emissions inevitably leads to large logarithmic corrections
in the running of \alphaS that are not properly accounted for.
A clustering-based dynamical scale choice resolves this by adapting this scale to the kinematics of each splitting.

The extent to which multi-jet merging improves the predictions depends on the observable under consideration. In general, parton showers excel at describing the \emph{evolution} of individual jets and observables
sensitive to multiple soft emissions,
such as the transverse momentum of the $Z$ boson, $\pT^Z$, in Drell--Yan production
for $\pT^Z \lesssim 20\,\GeV$.

By contrast, observables that probe the \emph{generation} of hard jets
are not well described by the parton-shower approximation alone.
A typical example is the scalar sum of jet transverse momenta, \HT, which is more accurately modelled using exact matrix elements.
However, each fixed jet multiplicity $n$
contributes predominantly in a different region of \HT,
so only a merged sample including several jet multiplicities can provide a reliable description over a wide kinematic range.

Further examples that benefit from multi-jet merging
are inter-jet observables, i.e.~quantities
involving two or more well-separated jets,
such as the azimuthal separation
(often referred to as angular decorrelation), $\Delta\phi$.

Since centrally produced event samples
are typically used for a wide range of analyses,
multi-jet merging techniques are generally employed -- where feasible -- to ensure a robust description across observables.

Several approaches exist for performing multi-jet merging
at tree-level: CKKW~\cite{Krauss:1999mt,Catani:2001cc},
the closely related CKKW-L~\cite{Lonnblad:2001iq},
and MLM~\cite{Mangano:2001xp,Mangano:2002ea}.
A variant of the CKKW(-L) approach is the UMEPS method~\cite{Lonnblad:2012ng,Platzer:2012bs},
which aims to improve the preservation of the parton shower evolution's unitarity
and to reduce the merging-cut dependency,
which can be sizeable in some cases, even though it is formally below the claimed accuracy.
CKKW(-L) has been generalised to NLO by integrating it with
the POWHEG and MC@NLO matching schemes
in~\cite{Hoche:2010kg} and
in~\cite{Gehrmann:2012yg,Hoeche:2012yf}, respectively.
For CKKW-L and UMEPS, this has been achieved for POWHEG matching in~\cite{Lonnblad:2012ix}, and for MC@NLO matching in~\cite{Platzer:2012bs,Bellm:2017ktr}. With that,
they have been dubbed NL$^3$ and UNLOPS, respectively.
Finally, MLM merging has been combined with NLO using POWHEG
matching in~\cite{Hamilton:2010wh}.
It is worth emphasising again that, at
both LO and NLO, all of these methods differ only in their treatment
of contributions that are formally of higher order.

Another widely used method is MINLO~\cite{Hamilton:2012np}, which is neither a matching nor a merging method.
Instead, it is a method to calculate processes with associated jet production at NLO
and to yield meaningful physical distributions also at low energies
by including Sudakov form factors.
This is done by treating the Born and the and real-emission terms
as multiplicities in a CKKW-like merging at LO.
When combined with a shower using
either POWHEG or MC@NLO as a matching method, it is referred to as MINLOPS.

\subsubsection{Overview of matching/merging algorithms in
general-purpose event generators}
\label{sec:matching_merging_in_event_generators}

\Pythia supports
POWHEG and MC@NLO matching
via interfaces to the \mc{POWHEG-BOX}~\cite{Alioli:2010xd}
and
\mc{MG5aMC}~\cite{Alwall:2014hca,Frederix:2018nkq} frameworks, respectively.
Multi-jet merging is directly available using the CKKW-L,
UMEPS, NL$^3$ and UNLOPS approaches.
The MLM approach is also available in combination with 
\mc{MG5aMC}, which additionally supports FxFx merging at NLO~\cite{Frederix:2012ps}, which uses MC@NLO-like matching
and combines elements of the CKKW, MINLO and MLM merging methods.

The event generator \Herwig
implements MC@NLO and POWHEG matching
for both its angular-ordered and its dipole parton showers.
When used in combination with colour-single processes, the latter
additionally supports the KrkNLO matching method~\cite{Jadach:2016qti,Sarmah:2024hdk}.
Multi-jet merging at LO and NLO is available
for the dipole shower,
using a unitarity-based merging algorithm similar to the UMEPS and UNLOPS methods,
respectively,
as described in~\cite{Platzer:2012bs,Bellm:2017ktr}.
When using
\mc{MG5aMC} as a matrix-element provider, 
one can use MLM merging at leading order, and additionally FxFx merging at NLO (see below).
\Herwig's matching and merging implementations
are closely integrated within its \mc{Matchbox} module.
For details, see~\cite{Bellm:2025pcw}.

Finally, \Sherpa
implements a variant of MC@NLO, S-MC@NLO~\cite{Hoeche:2011fd}.
The NLO multi-jet merging procedure implemented
in \Sherpa
is called MEPS@NLO~\cite{Gehrmann:2012yg,Hoeche:2012yf},
which uses S-MC@NLO and is based on the CKKW-L approach.

\subsubsection{Toward higher orders}
\label{sec:matching_and_merging_at_higher_orders}

NNLO matching (and, by extension, merging)
is much more involved compared to NLO,
and while many process-specific results exist,
there is currently no full generalisation and automation as in the NLO case.

An early approach was dubbed NNLOPS~\cite{Hamilton:2012rf}.
MINNLOPS~\cite{Monni:2019whf} is
an extension of the MINLOPS method, while
an extension of the UNLOPS method to NNLO accuracy
was proposed in~\cite{Lonnblad:2012ix}
and first implemented and applied in~\cite{Hoche:2014uhw}.
\mc{GENEVA}~\cite{Alioli:2013hqa} is another approach,
which works similar to UNLOPS and combines NNLO
with analytic (N)NLL results before adding shower emissions.
Since the original formulation of these methods,
they have been refined~\cite{Monni:2020nks,Alioli:2021qbf,Alioli:2023har,Gavardi:2025zpf}
and applied to many important LHC processes~\cite{Lombardi:2020wju,Lombardi:2021wug,Lombardi:2021rvg,Buonocore:2021fnj,Zanoli:2021iyp,Haisch:2022nwz,Gavardi:2022ixt,Lindert:2022qdd,Mazzitelli:2020jio,Mazzitelli:2021mmm,Mazzitelli:2024ura,Mazzitelli:2023znt,Niggetiedt:2024nmp,Biello:2024vdh,Biello:2024pgo,Biello:2026nhj,Alioli:2015toa,Alioli:2019qzz,Alioli:2020fzf,Alioli:2020qrd,Alioli:2021egp,Cridge:2021hfr,Alioli:2022dkj}.
More recently, a new approach has been presented for matching
sectorised parton showers~\cite{Campbell:2021svd,El-Menoufi:2024sys}.

There have also been steps toward extending UNLOPS to 3-loop order,
dubbed the TOMTE method~\cite{Prestel:2021vww,Bertone:2022hig}.



\subsubsection{Logarithmic accuracy}
\label{sec:shower_accuracy}

As explained in the introduction of \cref{sec:matching_and_merging}, it is in general not trivial to assess the logarithmic accuracy of a parton shower. 
Formal requirements to claim any logarithmic accuracy of shower algorithms have been written down by the \Panscales project~\cite{Dasgupta:2018nvj, vanBeekveld:2023ivn}, which state:
\begin{enumerate}
    \item Ensure that the shower produces the correct factorised matrix elements in the limits appropriate to acquire a certain logarithmic accuracy; \emph{and}
    \item numerically prove that a parton-shower algorithm reduces to the same results as a resummed (semi-)analytical calculation when all higher-order spurious corrections of the former are removed.
\end{enumerate}
By now, we know that any dipole shower with \emph{dipole-local} momentum conservation and transverse momentum ordering, i.e.\ basically all showers that are contained in public general-purpose event generators, are leading-logarithmic (LL) accurate.%
\footnote{Note that a matched LL shower is still LL. 
Logarithmic accuracy refers to the all-order accuracy of showers, whereas matching improves their fixed order expansion up to a certain power in the coupling.}
The reason is relatively simple to understand: these showers basically identify the wrong emitter, and therefore distribute their recoil in a non-local way that is not in agreement with predictions from resummation. 
A notable exception is the angular-ordered shower in \Herwig, which has next-to-leading logarithmic (NLL) accuracy for global-event shapes like thrust in $e^+e^-$ collisions (because the resummation for those observables directly follows from colour coherence), but not for non-global observables, which are observables that only depend on radiation in part of the phase space~\cite{Bewick:2019rbu}. 

The development of showers that are NLL accurate for the complete range of observables started by considering events of the type $e^+ e^- \to 2j$, for which various groups have found solutions for constructing dipole showers that are NLL accurate~\cite{Dasgupta:2020fwr,Forshaw:2020wrq,Nagy:2020dvz,Herren:2022jej,Assi:2023rbu,Preuss:2024vyu}. 
For simple $pp$ processes like $pp \to X$, where $X$ is a collection of final-state particles that are all colour-neutral, two groups currently have shower algorithms that are NLL accurate: \Panscales~\cite{vanBeekveld:2022ukn,vanBeekveld:2022zhl} (with two implementations, \mc{PanGlobal} and \mc{PanLocal}) and \mc{Alaric}~\cite{Hoche:2024dee} (which will become part of \Sherpa in the near future).
NLL accuracy for showers for processes with more than two coloured particles at Born level has not been explicitly proven by any group, but it is to be expected that the aforementioned algorithms are NLL accurate (at leading colour) for those processes as well. 
%
Steps towards NNLL showers,
which require new ingredients in the shower,
such as higher-order splitting kernels and adjustments to the virtual corrections
that are generated by the shower,
have been taken in \cite{Hoche:2017iem, Dulat:2018vuy, Gellersen:2021eci, FerrarioRavasio:2023kyg, vanBeekveld:2024wws}.

\subsubsection{Matching and merging for other resummation schemes}

The resummation of large logarithmic terms that appear at all orders
when approaching the soft and/or collinear limits
is of particular importance for event generation at hadron colliders.
The reason is that the numerical evaluation using parton showers
allows connecting the generated final state
to phenomenological models of soft physics such as hadronisation and thus eventually
enables the generation of events that can directly be compared to experimental data.
However, at the other end of the spectrum there is another limit
that is phenomenologically relevant,
e.g.\ for weak-boson scattering and weak-boson fusion production of a Higgs boson.
This limit is the high-energy limit
where logarithms of the large ratio $\hat{s}/\pT^2$ appear.
Here, $\hat s$ is the square of the partonic centre-of-mass energy
and \pT is the characteristic transverse momentum scale of the process.
As for the soft or collinear limits, these logarithms spoil the convergence
of the perturbative series, which can be cured by resumming them.

The specialised event generator framework \mc{HEJ}
evaluates this resummation numerically
and can generate fully differential event samples~\cite{Andersen:2019yzo,Andersen:2021qma,Andersen:2023kuj}.
The combination of both resummations, high-energy and soft or collinear,
has been described in Ref.~\cite{Andersen:2017sht}.

\section{From partons to hadrons}
\label{sec:hadrons}

\responsible{Andy}

The previously described perturbative event generation steps
produce partonic events whose final states are composed of quarks, gluons, leptons and photons. The latter two of these are effectively stable at this point and can be ignored in what follows. But these are not realistic events: QCD confinement tells us that the partonic final state cannot be the end of the story, and modelling is required to capture the transition to primary hadrons, followed by their decays. Furthermore, when colliding hadrons (or even quasi-real photons), events with more than a single parton-parton scattering can occur, as can soft ``collective'' physics effects that may involve the whole hadrons.

These aspects are significantly non-perturbative and hence are further from \emph{a priori} quantum field-theory calculations than the modelling discussed so far. Instead, their current treatment is typically handled by phenomenological models\footnote{Requiring \emph{tuning} -- see \cref{sec:tuning}.} led by the asymptotic behaviours and sometimes group structures of the theory. In the final state, the transition to hadrons also acts as a semantic barrier between the event as an intrinsically quantum mechanical object\footnote{Components of the partonic event record may not have unambiguous physical meaning -- they may reflect the internals of the calculation rather than observables, and give specific values to objects which represent integrals or admixtures.}, to an on-shell, semi-classical picture where essentially classical hadrons free stream and decay, and particle momenta are physically reliable. This distinction is central to modern approaches to event analysis, which emphasise minimal use of the pre-hadronisation partonic event in the interests of physical meaningfulness.




\begin{figure}[tp]
\centering
\begin{equation*}
  \begin{array}{ccc}
\mbox{{Fragmentation}} & = & 
\left\{
  \begin{array}{c} \\[-3mm]
    \mbox{(Final-state) parton showers}\\[1mm]
    \mbox{{Hadronisation}}\\[1mm]
    \mbox{{Prompt hadron (and $\tau$) decays}}\\[1mm]
    \mbox{\color{gray}\emph{Non-prompt decays}}\\[1.5mm]
\end{array}\right.\\[1mm]
\end{array}
\end{equation*}\vspace*{-2mm}
\caption{Unless otherwise stated, we use the term \emph{fragmentation} to cover the combined effect of (perturbative) showering and (non-perturbative) hadronisation of hard partons. Also included are the effects of hadron (and $\tau$) decays, divided somewhat arbitrarily into so-called prompt and non-prompt decays. The latter (e.g.\ $K_S^0\to \pi\pi$) may or may not be included in what one defines as \emph{fragmentation} and hence are shown grayed out here.\label{fig:frag}}
\end{figure}

\index{Fragmentation}\index{Hadronisation}Be aware that the terms \emph{fragmentation} and \emph{hadronisation} are often used interchangeably in the literature. Here, we will attempt to stick to the following more specific terminology -- illustrated in \cref{fig:frag} -- as much as possible: 
\begin{itemize}
\item We will use the term \index{Fragmentation}\emph{fragmentation} to refer to the combined effects of (final-state) parton showering plus hadronisation. This is consistent with how the term is used, e.g.\ in the context of parton-to-hadron fragmentation functions in fixed-order QCD.
\item We will use the term \index{Hadronisation}\emph{hadronisation} to refer to the non-perturbatively modelled process of (post-shower) partons turning into hadrons. This is consistent with how the term is used in the context of MC hadronisation models.
\end{itemize}
In the context of hadronisation models, we will also distinguish between \index{Primary hadrons}\index{Hadron production}\emph{primary hadrons} and \index{Secondary hadrons}\index{Hadron production}\emph{secondary hadrons}.  Primary hadrons refer to ones that are produced directly from the hadronising (i.e.\ post-shower) partons, while secondary hadrons refer to hadrons that are produced by decays of other (primary or secondary) hadrons.\footnote{As distinct from the experimental use of ``secondary particles'' to mean those produced by interactions of any particles from the beam collision with detector material.} This classification is specific to the hadronisation stage, where the relevant degrees of freedom are colour-singlet hadrons that have effectively decoupled from the perturbative part of the event.

In experimental contexts, a related but not identical terminology is used, based on the concepts of \emph{prompt} and \emph{non-prompt}. These labels refer to the spacetime properties of particle production and decay, and hence to what can be experimentally resolved. For decays, 
\begin{itemize}
\item \index{Prompt hadron decays}\index{Hadron decays}\emph{Promptly decaying hadrons} are ones that occur over distance scales that cannot be  resolved by any detector. The class of promptly decaying particles typically include ones that can decay strongly (e.g.\ $\rho \to \pi\pi$) or electromagnetically (e.g.\ $D^* \to D\gamma$). 
\item \index{Non-prompt hadron decays}\index{Hadron decays}\emph{Non-prompt decays} are ones that take place over distance scales that may be resolvable as displaced vertices and/or for which the decaying particle may interact with beam-pipe, detector, or shielding material before it decays. The class of non-promptly decaying particles typically include ones that can only undergo weak decays and hence have decay distances of order millimetres or more (e.g.\ kaons, hyperons, the lowest-lying charm and beauty hadrons, and potentially particles from beyond-SM models). 
\end{itemize}
For production, 
\begin{itemize}
\item \index{Hadron production}\index{Prompt hadron production}\emph{Promptly produced particles} include the ones that would be called primary hadrons in a modelling context, \emph{plus} any secondaries that are produced solely via prompt decays. For example, $D$ mesons created by $D^* \to D\gamma$ decays of prompt $D^*$ mesons, would be labelled prompt $D$ mesons.
\item \index{Displaced hadron production}\index{Hadron production}\emph{Non-promptly produced particles} are simply ones that are produced via non-prompt decays. For example, this label would be given to $D$ mesons produced from decays of $B$ mesons. In hadronisation studies, this class should ideally be treated separately, as it consists 100\% of secondaries and hence is not directly sensitive to the primary hadronisation dynamics. (It may of course still be of high interest for the reconstruction of decay chains of higher-lying promptly produced particles.)
\end{itemize}

In LHC experiments, a typical value for which particles are categorised as decaying promptly are ones with $c\tau \le 10~\mathrm{mm}$. There are also a few unstable particles which tend to be treated as \index{Hadron decays}\index{Stable hadrons}\emph{stable} by default in collider contexts since they typically decay outside the detector. These include charged pions, $K^0_L$, muons, and neutrons.

It is important to note that \emph{promptness} does not uniquely specify the origin of a particle in the event history. In particular, particles produced in hadron decays may still be classified as prompt if the decay occurs sufficiently close to the interaction point. For this reason, it is often useful to also introduce the notion of \emph{direct} particles, defined as those that do not originate from hadron decays and can therefore be more closely associated with the hard scattering process. This terminology is further discussed in the section on truth-level analysis, \cref{sec:truth-level-defns}.

\subsection{The physics of hadronisation}

Hadronisation imposes the effects of QCD confinement on post-shower partons. 

\index{Confinement}Confinement mandates that colour-charged particles can only exist in combinations that are overall colour neutral, when probed at distances larger than the \index{Confinement scale}confinement scale,
\begin{equation}
\Lambda_\mathrm{QCD} ~\sim~ 200~\mev~~~~~\leftrightarrow~~~~~ r_\mathrm{\Lambda} ~=~ \frac{\hbar c}{\Lambda_{\mathrm{QCD}}}~\sim~1~\mathrm{fm}~.
\label{eq:rLambda}
\end{equation}
It is important to note that we are talking about a non-perturbatively strong effect that coloured particles exert \emph{upon each other}. Consider a $b$-quark that undergoes hadronisation. There must be at least one other particle, somewhere, which has the ``opposite'' colour charge to our $b$-quark. It is that opposite colour charge which will exert a confinement force on the $b$-quark  (and vice versa), once the two become separated by more than $\sim$ one femtometre. Generalised to arbitrary systems of partons, this means that we can only meaningfully consider the hadronisation of parton systems which are \emph{overall colour neutral}, or equivalently whose colour charges -- considered as vectors in SU(3) colour space -- sum to zero. We say that such parton systems are in an overall \index{Colour-singlet state}\emph{colour-singlet state}. 

It is therefore unphysical to consider the hadronisation of a single quark or gluon. The simplest physical system we can consider is that of a quark-antiquark pair in an overall colour-singlet state. The textbook example is: 
\begin{equation}
    e^+e^- ~\to~ \gamma^*/Z ~\to~ q\bar{q} ~\to~ \mathrm{shower}~\to~\mathrm{hadrons}\,.
\end{equation} 
In this process, studied intensively at $e^+e^-$ colliders, the initial state is colour neutral. The final state must therefore also be in an overall colour-singlet state. \Eg, we can write the ``colour-space wave function'' of the hard $q\bar{q}$ pair as:\index{Colour-singlet state}
\begin{equation}
\frac{1}{\sqrt{3}}\Big(\left| \mathrm{R} \overline{\mathrm{R}}\right>+\left|\mathrm{G} {\overline{\mathrm{G}}}\right>+\left|\mathrm{B} \overline{\mathrm{B}}\right>\Big)\,.
\end{equation}
After the parton shower, we will generally have a much more complicated parton system -- a much more complicated set of colour charges. By colour conservation,  we know that this system will still have zero \emph{total} colour charge, and by explicitly tracing the colour flow through each parton-shower branching, we may also have at least partial information (typically limited to the so-called \index{Colour flow}\emph{leading-colour approximation}, see \cref{sec:showers}) about the \index{Colour correlations}\emph{colour correlations} between the individual partons that make up the post-shower state.
Any explicit hadronisation model must now address the following questions:
{\it
\begin{enumerate}
\item\label{item:whichPartons} Which partons will exert confinement forces upon one another?
\item\label{item:whatDynamics} What dynamical effects do these forces give rise to?
\item How do these  effects ``convert'' a system of post-shower partons into a system of hadrons?
\item\label{item:howMC} How can the resulting physics model be cast (as conveniently and efficiently as possible) as a numerical algorithm suitable for implementation in an MC event generator?
\end{enumerate}}
All current parton showers that can be interfaced to hadronisation models follow similar rules for \index{Colour flow}``colour flow'', which are based on the leading-colour (LC) approximation  outlined in \cref{sec:showers}. Efforts are underway to construct parton showers that can go beyond this approximation but the interface of such showers to hadronisation models is still at the proof-of-concept stage. 
In the LC approximation, each new colour-anticolour pair that is created is treated as distinct from all others\footnote{\index{Leading colour limit}\index{LC|see{Leading colour limit}}This is formally equivalent to letting the number of colours, \Nc, go to infinity, so that the stochastic probability of ever picking the same colour twice vanishes.}. This implies that each colour charge in a parton-level event will be matched by a single unique anticolour. Possible modifications to this picture will be discussed separately under colour reconnections, in \cref{sec:CR}, but for now we shall stick to the simplicity of leading colour, as that suffices for the present discussion and corrections to it are parametrically of order $1/\Nc^2\sim 10\%$.

Within the LC approximation, the answer to question \ref{item:whichPartons} above is simply that it is these uniquely assigned colour-anticolour pairs which should exert confinement forces upon each other. This aspect is common to both string and cluster hadronisation models. Before discussing the different approaches these models take to questions~\ref{item:whatDynamics} to~\ref{item:howMC} in \cref{sec:string,sec:cluster}, respectively, we make two more physical observations that are relevant to hadronisation and which are known model-independently: \emph{preconfinement} and \emph{linear confinement}. We note that cluster models are heavily inspired by the former, while string models are constructed to be explicit realisations of the latter. 

Incidentally, both preconfinement and linear confinement exhibit aspects of simple laws arising from complex microphysics. This is sometimes referred to as \index{Emergence}\emph{emergence}. 

\subsubsection{Preconfinement} 
\label{sec:preconfinement}
\index{Preconfinement}Preconfinement refers to an intriguing and quite beautiful property of parton showers~\cite{Amati:1979fg,Bassetto:1979vy}. When a gluon undergoes a $g\to q\bar{q}$ splitting during a parton shower, colour conservation implies that the produced \index{Colour- octet state}$q\bar{q}$ pair must be in a {colour-octet state (since the gluon is an octet). In SU(3) group-theory language, it must belong to the octet representation in  $3\otimes \bar{3} = 8 \oplus 1$, not the singlet.

Since the produced $q\bar{q}$ pair is \emph{not} in a colour-singlet state, there will \emph{not} be a confining force between the pair, even if they get separated by large distances.\footnote{In fact, any non-perturbative force between them would be mildly repulsive.} \index{Colour singlets}Thus, the $g\to q\bar{q}$ splitting effectively \emph{separates} what was originally a single overall colour-singlet system into two smaller, separate, colour-singlet subsystems.\footnote{Beyond LC, we note that there is a probability of order $1/\Nc^2 \sim 10\%$ that further perturbative gluon emissions could bring them back into a possible singlet state, ignored here.}

What do we mean by \emph{smaller}? Consider the hard  $q\bar{q}$ pair from our example process, $Z^0\to q\bar{q}$. From colour and 4-momentum conservation, respectively, we know that this $q\bar{q}$ pair is in an overall colour-singlet state and that it has a total invariant mass of $M_Z$. Suppose that this hard $q\bar{q}$ pair emits a gluon, which subsequently splits into another quark-antiquark pair, $q'\bar{q}'$. By the LC colour-flow rules, the system now consists of \emph{two} separate colour-singlet subsystems, $(q\bar{q}')$ and $(q'\bar{q})$, whose invariant masses are bounded by $m^2_{q\bar{q}'} + m^2_{q'\bar{q}} \le m_Z^2$. 

If we continued the parton shower without end, gluon emissions interspersed with occasional $g\to q\bar{q}$ splittings would continue to occur, stochastically, inside each of these two subsystems, splitting them further into ever more and ever smaller colour-singlet subsystems, in a process reminiscent of \index{Fractal structures}fractals. 

Eventually, we might suppose that more and more of these subsystems would have invariant masses small enough to be identified with hadrons. However, since the shower splitting kernels for $g\to q\bar{q}$ are derived perturbatively, they have no knowledge of the spectrum of hadrons. What we end up with, after the parton shower, is a set of colour-singlet subsystems that have a continuous mass distribution, with no hadronic resonance peaks. Nevertheless, we see that perturbation theory does ``break up'' an original high-mass colour-singlet system into many smaller ones. 

\index{Universalities}Furthermore, due to the quasi-fractal nature of this process, the invariant-mass spectrum of the resulting colour-singlet subsystems is approximately \emph{universal}, independent of the starting process. A larger invariant-mass starting point simply gives rise to \emph{more} colour-singlet subsystems after the shower, than a lower-invariant-mass starting point would. But the invariant-mass spectra of the colour-singlet subsystems they produce (normalised to the same area) are approximately the same, as long as the respective shower starting scales are not too close to the IR cutoff of the shower. This property of perturbative cascades to break high-mass colour-singlet parton systems into ever smaller colour-singlet subsystems, with a universal invariant-mass spectrum, is called \index{Preconfinement}\emph{preconfinement}. 

Note that, although it contains the word \emph{confinement}, preconfinement is an entirely perturbative effect which, a priori, has no non-perturbative input. Thus, to turn this into a hadronisation model, more input will be needed. This is the starting point for  \index{Cluster models}\index{Hadronisation!Cluster models}cluster models of hadronisation, discussed in \cref{sec:cluster}.

\subsubsection{Linear confinement} 
\label{sec:linearConf}

\index{Linear confinement}\index{Confinement!Linear confinement}Linear confinement refers to the assumption that the non-perturbative confinement force between two oppositely colour-charged particles can be modelled by a linearly rising term in the effective potential between the charges. The so-called \index{Cornell potential}Cornell potential~\cite{Eichten:1974af} is an empirical formula which expresses this:
\begin{equation}
V_{q\bar{q}}(r) ~=~ -\frac{a}{r} \,+\, {\kappa r} \,+\, V_0\,, \label{eq:Cornell}
\end{equation}
where $r$ is the distance between the two colour charges, $a$ is proportional to an effective strong coupling which represents the strength of the (short-distance) Coulomb part of the potential, $\kappa$ is the slope of the linear confinement term, and $V_0$ is a constant that allows for a ground-state-energy offset.

As it stands, the Cornell potential is just a formula. How can we know that it has anything to do with the forbiddingly complicated non-perturbative dynamics of QCD confinement? 

The main first-principles theoretical reason  comes from \index{Lattice QCD}lattice QCD. On the lattice, one can place two fixed colour sources a distance $r$ apart, and then compute the potential energy between them directly. Repeating this for many different distances, one can ``map out'' the shape of the potential between them. Such studies (e.g.~\cite{Bali:2000un}) exhibit excellent agreement with \cref{eq:Cornell} and yield a value for $\kappa$ of typical hadronic dimensions, 
\index{String tension}%
\begin{equation}
{\kappa_\text{Lattice}}~\sim~  0.94\,\gev/\fm~\sim~0.185\,\gev^2
~~~~~\leftrightarrow~~~~~
\sqrt{\kappa_\text{Lattice}} ~\sim~ \SI{430}{\MeV}~(\pm\sim\SI{10}{\MeV})\,.
\end{equation}
Two contextual remarks are in order: 
\begin{itemize}
\index{Quenched approximation}\item Many of these lattice studies are done in the so-called ``quenched'' lattice approximation, without dynamical quarks. More advanced studies (such as \cite{Bali:2005fu}), however, do not appear to change the fundamental conclusion that the effective confinement potential is linear.
\item The physical situation simulated on the lattice, with two fixed (immovable) colour sources a fixed distance $r$ apart, is not quite the same as the process of hadronisation, in which the colour sources are flying apart close to the speed of light. 
\end{itemize}
Especially the latter -- the difference between the static, steady-state situation simulated on the lattice and the dynamic, time-dependent process of hadronisation of relativistic partons -- may provide guidance for future revisions of linear-confinement-based hadronisation models (see e.g.~\cite{Hunt-Smith:2020lul}), but for now, we merely note it as a potential caveat.

Other pieces of evidence for linear confinement come from the observation of hadronic mass spectra (e.g.~for charmonium states) that are consistent with the state spectrum of \cref{eq:Cornell}; and from the observation of linear relations between hadronic mass squares and spins -- called \index{Regge trajectories}Regge trajectories.  

More complicated to interpret but more directly sensitive to the \emph{dynamics} of hadronisation, are observations of so-called \index{Rapidity plateaus}rapidity plateaus  of roughly constant particle densities being produced between colour-connected partons (along the rapidity axis spanned by them) in collider experiments -- interpreted as evidence of a constant energy density between them, as would be the case for a constant slope, $\kappa$. 

\index{JADE effect}\index{String effect}The observation of the so-called JADE (or ``string'') effect~\cite{JADE:1981ofk,JADE:1983ihf,Azimov:1985zta} also deserves mention in this context. Briefly stated, this was an observation of higher particle densities between colour-connected partons than between colour-non-connected ones. This mainly demonstrates that confinement is indeed an effect that colour-connected partons have \emph{upon each other} -- as opposed to hadronising independently of each other. It does not follow directly from this observation that the effective confinement potential \emph{has} to be  \emph{linear}, but hadronisation models based upon this assumption are in good agreement with data.

The assumption of linear confinement provides an immediate answer to question~\ref{item:whatDynamics} in our list above: \emph{what dynamics do the confinement forces give rise to?} 
The physical system that is described by a linear potential is called a \emph{string}. In this context, the linear slope, $\kappa$, is called the \index{String tension}\emph{string tension}. This is the starting point for \index{String models}\index{Hadronisation!String models}string models of hadronisation. 

The gradient of the linear term in \cref{eq:Cornell} is just the constant, $\kappa$. Hence, modulo the (short-distance) Coulomb term in \cref{eq:Cornell}, we see that hadronising partons will experience a constant force, oriented in the direction of their colour partner. Since the partons are separating, this amounts to the partons exerting a constant deceleration upon each other.

For ultra-relativistic partons (with $E \gg m$ as is typical in high-energy collider contexts), their velocities will remain close to $c$ until they have shed almost all their momentum. Hence, the ``deceleration'' of these partons by the confinement field should not be thought of so much as producing significant changes to their actual velocities but as producing a constant energy loss per unit time. For the simple example of a colour-singlet $q\bar{q}$ pair,
\begin{equation}
\frac{\d V_{q\bar{q}}(r)}{\d r} ~\stackrel{{\color{midteal}r\gtrsim r_\Lambda}}{\sim}~  {\kappa}~~~~~{\Rightarrow}~~~~~\frac{\d E_q}{\d t}~\stackrel{{\color{midteal}E_q\gg m_q}}\sim~ {-\kappa}\thinspace c\,,
\label{eq:stringEOM}
\end{equation}
where $r$ is the distance between the quark and antiquark which we take to be larger than the confinement scale $r_\Lambda\sim1~\mathrm{fm}$ defined in \cref{eq:rLambda}, $E_q$ ($m_q$) is the energy (mass) of one of the quarks, and we have inserted an explicit factor of $c$ in the last expression, as a reminder of the quark velocity, which we take to be $v_q \approx c = 1$ in natural units. 

Partons losing energy while remaining at near-constant (speed-of-light) velocity may seem counter-intuitive. If so, it may help to think of the constant energy loss as producing an increasing redshifting of the partons with time, $\frac{1}{\lambda}\d\lambda/\d t \propto {\kappa} \lambda$.\,\footnote{For comparison, standard Hubble expansion would translate to a wavelength-\emph{independent} $\frac{1}{\lambda}\d\lambda/\d t$.}

So far, we have focused on a simple string spanned directly between a quark and an antiquark endpoint. Perturbative processes obviously also produce gluons. How are they handled in a string context? At leading colour, gluons can be thought of as carrying one colour and one anticolour charge, cf.\ the discussion in \secRef{sec:showers}. Group theoretically, this can be seen by considering the direct product of a colour triplet and an anticolour triplet, $3 \otimes \bar{3} = 8 \oplus 1$. Neglecting the singlet (leading colour), we see that we can express an octet (i.e.\ gluon) colour charge as $8 \approx 3\otimes \bar{3}$. Thus, a gluon created during the perturbative phase of the event evolution can be thought of as the endpoint of one piece of string and the beginning of another. That is to say that gluons can be identified with ``kinks'' on strings. The space-time evolution of a string configuration for a $e^+e^- \to qg\bar{q}$ event is pictured in \figRef{fig:kink}.
\begin{figure}[t]
\centering
\includegraphics*[width=\textwidth]{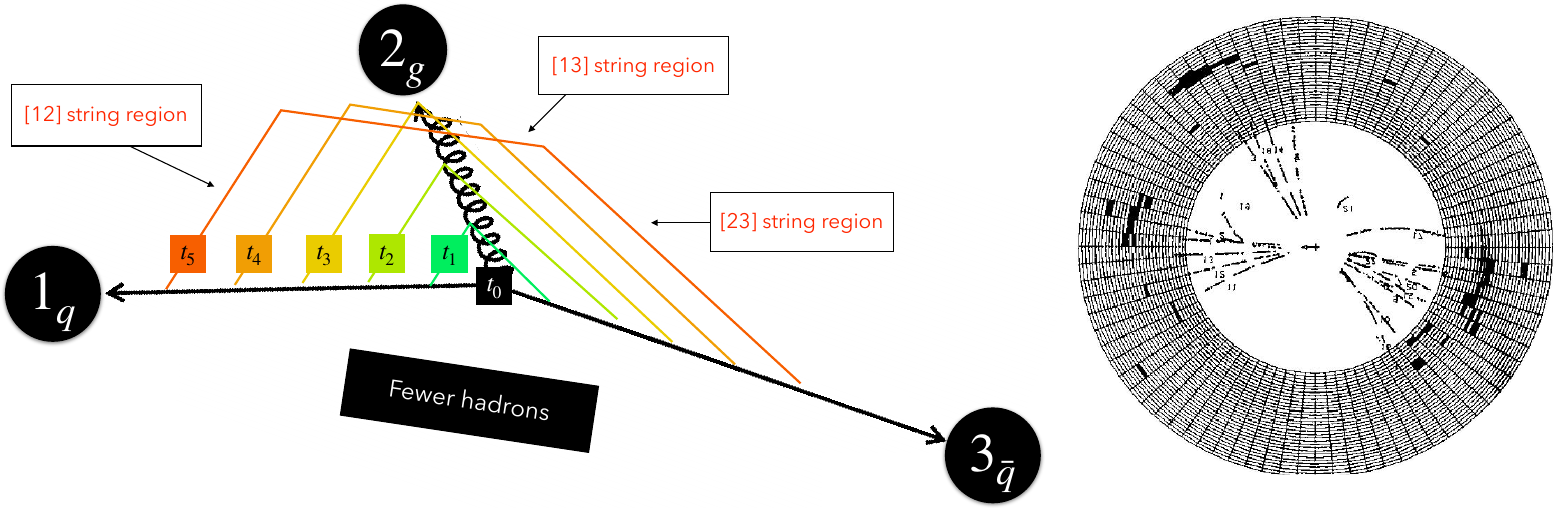}
\caption{{\sl Left:} Snapshots at different times ($t_0$, $t_1$, $\ldots$, $t_5$) of a string configuration corresponding to an $e^+e^-\to qg\bar{q}$ event. Fewer hadrons will be produced in the angular region between the $q$ and $\bar{q}$ jets than between the quark and gluon jet pairs. {\sl Right:} a real 3-jet event recorded by the JADE experiment~\cite{JADE:1981ofk,JADE:1983ihf,Bethke:2022cfc}.
\label{fig:kink}}
\end{figure}
Note that since gluons are connected to \emph{two} string pieces, one on each ``side'', they lose energy at twice the rate that quarks do~\cite{Andersson:1980vk}, with 
\begin{equation}
\frac{\d E_g}{\d t} ~\sim~ -\,\frac{C_A}{C_F}\,\kappa\, c ~\sim~ -2\,\kappa\,c~. 
\end{equation}
The ratio of the gluon to quark colour factor, $C_A/C_F$, is a special case of a property referred to as \index{Casimir scaling}\emph{Casimir scaling}~\cite{Bali:2000un}: the eigenvalues of the so-called quadratic Casimir operator measure the magnitude of the effective colour charges for each different representation of QCD. Quarks (triplets) have colour charge $C_F = 4/3$, while gluons (octets) have $C_A = N_C = 3$. 

This is about as far as the assumption of linear confinement by itself can take us. Based on it, we could also write down (classical) equations of motion for the hadronising partons, but this would still not tell us explicitly how a system of partons connected by such potentials turns into jets of (free) hadrons. 

Just as was the case for preconfinement, more input is required to turn linear confinement into a hadronisation model. 
This will be the topic of \cref{sec:string} below.

\subsection{MC models of hadronisation}
\label{sec:MChad}

In \secRef{sec:string}, we discuss string models of hadronisation, focusing on the implementation in \Pythia. Next, we introduce cluster hadronisation models in \secRef{sec:cluster}, followed by remarks specific to their implementations in \Herwig and \Sherpa in \secsRef{sec:herwigCluster} and \ref{sec:sherpaCluster} respectively.

\subsubsection{String models of hadronisation}
\label{sec:string}\index{String models}
\index{Lund model}\index{Hadronisation!Lund model}

String-based models of hadronisation, such as the Lund model~\cite{Andersson:1983ia,Andersson:1997xwk} implemented in \Pythia~\cite{Sjostrand:2006za,Bierlich:2022pfr} or the Caltech-II model~\cite{Gottschalk:1986bv,Morris:1987kj} implemented in \mc{Epos}~\cite{Pierog:2013ria}, proceed in two main steps:
\begin{enumerate}
\item\label{item:string1} Each colour-singlet post-shower parton system is replaced by an equivalent string system. Quarks and antiquarks from the shower become \index{String endpoints}\emph{string endpoints}, while gluons become \index{Transverse kinks}\emph{transverse kinks}. Total four-momentum is conserved, so the total invariant mass of each string system ($m_\mathrm{s}$) equals that of its parent colour-singlet parton system.\footnote{Up to small momentum reshufflings for very low-mass systems.}
\item\label{item:string2} Each string system hadronises into a number of \index{Primary hadrons}\emph{primary hadrons} (which in turn may \index{Hadron decays}decay to \index{Secondary hadrons}\emph{secondaries}). As a rule of thumb, the total number of primary hadrons produced by a hadronising string grows logarithmically with its invariant mass,  $n_\mathrm{had}\propto \ln(m^2_\mathrm{s})$, though heavy-flavour endpoints and/or hard gluon kinks result in a more complicated dependence.
\end{enumerate}
In the following, we focus on the Lund model unless otherwise stated.

Step \ref{item:string1} above follows directly from the rules for colour flow invoked during the parton shower (\cref{sec:showers}), and from the assumption of linear confinement (\cref{sec:linearConf}) by which 
linear potentials are assumed to arise non-perturbatively between colour-connected partons, at separation distances $r \gtrsim r_\Lambda \sim 1~\mathrm{fm}$. After this mapping, the further evolution of the combined system of endpoints, kinks, and string segments can then be taken to be determined by relativistic string dynamics, yielding (classical) equations of motion such as~\cref{eq:stringEOM}.

For step \ref{item:string2}, the crucial additional assumption is that new quark-antiquark pairs can be non-perturbatively produced along the strings, in colour states that \index{Colour screening}\emph{screen} the endpoint charges. Consider a post-shower $q\bar{q}$ pair with a string spanned between them. This is illustrated in \cref{fig:stringA}. Suppose a new, screening, quark-antiquark pair, $\bar{q}'q'$, is spontaneously created somewhere along the string, and suppose further that they \index{Quantum tunneling}\emph{tunnel} a distance $\Delta\ell$ apart from each other.

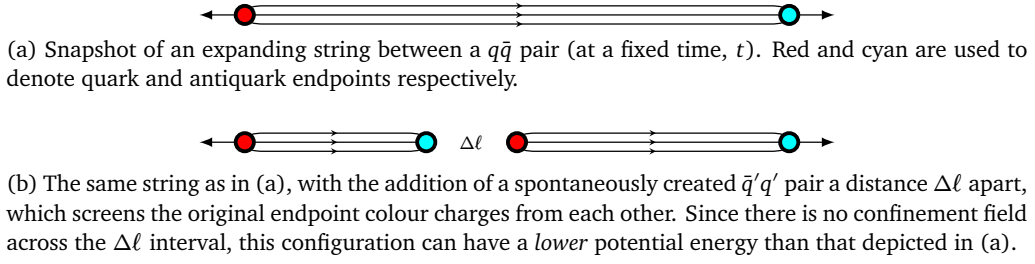
\begin{figure}[t]
\centering
\subfloat[Snapshot of an expanding string between a $q\bar{q}$ pair (at a fixed time, $t$). Red and cyan are used to denote quark and antiquark endpoints respectively.\label{fig:stringA}]{
\scalebox{1.2}{\begin{tikzpicture}
   \draw (0,0) to [out=90,in=180] (0.2,0.1)  [decoration={markings,
mark=at position 0.51 with {\arrow[scale=0.65]{stealth}}},postaction={decorate}] -- (5.8,0.1)
to [out=0,in=90] (6.0,0.0);
   \draw (0,0) to (0.2,0.0)  [decoration={markings,
mark=at position 0.51 with {\arrow[scale=0.65]{stealth}}},postaction={decorate}] -- (5.8,0.0)
to  (6.0,0.0);
   \draw (0,0) to [out=270,in=180] (0.2,-0.1)  [decoration={markings,
mark=at position 0.51 with {\arrow[scale=0.65]{stealth}}},postaction={decorate}] -- (5.8,-0.1)
to [out=0,in=270] (6.0,0.0);
  \draw [-latex] (0,0) -- (-0.5,0);
  \draw [-latex] (6,0) -- (6.5,0);
  \draw [very thick,fill=red] (0,0) circle (0.1);
  \draw [very thick,fill=cyan] (6,0) circle (0.1); 
\end{tikzpicture}}
}\\
\subfloat[The same string as in (a), with the addition of a spontaneously created $\bar{q}'q'$ pair a distance $\Delta\ell$ apart, which screens the original endpoint colour charges from each other. Since there is no confinement field across the  $\Delta\ell$ interval, this configuration can have a \emph{lower} potential energy than that depicted in (a).\label{fig:stringB}]{
\scalebox{1.2}{\begin{tikzpicture}
   \draw (0,0) to [out=90,in=180] (0.2,0.1)  [decoration={markings,
mark=at position 0.51 with {\arrow[scale=0.65]{stealth}}},postaction={decorate}] -- (1.8,0.1)
to [out=0,in=90] (2.0,0.0);
   \draw (0,0) to (0.2,0.0)  [decoration={markings,
mark=at position 0.51 with {\arrow[scale=0.65]{stealth}}},postaction={decorate}] -- (1.8,0.0)
to  (2.0,0.0);
   \draw (0,0) to [out=270,in=180] (0.2,-0.1)  [decoration={markings,
mark=at position 0.51 with {\arrow[scale=0.65]{stealth}}},postaction={decorate}] -- (1.8,-0.1)
to [out=0,in=270] (2.0,0.0);
   \draw (3,0) to [out=90,in=180] (3.2,0.1)  [decoration={markings,
mark=at position 0.51 with {\arrow[scale=0.65]{stealth}}},postaction={decorate}] -- (5.8,0.1)
to [out=0,in=90] (6.0,0.0);
   \draw (3,0) to (3.2,0.0)  [decoration={markings,
mark=at position 0.51 with {\arrow[scale=0.65]{stealth}}},postaction={decorate}] -- (5.8,0.0)
to  (6.0,0.0);
   \draw (3,0) to [out=270,in=180] (3.2,-0.1)  [decoration={markings,
mark=at position 0.51 with {\arrow[scale=0.65]{stealth}}},postaction={decorate}] -- (5.8,-0.1)
to [out=0,in=270] (6.0,0.0);
  \draw [very thick,fill=cyan] (2,0.0) circle (0.1);
  \draw [very thick,fill=red] (3,0.0) circle (0.1);
  \node at (2.5,0) {\tiny$\Delta\ell$};
  \draw [-latex] (0,0) -- (-0.5,0);
  \draw [-latex] (6,0) -- (6.5,0);
  \draw [very thick,fill=red] (0,0) circle (0.1);
  \draw [very thick,fill=cyan] (6,0) circle (0.1); \end{tikzpicture}}
}
\caption{
Illustration of string formation and breaking in the fragmentation of a $q\bar{q}$ system.
\label{fig:stringBreak}}
\end{figure}

The resulting string configuration, illustrated in \cref{fig:stringB}, is energetically favourable if the rest-mass equivalent of the ``missing'' string piece across the $\Delta\ell$ interval is greater than the combined mass equivalents of the newly created $q'$ and $\bar{q}'$ endpoints.
When allowing for transverse momentum to be imparted to the newly created quark and antiquark as well, the condition for (b) to be energetically favourable is 
\begin{equation}
\kappa \Delta \ell ~\,\ge~\, 2m_{\mathrm{T}, q'} ~=~ 2\sqrt{m_{q'}^2+p_{\mathrm{T}, q'}^2}\,,
\end{equation} 
where 
$m_{q'}$ is the rest mass of each of the new quarks, and 
$p_{\mathrm{T}, q'}$ is its transverse momentum relative to the string axis. The quantity
\begin{equation}
m^2_\mathrm{T} = m^2+\pT^2    
\end{equation}
is denoted the (squared) \index{Transverse mass}transverse mass. Note also that the Lund model assumes that the newly produced quark and antiquark have the same effective masses, $m_{q'} = m_{\bar{q}'}$ and that the vector sum of their transverse momenta (relative to the string axis) is zero: $\vec{p}_{\mathrm{T}, q'} = -\vec{p}_{\mathrm{T}, \bar{q}'}$.

The picture of spontaneous pair creation by quantum tunnelling leads to an expectation that non-perturbative production of quarks with large transverse masses will be suppressed. This underpins \index{Strangeness suppression}\emph{strangeness suppression} in \Pythia (and vanishing rates for non-perturbative $c$ and $b$ quark production) as well as the default assumption  of \index{Gaussian \pT kicks}Gaussian \pT spectra, via an analogy with the so-called \index{Schwinger mechanism}Schwinger mechanism in QED~\cite{Schwinger:1951nm}, with the approximate form:
\begin{equation}
\mathrm{Prob}(m_\mathrm{T}^2) ~\propto~ \exp\left(
\frac{-m_\mathrm{T}^2}{\pi \kappa}
\right)~=~\exp\left(
\frac{-m_q^2}{\pi \kappa}
\right)\exp\left(
\frac{-p_{\mathrm{T}, q}^2}{\pi \kappa}
\right)\,.\label{eq:Schwinger}
\end{equation}
Note that since light ($u$, $d$, $s$) quark masses are difficult to define precisely in a non-perturbative context, the quark-mass suppression factor above is effectively tuned directly to experimental data, independently of the quark masses. In \Pythia, the main strangeness suppression parameter is \texttt{StringFlav:ProbStoUD}, while the width of the Gaussian \pT distribution is given by \texttt{StringPT:sigma}.

An alternative, also available in \Pythia, is that of a \index{Thermal hadron production}\emph{thermal spectrum} of hadron production, in which the Gaussian $m_\mathrm{T}$ distribution above is replaced by a thermal distribution with an \index{Exponential \pT spectra}exponential $m_\mathrm{T}$ dependence~\cite{Fischer:2016zzs}.

With lower probabilities, strings can also be allowed to break via the production of so-called \index{Diquarks}diquark-antidiquark pairs. A diquark denotes two quarks in an overall colour-antitriplet state, the antitriplet in the product
$3 \otimes 3 ~=~ 6_\mathrm{S} + \overline{3}_\mathrm{A}$, where the subscripts S and A emphasise that the corresponding colour wave functions are symmetric and antisymmetric, respectively, under interchange of the two quarks. If colour algebra is not your thing, think of combining, e.g.\ a blue quark and a green quark in an overall cyan (= anti-red) state. The subtlety in QCD is that there are two ways to do this, one that results in a ``total'' colour charge of two (the symmetric one) and one that results in a total colour charge of minus one (the antisymmetric one). 

Since the magnitude of one antitriplet colour charge is \emph{smaller} than that of the incoherent sum of two triplets, the non-perturbative potential of the $\overline{3}$ quark-quark state is attractive, and hence it is not unreasonable to think of diquarks as effective (loosely) bound two-quark states. 

In \Pythia, string breaks involving diquarks underpin the mechanism for \index{Baryons}baryon production, via the parameter \texttt{StringFlav:ProbQQtoQ}, with further parameters related to spin and flavour composition and the so-called \index{Popcorn}\emph{popcorn} mechanism. Briefly stated, the latter effectively allows to decorrelate baryon-antibaryon pairs in momentum and flavour space, cf.~Ref.~\cite{Bierlich:2022pfr}. 

Diquarks (and, consequently, additional final-state baryons) can also result from the formation of \index{String junctions}\emph{string junctions}~\cite{Sjostrand:2002ip,Altmann:2024odn}, which can be created either directly via baryon-number violating BSM processes~\cite{Sjostrand:2002ip} or via QCD \index{Colour reconnections}\emph{colour reconnections}~\cite{Christiansen:2015yqa}. Recent measurements of baryon-to-meson ratios at the LHC indicate that the latter may be quite significant there, cf.\ e.g.~\cite{Altmann:2024kwx,Altmann:2025afh}. 

For both quarks and diquarks, \cref{eq:Schwinger} provides a well-motivated form for the mass and \pT dependence of each string break. There remains to determine \emph{how many} breaks will occur on a given string (equivalent to how many hadrons it will produce), how much longitudinal momentum each hadron should take, and what spin and isospin states to assign the hadrons.\footnote{\Eg, spin 0 or spin 1 for mesons, and spin 1/2 or spin 3/2 for baryons, and whether or not to include possible higher excited hadronic states as explicit primary hadrons.} A further nontrivial complication is how to ensure that the full string-breakup process conserves the total energy and momentum (invariant mass) of the string, while also producing only physical hadrons with correct (on-shell) masses.

The questions of longitudinal momenta and physical hadron masses are addressed simultaneously in the Lund model. Noting that individual string breaks must be separated by spacelike distances, their relative time ordering cannot play a physical role in a Lorentz invariant modelling. The Lund model exploits this to consider the breaks iteratively, ``from the outside in'', splitting off one on-shell hadron at a time from either end of the string (chosen randomly~\cite{Sjostrand:1982fn}). In an asymptotically long chain of such breaks, the momentum taken by a given hadron must not depend on whether the last break happened to be taken ``from the left'' or ``from the right''. Denoting longitudinal momentum by $p_L$ and \index{Lightcone momentum}\emph{lightcone momentum} by $p_{\pm} \equiv E\pm p_L$, this ``left-right symmetry'' constrains the probability density function for the longitudinal momentum taken by each hadron to be of the form~\cite{Andersson:1983ia} 
\begin{equation}
f_\mathrm{Lund}(z)~=~N\, \frac{(1-z)^{\color{midteal}a}}{z}\,\exp\left(\frac{-{\color{midteal}b}\hspace*{0.2mm}m_{\mathrm{T}, h}^2}{z} \right)\,,
\end{equation}
where $z$ is the fraction the hadron takes of the remaining lightcone momentum ($p_+$ or $p_-$ according to which end it is split off from) after an arbitrary number of previous hadrons have been split off from the same end, $N$ is a normalisation factor fixed by requiring the function to integrate to unity over $0<z<1$, and $m_{\mathrm{T}, h}$ is the transverse mass of the produced hadron. 
This function is known as the \index{Lund Symmetric Fragmentation Function}Lund Symmetric Fragmentation Function, here reproduced only in its simplest form with two free parameters, $\color{midteal}a$ and $\color{midteal}b$, which are fit to experimentally measured hadron momentum spectra. In \Pythia, these are \texttt{StringZ:aLund} and \texttt{StringZ:bLund}, with further parameters allowing to modify the form of the fragmentation function for diquarks, strange quarks, and heavy quarks. To help decorrelate the parameters in fits, several alternative parameterisations of the same function are also available, e.g.\ using the mean of the fragmentation function, $\color{midteal}\left<f\right>$, instead of the $\color{midteal}b$ parameter as in~\cite{Amoroso:2018qga}. 

The fact that $f_\mathrm{Lund}(z)$ does not depend on any global properties of the fragmenting string but only on the \emph{relative} lightcone momentum fraction, $z$, leads us to a central property of string-based models, called \index{Lightcone scaling}\emph{lightcone scaling}, which predicts that, modulo endpoint effects:
\begin{itemize}
\item A fragmenting string, of constant tension, produces a constant average density of hadrons per unit rapidity along the string axis. 
\item Relative rates of each hadron species (also known as \index{Hadrochemistry}\emph{hadrochemistry}), \pT spectra, and any other boost-invariant measurable properties, are also constant in rapidity along the string axis.
\end{itemize}
These are the \index{Rapidity plateaus}rapidity plateaus that were mentioned in the discussion of linear confinement in \cref{sec:linearConf}. They are the post-fragmentation manifestations of the longitudinal boost invariance of the original string. 

Note that, in the current implementation in \Pythia (version 8.317 at the time of writing), the way total energy and momentum conservation is imposed can lead to some departures from the ideal of rapidity-invariant distributions. Although some of these properties can be restored at least on average~\cite{Sjostrand:1982fn}, a more rigorous treatment of lightcone scaling for finite-energy strings remains an open research topic. 

On the question of which \emph{spin states} to assign to the produced hadrons -- e.g.\ does a $u\bar{d}$ combination turn into a $\pi^+$ or a $\rho^+$? -- the baseline Lund string model is not predictive. By spin counting, there are three vector-meson states for each pseudoscalar one, and four spin-3/2 baryon states for each two spin-1/2 ones. But since the higher spin states have higher masses (especially for the lightest-flavour hadrons), one expects them to be suppressed relative to the naive spin counting. In \Pythia, these suppression factors are cast as purely empirical parameters which are constrained by fits to data. A recent work introduced additional freedom to adjust the details of multistrange hadron rates~\cite{Bierlich:2022vdf}. It is also worth noting that the alternative UCLA string fragmentation model~\cite{Abachi:2006qf} sought to exploit the string area law to be more predictive in this regard, but so far this line of thought has not been pursued in the baseline \Pythia model.

By default, the primary hadrons that can be produced directly from strings in \Pythia include the two lowest-lying meson multiplets (spin-0 pseudoscalar mesons and spin-1 vector mesons) and the two lowest-lying baryon ones (spin-1/2 and spin-3/2 baryons). 

Optionally, these can be extended to include the four lowest-lying $L=1$ meson multiplets as well, though one should note that many of these are associated with substantial uncertainties on masses and decay branching fractions for which the model can only attempt to make some reasonable default assumptions. It is therefore not obvious that their inclusion necessarily improves the overall modelling (certainly not without careful cross checks and retuning of the hadronisation parameters), and they are left out by default. This is equivalent to modelling their overall effects via the string continuum, rather than in terms of distinct mass eigenstates. Note that when $L=1$ mesons are not included, states such as the $f_0(1370)$ and $a_0(1450)$ scalar mesons will only appear as explicit distinct particles in \Pythia event records when they occur as \emph{secondaries} (i.e.\ in decays of other hadrons). Any would-be primary ones are  only accounted for implicitly, via the string continuum. 

Of special interest to specific studies are \index{Quarkonia}\emph{quarkonia} (i.e.\ $J/\psi$, $\eta_c$, $\Upsilon$, $\eta_b$, etc.), for which \Pythia incorporates several dedicated modelling options. These, and multiply-heavy hadrons in general (e.g.\ $B_c$), involve additional physics aspects which go beyond the scope of this review, hence we refer to the physics manual~\cite{Bierlich:2022pfr} and dedicated studies~\cite{Egede:2022lws,Cooke:2023ldt} for details.

The hadronisation of \index{Beam remnants}\emph{beam remnants} in the far forward direction also has ambiguities  -- and associated modelling options -- of its own. We refer to \cite{Sjostrand:2004pf,Albrecht:2025kbb} for more on these aspects.

Finally, we note that \Pythia does not currently include any models for the production and decays of exotic hadronic states like tetraquarks, pentaquarks, etc. 

\subsubsection{Cluster models of hadronisation}
\label{sec:cluster}\index{Cluster models}\index{Hadronisation!Cluster models}

Cluster models of hadronisation
take the preconfinement feature
of parton showers as their starting point,
which is explained in \cref{sec:preconfinement}.
As a reminder, it describes the tendency
that the partons generated by the shower
in the large-\Nc limit
are eventually
arranged in localised colour-singlet clusters,
with a universal (i.e.\ independent of the hard process)
invariant mass distribution.

Variants of cluster hadronisation are implemented in the \Sherpa
and \Herwig event generators.
Before going into detail about their individual choices,
we will explain the general features,
as introduced in the original model by Field and Wolfram~\cite{Field:1982dg}.
We follow the description by Webber in~\cite{Webber:1983if},
as illustrated in \cref{fig:clusterfrag}.

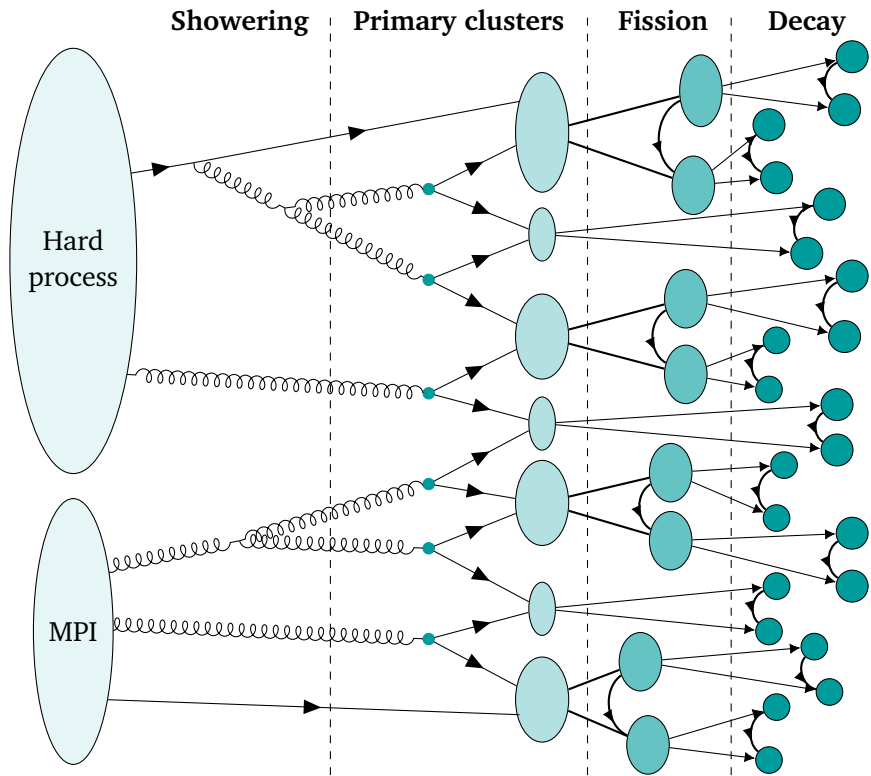
\begin{figure}[t]
\centering
\scalebox{1.0}{\begin{tikzpicture}[
    hardscatter/.style   = {ellipse, draw=black, fill=myteal!10,
                            minimum width=40pt, minimum height=160pt, inner sep=0pt},
    mpiscatter/.style    = {ellipse, draw=black, fill=myteal!10,
                            minimum width=30pt, minimum height=100pt, inner sep=0pt},
    largecluster/.style  = {ellipse, draw=black, fill=myteal!30,
                            minimum width=20pt, minimum height=45pt, inner sep=0pt},
    cluster/.style       = {ellipse, draw=black, fill=myteal!30,
                            minimum width=20pt, minimum height=32pt, inner sep=0pt},
    smallcluster/.style  = {ellipse, draw=black, fill=myteal!30,
                            minimum width=10pt, minimum height=20pt, inner sep=0pt},
    mediumcluster/.style = {ellipse, draw=black, fill=myteal!60,
                            minimum width=16pt, minimum height=27pt, inner sep=0pt},
    lightcluster/.style  = {ellipse, draw=black, fill=myteal!60,
                            minimum width=16pt, minimum height=22pt, inner sep=0pt},
    hadron/.style        = {circle, draw=black, fill=myteal,
                            minimum size=12pt, inner sep=0pt},
    smallhadron/.style   = {circle, draw=black, fill=myteal,
                            minimum size=10pt, inner sep=0pt},
    fission/.style       = {draw=black, thick},
    decay/.style         = {draw=black, -{Latex[length=4pt,width=4pt]}},
    collabel/.style      = {font=\bfseries},
    bendarrow/.style     = {bend right=60,thick, postaction={
                             decorate, decoration={markings,mark=at position 0.75 with {\arrow{latex}}
                           }}}
  ]

  \foreach \x in {3.2, 6.6, 8.5} { \draw[dashed] (\x,-5.8) -- (\x,4.25); }
  \node[collabel,anchor=base] at (2.0,4.05) {Showering};
  \node[collabel,anchor=base] at (4.9,4.05) {Primary clusters};
  \node[collabel,anchor=base] at (7.6,4.05) {Fission};
  \node[collabel,anchor=base] at (9.5,4.05) {Decay};

  \begin{feynman}
    \vertex (v0)  at (0.5, 2.15);
    \vertex (qa) at (1.4, 2.3);
    \vertex (qb) at (5.7, 3.11);

    \vertex (ba) at (0.2,-3.05);
    \vertex (bb) at (2.0,-2.7);
    \vertex (bc) at (5.7,-2.55);

    \vertex (ma) at (0.2,-4.8); 
    \vertex (mb) at (5.7,-5.0);
    \vertex (mc) at (0.2,-3.8); 
    \vertex (md) at (5.65,-3.8);
    \vertex (me) at (5.65,-4.8);

    \vertex (g1) at (2.6, 1.7);
    \vertex (g2) at (0.5, -0.5);

    \vertex [dot, myteal] (s1) at (4.5, 1.95) {};
    \vertex [dot, myteal] (s2) at (4.5, 0.75) {};
    \vertex [dot, myteal] (s3) at (4.5,-0.75) {};
    \vertex [dot, myteal] (s4) at (4.5,-1.95) {};
    \vertex [dot, myteal] (s5) at (4.5,-2.8) {};
    \vertex [dot, myteal] (s6) at (4.5,-4.0) {};

    \vertex [largecluster] (C1) at (6, 2.70) {};
    \vertex [smallcluster] (C2) at (6, 1.35) {};
    \vertex [cluster] (C3) at (6, 0.00) {};
    \vertex [smallcluster] (C4) at (6,-1.15) {};
    \vertex [cluster] (C5) at (6,-2.20) {};
    \vertex [smallcluster] (C6) at (6,-3.60) {};
    \vertex [cluster] (C7) at (6,-4.80) {};

    \vertex [mediumcluster] (C1a) at (8.1, 3.25) {};
    \vertex [lightcluster] (C1b) at (8.0, 2.00) {};
    \vertex [lightcluster] (C3a) at (7.9, 0.50) {};
    \vertex [lightcluster] (C3b) at (7.9,-0.50) {};
    \vertex [lightcluster] (C5a) at (7.7,-2.7) {};
    \vertex [lightcluster] (C5b) at (7.7,-1.8) {};
    \vertex [lightcluster] (C6a) at (7.3,-4.3) {};
    \vertex [lightcluster] (C6b) at (7.4,-5.4) {};

    \vertex [hadron] (h1)  at (10.1, 3.7) {};
    \vertex [hadron] (h2)  at (10.0, 3.0) {};
    \vertex [hadron] (h3)  at (9.0, 2.80) {};
    \vertex [hadron] (h4)  at (9.1, 2.10) {};
    \vertex [hadron] (h5)  at (9.8, 1.75) {};
    \vertex [hadron] (h6)  at (9.5, 1.10) {};
    \vertex [hadron] (h7a)  at (10.1, 0.80) {};
    \vertex [hadron] (h7b)  at (10.0, 0.00) {};
    \vertex [smallhadron] (h8a)  at (9.1,-0.05) {};
    \vertex [smallhadron] (h8b)  at (9.0,-0.7) {};
    \vertex [hadron] (h9)  at (9.9,-0.9) {};
    \vertex [hadron] (h10) at (9.9,-1.5) {};
    \vertex [smallhadron] (h11) at (9.2,-1.7) {};
    \vertex [smallhadron] (h12) at (9.1,-2.4) {};
    \vertex [hadron] (h13) at (10.1,-2.60) {};
    \vertex [hadron] (h14) at (10.1,-3.30) {};
    \vertex [smallhadron] (h15) at (9.1,-3.30) {};
    \vertex [smallhadron] (h16) at (9.0,-3.90) {};
    \vertex [smallhadron] (h17) at (9.6,-4.10) {};
    \vertex [smallhadron] (h18) at (9.8,-4.70) {};
    \vertex [smallhadron] (h19) at (9.1,-4.90) {};
    \vertex [smallhadron] (h20) at (9.0,-5.6) {};

    \draw[bendarrow] (C1a) to (C1b);
    \draw[bendarrow] (C3a) to (C3b);
    \draw[bendarrow] (C5b) to (C5a);
    \draw[bendarrow] (C6a) to (C6b);
    \draw[bendarrow] (h1) to (h2);
    \draw[bendarrow] (h3) to (h4);
    \draw[bendarrow] (h5) to (h6);
    \draw[bendarrow] (h7a) to (h7b);
    \draw[bendarrow] (h8a) to (h8b);
    \draw[bendarrow] (h9) to (h10);
    \draw[bendarrow] (h11) to (h12);
    \draw[bendarrow] (h13) to (h14);
    \draw[bendarrow] (h15) to (h16);
    \draw[bendarrow] (h17) to (h18);
    \draw[bendarrow] (h19) to (h20);

    \diagram* {
      (v0) -- [fermion] (qa) -- [fermion] (qb),
      (qa) -- [gluon] (g1) -- [gluon] (s2),
      (s1) -- [gluon] (g1),
      (ba) -- [gluon] (bb) -- [gluon] (s5),
      (s3) -- [gluon] (g2),
      (s4) -- [gluon] (bb),
      (ma) -- [fermion] (mb),
      (mc) -- [gluon] (s6),
      (s1) -- [fermion] (C1), (s1) -- [fermion] (C2),
      (s2) -- [fermion] (C2), (s2) -- [fermion] (C3),
      (s3) -- [fermion] (C3), (s3) -- [fermion] (C4),
      (s4) -- [fermion] (C4), (s4) -- [fermion] (C5),
      (s5) -- [fermion] (C5), (s5) -- [fermion] (C6),
      (s6) -- [fermion] (C6), (s6) -- [fermion] (C7),
    };

    \vertex [hardscatter, align=center] (HS) at (-0.2, 1.0) {Hard\\ process};
    \vertex [mpiscatter] (MPI) at (-0.2, -3.9) {MPI};
    
  \end{feynman}

  \draw[fission] (C1) -- (C1a);
  \draw[fission] (C1) -- (C1b);
  \draw[fission] (C3) -- (C3a);
  \draw[fission] (C3) -- (C3b);
  \draw[fission] (C5) -- (C5a);
  \draw[fission] (C5) -- (C5b);
  \draw[fission] (C7) -- (C6a);
  \draw[fission] (C7) -- (C6b);

  \draw[decay] (C1a) -- (h1);  \draw[decay] (C1a) -- (h2);
  \draw[decay] (C1b) -- (h3);  \draw[decay] (C1b) -- (h4);
  \draw[decay] (C2)  -- (h5);  \draw[decay] (C2)  -- (h6);
  \draw[decay] (C3a) -- (h7a); \draw[decay] (C3a) -- (h7b);
  \draw[decay] (C3b) -- (h8a); \draw[decay] (C3b) -- (h8b);
  \draw[decay] (C4)  -- (h9);  \draw[decay] (C4)  -- (h10);
  \draw[decay] (C5b) -- (h11); \draw[decay] (C5b) -- (h12);
  \draw[decay] (C5a) -- (h13); \draw[decay] (C5a) -- (h14);
  \draw[decay] (C6)  -- (h15); \draw[decay] (C6) -- (h16);
  \draw[decay] (C6a) -- (h17); \draw[decay] (C6a) -- (h18);
  \draw[decay] (C6b) -- (h19); \draw[decay] (C6b) -- (h20);

\end{tikzpicture}}
\caption{Illustration of cluster fragmentation, proceeding from left to right. The parton shower adds emissions to the hard partons created in the hard processes and through the MPI model and stops at the parton shower cut-off scale $Q_0$ (indicated by the dashed vertical line to the left). Gluons then undergo a forced non-perturbative splittings. Color-connected pairs of partons then form the primary clusters. These then undergo a series of fissions, until their invariant masses fall below a fission cut-off scale (indicated by the dashed vertical line to the right). The resulting clusters then decay into hadrons. Based on Fig.~1 in \cite{Gieseke:2025mcy}.\label{fig:clusterfrag}}
\end{figure}

The first step of the model
-- applied once the shower terminates at $Q_0$ --
is to render the splitting
into localised colour-singlet subsystems 
explicit. This is achieved through forced non-perturbative $g \to q \bar q$ splittings,
resulting in a set of colour-singlet $q \bar q'$ clusters.
These are referred to as ``primary'' clusters.

It is then assumed that their universal invariant mass
distribution can be regarded as a smeared version
of the spectrum of ``primordial'' hadronic resonances
that would form in the early stages of the dynamical process
of confinement in real jets.
From that point of view, it seems reasonable
to model the transition of the clusters
into such hadronic resonances
as an averaged resonance decay process.
For that, only the (dominant) two-body decays are considered.
The branching ratios are determined entirely by the density of states,
i.e.\ the available phase-space volume times the number of spin states.
Moreover, the decays are considered isotropic.

The decay of a $q_1 \bar q_2$ cluster is selected by first 
choosing randomly another flavour $q_3$ ($u$, $d$, $s$).
The flavours of the two resulting mesons of the decay
are then given by the combinations $q_1 \bar q_3$ and $q_3 \bar q_2$.

To allow for decays into baryons,
$q_3$ can also be chosen to be one of six ``diquarks'' $d_3$,
i.e.\ pair combinations of the three light-quark flavours $u$, $d$, $s$ and their anti-quark flavours.
In that case, the flavours of the two resulting baryons of the decay
are given by the combinations $q_1 d_3$ and $\bar d_3 \bar q_2$.

Another building block of the model is cluster fission.
The motivation for it is that
for clusters with large invariant masses,
and hence a lot of kinetic energy available in their rest system,
the assumption of an isotropic decay is less reasonable.
The idea is to first split large-mass clusters into smaller-mass ones,
by the production of a quark-antiquark pair in the middle of the string
formed by the two quarks of the original cluster.
This process is repeated until all clusters are below
the fission threshold parameter $M_f$.

This original model has only two free parameters:
The gluon mass cutoff $Q_0$, which controls
the non-perturbative gluon mass for the forced $g \to q\bar q$ decays,
and the fission threshold $M_f$.
$Q_0$ is set to the cut-off scale of the parton shower.
The up and down quarks are assigned constituent masses
during the cluster hadronisation,
but they are fixed as $m_u = m_d = Q_0 / 2$.
As an example, $Q_0 = 0.6\,\GeV$ and $M_f = 4\,\GeV$
are quoted in~\cite{Webber:1983if} as a reasonable choice,
and compared rather successfully to $e^+ e^- \to \text{jets}$ data.
Only a mild dependence on the parameters is observed.

\subsubsection{Cluster hadronisation in Herwig \label{sec:herwigCluster}}

The cluster hadronisation model as implemented in \Herwig
incorporates several refinements with respect to the original formulations
by Field, Wolfram and Webber described above~\cite{Bahr:2008pv}.
Arguably the most notable addition is the incorporation
of colour reconnections, which is described in \cref{sec:CR}.

The cluster fission criterion is refined
and takes into account not only the invariant mass of the cluster
but also its constituent quark masses,
and allows for additional control by flavour-dependent parameters
to fine-tune the behaviour,
which is in particular used to improve the bottom and charm hadron rates.
The quark-antiquark or diquark-antidiquark pairs created during the fission
can also be selected in a flavour-dependent manner with tunable parameters
for the newly created cluster masses.
There is also a special treatment for clusters containing beam remnants.

The cluster decay procedure follows~\cite{Field:1982dg,Webber:1983if},
with a correction
to address isospin violation present in the original model,
and introduces a balancing parameter to regulate
the relative rates of mesonic and baryonic decays.
Decay probabilities can be adjusted individually for each meson multiplet
and separately for singlet and decuplet baryons.
Decays are treated as isotropic,
with the exception of hadrons that contain a parton originating
from the perturbative stage of event generation,
for which the parton's own direction is used instead,
subject to Gaussian smearing.
The smearing width is controlled by independent parameters
for light, charm, and bottom quarks.
In special cases where the cluster mass falls
below a viable two-body decay threshold, a forced single-hadron decay is applied.

With these refinements and more fine-tuned control,
the full model is characterised by about 30 tuneable parameters,
listed in Table 10 of Ref.~\cite{Bahr:2008pv}.

In~\cite{Gieseke:2017clv},
non-perturbative gluon splitting into strange
quark-antiquark pairs was added to the model,
resulting in a better description of Kaon $\pT$ spectra,
and, when combined with the updated colour reconnection,
also improving the description of the production of heavy baryons $\Delta$ and $\Xi^-$.

More recent developments
include i) an enhancement of the transfer of heavy-quark polarisation
to the daughter hadron, inspired by Heavy Quark Effective Theory (HQET),
as well as a more finely controllable
cluster fission decision process~\cite{Masouminia:2023zhb},
ii) an investigation
of the parton shower's IR cutoff as a factorisation scale
from the perspective of the cluster hadronisation model~\cite{Hoang:2024nqi},
and, related to that, iii) a proposal of a new set of building blocks
to describe cluster fission and decays~\cite{Gieseke:2025mcy}.

\subsubsection{Cluster hadronisation in Sherpa \label{sec:sherpaCluster}}

The \Sherpa hadronisation model~\cite{Winter:2003tt,Chahal:2022rid}
is also based on the cluster model,
and also includes a colour reconnection model~\cite{Chahal:2022rid},
but makes different choices
compared to the original formulation,
and to the one implemented in \Herwig.


The first difference in the forced gluon splitting phase
is the absence of a non-perturbative gluon mass.
Instead, the mass and the momenta of the flavours
produced in the $g \to f \bar f$ splitting
are taken from the four-momenta of the directly neighbouring particle
in the colour-connected string called the ``spectator''.
The other notable difference is that \Sherpa
allows gluons to split into diquark pairs,
thus already generating seeds for the production of baryon pairs
at this early stage.
This leads to a softer correlation
of these baryon pairs, similar to the popcorn mechanism
used in String models, see \cref{sec:string}.

In cluster fission, instead of an isotropic decay,
and selecting the kinematics according to the masses of the new clusters,
the kinematics is chosen according to light-cone momentum fractions,
and a Gaussian \pT distribution.

When clusters are identified that are too light to decay into hadrons,
a momentum reshuffling is  employed, taking momentum from another cluster,
such that a decay to the lightest allowed hadron becomes possible.

Tables 3--6 of \cite{Chahal:2022rid} list the tunable parameters
of the most recent implementation of the \Sherpa cluster model,
amounting to about 50 parameters.

\subsection{The underlying event in hadron collisions}
\label{sec:mpi}\index{Underlying event}
\responsible{Peter}
\index{Pileup}
\index{Beam remnants}

\begin{quote}{\it 
``We distinguish two distinct regions in the average jet, the hard core of the jet ($\Delta \eta < 0.2$), and the
wings ($0.2 \lesssim \Delta \eta < 1.0$).} [...] {\it 
 Outside the jet\footnote{Paraphrased from the original article which used the word enhancement here rather than jet.},
for $\Delta\eta > 1.0$, a constant 
$E_T$ plateau is observed,
whose height is independent of the jet $E_T$. Its value is
substantially higher than the one observed for minimum bias events.''}
\flushright -- The UA1 Collaboration~\cite{UA1:1983hhd}.
\end{quote}
\index{Pedestal effect}
\index{Minimum bias}

The quote above comes from a 1983 study of 100-GeV jets~\cite{UA1:1983hhd} which plausibly can be called the discovery of the underlying event (UE). The quote remains valid today and succinctly summarises the main observed characteristics of the UE:
\begin{itemize}
\item In hadron-hadron collision events with a hard (i.e.\ high-\pT) trigger, the high-\pT objects themselves are accompanied by substantially increased hadron production \emph{all over the event}, relative to when no hard trigger is imposed (``minimum bias'').
\item For trigger-object $\pT \gtrsim 10$ GeV, the \emph{level} of this additional activity (e.g.\ in terms of $E_\mathrm{T}$ and/or charged-track densities) is almost independent of the trigger \pT.
\item The \emph{distribution} of this additional activity is roughly uniform in $\eta$ and $\phi$ (on average) and consists mainly of soft hadrons with $\pT \lesssim$ a few GeV.
\item By now, we also know that the level of the UE increases with increasing collider CM energy, cf.\ e.g.~\cite{ALICE:2011ac,CDF:2015txs,Brooks:2018tgf,Ortiz:2021gcr}.
\end{itemize}

In all modern general-purpose event generators (including also EPOS~\cite{Pierog:2013ria}), the way this additional hadron production is modelled is via multiple parton-parton interactions (MPI).
\index{Multi-parton interactions}
\index{MPI|see{Multi-parton interactions}}
The first explicit such MC model~\cite{Sjostrand:1987su} was proposed shortly after the UA1 measurement, and will form the basis of our discussion here. Where relevant, we will point out some later refinements as well, and highlight when different MC generators take different approaches to various ambiguous aspects.

The notion that multiple parton-parton interactions can occur within a single hadron-hadron collision should not be totally surprising. Hadrons are composite; at wavelengths short enough to resolve their internal structure, each hadron can be thought of as a beam of partons. Hence, when two hadrons collide, we can think of that as two beams of partons colliding. While the probability for a high-\pT interaction to occur is low and calculable using standard perturbation theory, the probability for low-\pT scatterings is very large -- dominated by $t$-channel gluon exchanges with
\begin{equation}
   \frac{\d \sigma_\mathrm{parton-parton}}{\d \pT^2} \propto N_C^2\frac{\alpha_s^2(\pT^2)}{\pT^4}~,
   \label{eq:dSigmaMPI}
\end{equation}
where the colour factor $N_C^2$ is appropriate for $gg\to gg$ scattering.
The average \emph{number} of such scatterings per hadron-hadron collision, can be estimated by the simple ratio,
\begin{equation}
\left<n_\mathrm{parton-parton}\right> \sim 
\frac{\sigma_\mathrm{parton-parton}}{\sigma_\mathrm{hadron-hadron}}\label{eq:nMPI}
\end{equation}
If this number is larger than unity, then the average hadron-hadron collision will contain more than one parton-parton interaction. In any case, we will expect roughly Poissonian fluctuations around this mean value, hence some collisions should contain  many parton-parton interactions while others would contain only one. Turning this basic observation into a realistic MPI model involves addressing the following aspects:
\begin{enumerate}
\item To compute the numerator of \cref{eq:nMPI}, one needs to integrate \cref{eq:dSigmaMPI} (or really, the full QCD $2\to 2$ differential cross section) over \pT. If we let the lower limit of that integral extend to $\pT \to 0$, this is obviously divergent. We will address this point below. 
\item We will need to ensure that the summed $x$ values of the MPI initiators in each hadron does not exceed unity. More generally, we will need a set of at least semi-realistic \index{Multi-parton PDFs} multi-parton PDFs to multiply onto the differential cross section in \cref{eq:dSigmaMPI}. In this review, we shall not delve into the details of multi-parton PDFs beyond noting that MC models are constructed to explicitly conserve energy and momentum in each hadron, and e.g.\ \Pythia also imposes some flavour sum rules~\cite{Sjostrand:2004pf}. The topic of multi-parton PDFs and their evolution is an active research area, c.f.\ e.g.~\cite{Gaunt:2009re,Fedkevych:2025lgp}. 
\item\label{item:impact_parameter} A realistic modelling should also account for the \index{Impact parameter}impact parameter at which the two hadrons hit each other, which will fluctuate from event to event. This point will be addressed below.
\item One needs to decide on the precise number to use for the hadron-hadron cross section in the denominator of \cref{eq:nMPI}. The original choice in \cite{Sjostrand:1987su} was to use the non-diffractive component of the total inelastic cross section, and this remains the baseline choice in \Pythia. But this choice is not unique (partly since the definition of diffraction is not itself totally unambiguous), and other choices are possible. As a minimum, any scattering events that are included in $\sigma_\mathrm{parton-parton}$ must also be present in one's definition of $\sigma_\mathrm{hadron-hadron}$. 
\item While low-\pT MPI do not generate big momentum transfers by themselves, they do represent ``colour sparks'': each MPI produces a colour exchange between the two colliding hadrons. These additional colour exchanges will obviously have to be taken into account in the modelling of how the total hadron-hadron collision system hadronises. In fact, within the MPI modelling paradigm, it is these additional colour exchanges which generate the increased amounts of hadron production that constitute the UE. 
\item While parton showers off the additional MPI scatterings were not included in \cite{Sjostrand:1987su}, modern formulations such as \cite{Sjostrand:2004ef,DeRoeck:711179,Bahr:2008dy} do include such showers; any high-\pT MPI should thus generate realistic-looking ``mini-jets''.  
\end{enumerate}

As mentioned under point 1 in the list above, the perturbative parton-parton cross section diverges at low \pT and needs to be regulated. Physically, a gluon  with $\pT \sim 0$ would not be able to resolve individual partons in the other hadron but would only see the hadron itself, which is of course overall colour neutral. This suggests that, at the very least we should only integrate the perturbative cross section down to a scale $\pT^\mathrm{min} \sim 1/r_h \sim {\cal O}(\Lambda_\mathrm{QCD})$, where $r_h$ is the hadron radius. However, this really only furnishes an absolute lower bound. In practice, one chooses a larger regularisation scale ${\color{midteal}p_\mathrm{T,0}} \gtrsim {\cal O}(1\,\mathrm{GeV})$. This can be seen as being representative of an average hadronic colour-screening scale. In \Pythia, the default is to impose this as a smooth dampening~\cite{Sjostrand:1987su}, 
\begin{equation}
\frac{\alpha_s^2(\pT^2)}{\pT^4} ~\to~ \frac{\alpha_s^2(\pT^2 + {\color{midteal}p_{\mathrm{T},0}^2})}{(\pT^2 + {\color{midteal}{p_{\mathrm{T},0}^2}})^2}~,
\end{equation}
whereas in \Herwig and \Sherpa it is imposed as a step function.

Regardless of whether it is imposed as a smooth dampening or as a step function,  this infrared regularisation scale, $\color{midteal}p_{\mathrm{T},0}$, constitutes the first and main tuning parameter of any MPI model, with lower values of $\color{midteal}p_{\mathrm{T}, 0}$ giving a higher $\left< n_\mathrm{parton-parton}\right>$ (i.e.\ more MPI) and consequently a higher UE, while higher $\color{midteal}p_{\mathrm{T},0}$ values produce a lower UE. 

A second aspect is whether to take $\color{midteal}p_{\mathrm{T},0}$ to be a universal constant, or to let it be a function of the collision parameters. Going back to its interpretation as (the inverse of) an effective colour-screening distance, this could certainly depend on collider CM energy, with higher CM energies resolving more densely packed lower-$x$ partons which presumably would lead to smaller effective colour-screening distance scales and hence larger $\color{midteal}p_{\mathrm{T},0}$ values. It could also in principle depend on event-by-event quantities such as actual $x$ fractions and/or hadron-hadron impact parameter. In \Herwig, \Pythia and \Sherpa, the baseline choice is a power-law dependence on the collider CM energy, $\sqrt{s}$:
\begin{equation}
p_{\mathrm{T},0}(\sqrt{s})~=~{\color{midteal}p_{\mathrm{T},0}^\mathrm{ref}} \left( \frac{\sqrt{s}}{\color{midteal}E^\mathrm{ref}_\mathrm{CM}}\right)^{\color{midteal}\alpha}\,,
\end{equation}
where $\color{midteal}E^\mathrm{ref}_\mathrm{CM}$ is an arbitrary reference CM energy scale at which $\color{midteal}p_{\mathrm{T},0}(\sqrt{s}) \equiv p_{\mathrm{T},0}^\mathrm{ref}$, and the power $\color{midteal}\alpha$ is the main tuning parameter governing the scaling of $p_{\mathrm{T},0}(\sqrt{s})$ away from $\sqrt{s} = E^\mathrm{ref}_\mathrm{CM}$. See \cite{Brooks:2018tgf,Ortiz:2021gcr,Gieseke:2012ft} for studies. In \Pythia, an option for a logarithmic $\sqrt{s}$ dependence is also available.

As mentioned under aspect~\ref{item:impact_parameter} in the list above, a realistic model must also account for the impact parameter at which the two hadrons hit each other. Denoting the magnitude of the hadron-hadron impact parameter by $b$ and given an assumed form for the shape of the parton distributions in transverse space, $f(x) \to f(x,\vec{r}_\mathrm{T})$, one can compute ``overlap factors'', ${\cal  O}(b)$, which account for the effective matter overlap between the two hadrons at a given $b$~\cite{Sjostrand:1987su}. These overlap factors multiply the parton-parton cross sections in \cref{eq:dSigmaMPI,eq:nMPI}, yielding small $\left< n_\mathrm{parton-parton}\right>$ in peripheral collisions (large $b$) and large $\left< n_\mathrm{parton-parton}\right>$ in central collisions (small $b$). 

The impact-parameter dependence plays an important role in  explaining why hadron densities and $E_T$ sums in the UE are ``substantially higher''~\cite{UA1:1983hhd} than in events without a hard trigger (``minimum bias'') at the same CM energy. Since each parton-parton interaction has a given (small) chance to produce a high-\pT interaction,  events with low $n_\mathrm{parton-parton}$ have ``fewer chances'' to pass a hard trigger requirement than ones with high  $n_\mathrm{parton-parton}$. Triggering on a hard process will therefore bias the event selection towards collisions with large $n_\mathrm{parton-parton}$. In an impact-parameter-dependent model, this corresponds to events with small impact parameters. Thus, events with a hard trigger requirement are more central than ones without such a requirement.

The impact-parameter dependence also explains why the UE effectively becomes constant and independent of trigger \pT for trigger-\pT values above $\sim$ 10 GeV. Around this scale, the selected sample is already close to maximally biased, with $\left<b\right> \sim 0$. It's not possible to be more central than fully central. Hence, above this trigger scale, the UE becomes roughly constant, reflecting the transition from a ``minimum bias'' event sample (without a trigger requirement) to a ``maximum bias'' one (with trigger $\pT \gtrsim 10~\GeV$).

At a given collider energy, the ratio between the UE plateau and the minimum-bias (MB) level is therefore also sensitive to the assumed shape of the hadronic matter profile. Sharply peaked shapes for the assumed hadronic matter profiles give rise to higher UE/MB ratios, while more homogenous mass-distribution assumptions predict lower UE/MB ratios. 

The case of \emph{two} high-\pT parton-parton interactions within a single hadron-hadron collision is sometimes of special interest, a classic example being same-sign $W^\pm W^\pm$ production. This is called \index{Double parton scattering}double parton scattering (DPS). It can obviously be regarded as a special case of MPI, but the requirement that two of the parton-parton interactions both have a perturbatively high scale also makes more specialised perturbative treatments possible. Combining these with event generators is a topic of active research~\cite{Fedkevych:2025lgp,Cabouat:2019gtm}. So far, \Pythia implements a fairly simple framework for requesting a specific second hard process explicitly, and combining this with the general MPI framework. 

\subsection{Colour reconnections and collective phenomena}
\label{sec:CR}\index{Colour reconnections}\index{CR|see{Colour reconnections}}

Colour reconnections and collective phenomena are both very broad terms, each covering multiple possible physics effects, and there are also some areas of overlap between them. In \secRef{sec:CRdet}, we give a low-level overview of the topic of colour reconnections and its implementation in MC models. In \secRef{sec:collective}, we do the same for collective effects.

\subsubsection{Colour reconnections}
\label{sec:CRdet}

From a theoretical standpoint, there are several potential effects which could be called ``colour reconnections'' (CR) in that they either operate solely or mainly in colour space, and/or could cause nontrivial rearrangements of the hadronising colour-singlet systems. Ordering them roughly from early (pre-hadronisation) to late (post-hadronisation) effects, they include:
\begin{itemize}
\item During the parton shower, general-purpose event generators rely heavily on the leading-colour approximation. This means that subleading-colour interference effects are neglected. The colour correlations between partons in the resulting post-shower state are thus only unambiguous to leading colour. Studies of colour rearrangements during the perturbative shower process include Refs.~\cite{Lonnblad:1995yk,Bellm:2018wwz}. 
\item Each MPI system (see \cref{sec:mpi}) is treated separately in colour space. Since multi-parton PDFs do not come with colour correlations, MC models must make assumptions about how the initiators of each MPI system are correlated with each other in colour space (see e.g.~\cite{Sjostrand:2004pf}). These assumptions could be imperfect.
\item Resonance decays are typically treated independently of the rest of the event (and of each other). However, finite widths $\Gamma \gtrsim {\cal O}(\Lambda_\mathrm{QCD})$ could induce shower radiation patterns involving colour connections to partons outside the resonance-decay system (cf.\ e.g.~\cite{Sjostrand:1993rb,Brooks:2021kji}). In top-quark decays, this would be a leading-colour effect, while it is subleading in decays of colour-neutral resonances. (For decays of SM Higgs bosons it is irrelevant, since $\Gamma_H \ll \Lambda_\mathrm{QCD}$.) Allowing top-quark decay products to radiate against partons outside of the top decay could be labelled a ``colour-reconnection'' effect.
\item During the formation of confining potentials, soft gluon exchanges between different colour-singlet subsystems, and/or between different segments of each subsystem, could modify the hadronising colour topology (see e.g.~\cite{Gieseke:2018gff}).
\item After the formation of confining potentials but before hadron production, strings or clusters could have non-trivial space-time overlaps, which could allow them to exchange constituents with each other and/or collide inelastically, effectively rearranging them into new structures. Explicit examples of this type of CR modelling are the SK-I and SK-II models~\cite{Sjostrand:1993rb,Sjostrand:1993hi}, based on analogies with flux tubes in Type-I and Type-II superconductors, respectively.
\end{itemize}

Thus, there are certainly plenty of motivations to consider possible modifications to colour configurations created by LC parton showers. The main challenge is to determine what physical principles to base such modifications on. Most modern CR models seek to minimise a measure of the total potential energy of the hadronising system. String CR models typically use a measure of total string length~\cite{Rathsman:1998tp,Skands:2007zg,Argyropoulos:2014zoa,Christiansen:2015yqa,Sjostrand:2017ele}, while cluster CR models typically use (summed) cluster masses~\cite{Gieseke:2012ft,Gieseke:2017clv,Chahal:2022rid,Sherpa:2024mfk}\footnote{CR models that act in configuration space (space-time) have also been constructed~\cite{Bellm:2019wrh}.}.

To see why this makes sense, consider the $q_1\bar{q}_2q_3\bar{q}_4$ state illustrated in \figRef{fig:CR}. 
\begin{figure}[t]
\centering
\begin{tabular}{c}
\includegraphics*[width=0.33\textwidth]{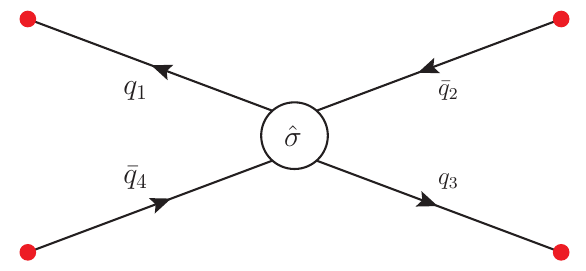}\end{tabular}
\begin{tabular}{c}
\huge ~$\longrightarrow$~
\end{tabular}
\begin{tabular}{c}
\includegraphics*[width=0.33\textwidth]{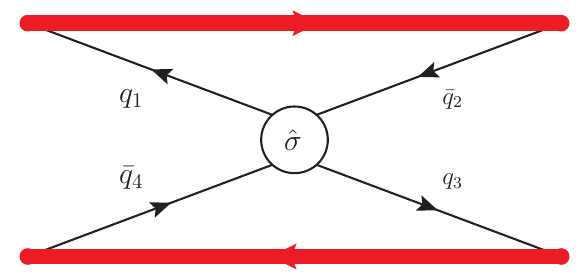}\\
or \\
\includegraphics*[width=0.33\textwidth]{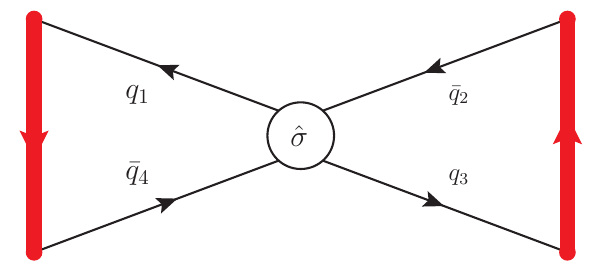}
\end{tabular}
\caption{Illustration of ambiguity in defining the confining colour singlets for a hard process, $\hat{\sigma}$, that produces 4 quarks, $q_1\bar{q}_2 q_3\bar{q}_4$ in a ``colour quadrupole'' state in which all of the quark-antiquark combinations have zero total colour charge. This could result in either of the two colour-singlet configurations illustrated on the right. A CR model driven by potential-energy minimisation would be more likely to choose the lower of the two configurations, than the higher one.
\label{fig:CR}}
\end{figure}Assuming all of the possible quark-antiquark combinations can have zero total colour charge (either due to perturbative ambiguities or due to soft gluon exchanges), a CR model driven by potential-energy minimisation would be more likely to choose the lower of the two configurations on the right, as this would correspond to a lower total cluster mass or string length.

The net result of such minimisations is that somewhat fewer hadrons will be produced (than would be the case without CR). Due to energy and momentum conservation, each of these hadrons must then carry a greater \emph{fraction} of the total partonic energy (than they would without CR), leading to a hardening of the hadronic spectra. 

Empirically, this hardening of hadronic spectra appears to agree with measurements, for example of correlations between particle momenta and multiplicities in hadron-hadron collisions. A well-studied example is the average charged-hadron transverse momentum versus charged-hadron multiplicity, $\left< \pT \right> (N_\mathrm{ch})$, in minimum-bias events, which shows a characteristic rising 
trend~\cite{Ames-Bologna-CERN-Dortmund-Heidelberg-Warsaw:1986amw,Sjostrand:1987su,Alexopoulos:1988na,E735:1994joq,Field:2006gq,Skands:2010ak,ALICE:2013rdo,Skands:2014pea,CMS:2022awf}. In \figRef{fig:meanPtVsNch}, we show a comparison of a CDF measurement of this observable~\cite{CDF:2009cxa} to \Sherpa (red), \Herwig (orange), and \Pythia with its default CR model switched on (bright blue) and off (faded blue). The latter does not describe the rising trend seen in the data.
\begin{figure}[tp]
\centering
\includegraphics*[width=0.56\textwidth]{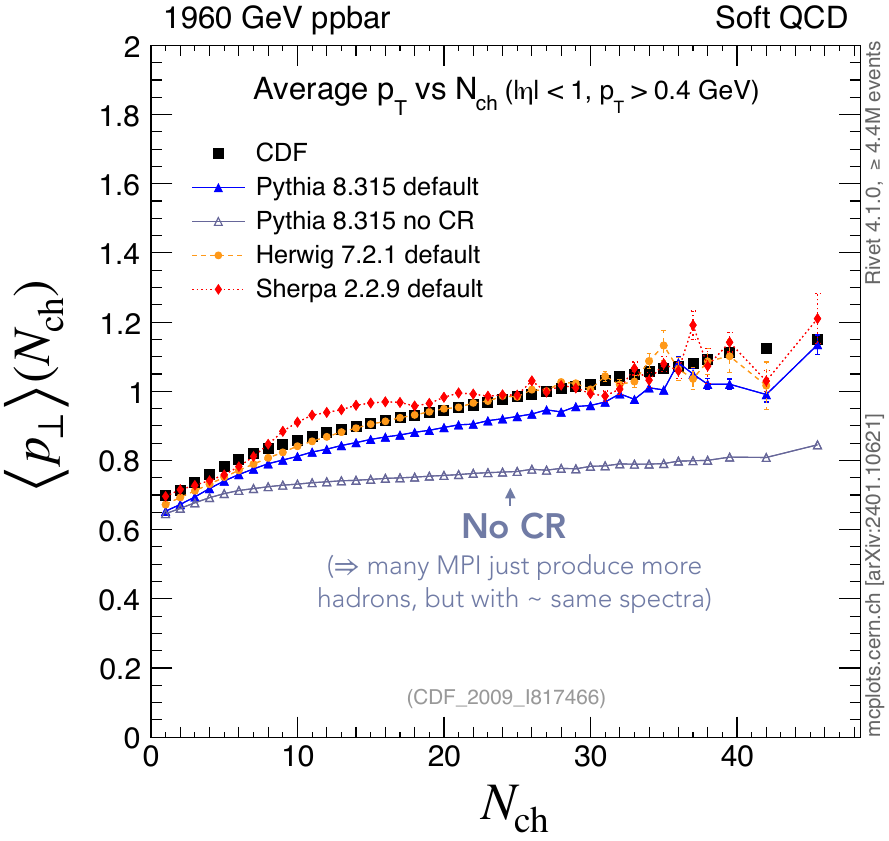}
\caption{Average charged-particle \pT versus charged multiplicity $N_\mathrm{ch}$ in Tevatron minimum-bias proton-antiproton collisions at $\sqrt{s} = 1960~\GeV$, as measured by the CDF experiment~\cite{CDF:2009cxa}, compared to default MC models and to \Pythia without CR. (From \url{mcplots.cern.ch}~\cite{Korneeva:2024oho}.)\label{fig:meanPtVsNch}}
\end{figure}
 
The same basic mechanism -- a relative hardening of the hadronic spectra -- is also active in the UE, for which the ratio of \pT densities to particle densitities is sensitive to CR. 
CR effects may also play a role in the formation of multiply-heavy hadrons~\cite{Egede:2022lws,ALICE:2026vjy}, and can represent a significant source of nonperturbative uncertainty in precision QCD measurements such as (hadronic) $W$ and top-quark mass determinations~\cite{Sjostrand:1993rb,Skands:2007zg,Argyropoulos:2014zoa,Christiansen:2015yqa}. 

Finally, CR effects may also lead to enhanced rates of baryon (and antibaryon) production in hadron collisions. 
Consider the partonic final state depicted on the left-hand side of \figRef{fig:CRJ}.
\begin{figure}[tp]
\centering
\begin{tabular}{c}
\includegraphics*[width=0.33\textwidth]{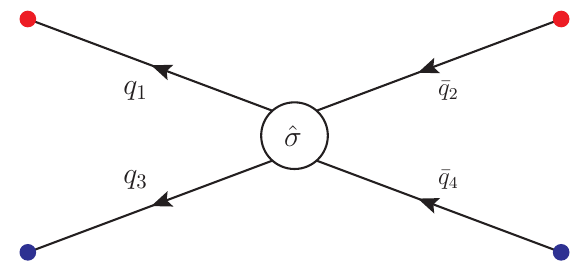}\end{tabular}
\begin{tabular}{c}
\huge ~$\longrightarrow$~
\end{tabular}
\begin{tabular}{c}
\includegraphics*[width=0.33\textwidth]{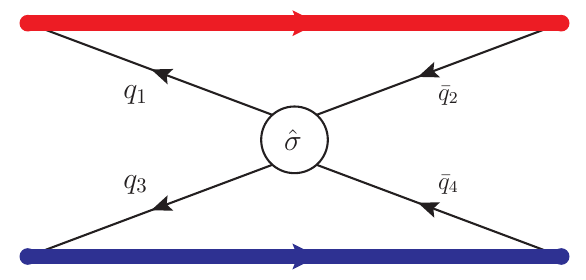}\\
or \\
\includegraphics*[width=0.33\textwidth]{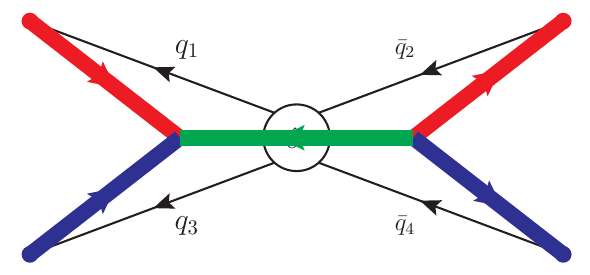}
\end{tabular}
\caption{Illustration of ambiguity in defining the confining colour singlets for a hard process, $\hat{\sigma}$, that produces 4 quarks, $q_1\bar{q}_2 q_3\bar{q}_4$ in a state in which only the $q_1\bar{q}_2$ and $q_3\bar{q}_4$ quark-antiquark combinations have zero net colour charge, and the $q_1 q_3$ and $\bar{q}_2\bar{q}_4$ combinations have colour charge -1 and 1 respectively. This could result in either of the two colour-singlet configurations illustrated on the right. A CR model driven by potential-energy minimisation would choose the one that minimises its corresponding measure.
\label{fig:CRJ}}
\end{figure}
This could lead to either of the hadronising topologies shown on the right. Not all CR and hadronisation models include the possibility shown in the lower right-hand diagram, but those that do~\cite{Christiansen:2015yqa,Gieseke:2017clv} would choose whichever minimises their corresponding measure of total potential energy. When the lower right-hand topology is selected, an extra baryon (and antibaryon) will be produced around the Y-shaped ``colour junctions''. This possibility has  received a lot of attention lately due to observations of substantially increased rates of low-\pT heavy-flavour baryons at the LHC~\cite{ALICE:2021rzj,LHCb:2023wbo,Altmann:2024kwx,Altmann:2024odn}.

We round off by highlighting that CR can also produce effects that are reminiscent of what is called collective flow. The basic mechanism behind this is that CR tends to produce smaller, more boosted, hadronising systems. Since a boost equates to a common velocity for the hadrons produced from a given system, this can mimic the effect of a collective flow~\cite{OrtizVelasquez:2013ofg}.

\subsubsection{Collective effects}
\label{sec:collective}

Collective effects is a term mostly associated with heavy-ion physics. However, it is by now clear that analogous effects are seen in hadron-hadron collisions, albeit with smaller magnitudes. Here, we focus on the phenomenology of collective effects in ``small'' systems, i.e.\ hadron-hadron collisions. These include:
\begin{itemize}
\item Bose-Einstein correlations (and Fermi-Dirac ones, for fermions),
\item collective flow, and
\item strangeness enhancement.
\end{itemize}
As mentioned above, colour reconnections could also be regarded as a ``collective'' effect. However, since this was discussed in \secRef{sec:CRdet}, we do not discuss it further here.

\index{Bose-Einstein correlations}\index{Fermi-Dirac correlations}\paragraph{Bose-Einstein (and Fermi-Dirac) correlations:} identical bosons (e.g.\ two or more $\pi^0$ me\-sons) are quantum mechanically indistinguishable. This means that their production amplitudes should be summed before squaring. At the cross-section level, this gives rise to interference terms which, for bosons, tend to be positive:
\begin{equation}
\left\vert \sum_{i=1}^n {\cal M}^\mathrm{boson}_i \right\vert^2 ~\ge~ \sum_{i=1}^n \left\vert {\cal M}^\mathrm{boson}_i \right\vert^2~,\label{eq:BE}
\end{equation}
while for fermions the corresponding (Fermi-Dirac) interference terms tend to be negative, reflecting the exclusion principle. 
These interference terms are especially important for identical particles that are ``close'' in phase space (small opening angles and small energy differences, low combined invariant masses), such that their wave functions have large integrated overlaps. 

In MC generators, the sequential production of individual hadrons does not easily admit to include these interference terms. None of the general-purpose event generators currently implement the absolute enhancement of \cref{eq:BE}. However, \Pythia offers a simplified treatment applied as an afterburner to the hadronisation process, which shifts identical-particle pairs closer in phase space, and pushes others apart~\cite{Lonnblad:1995mr,Lonnblad:1997kk}. This option is also available in \Sherpa via its interface to \Pythia's string fragmentation.

\index{Collective flow}\paragraph{Collective flow:} 
empirically, collective flow refers to observations of \emph{long-range angular correlations}; especially  positive correlations between particles that are travelling in the same direction in  azimuth, $\Delta\phi \sim 0$, but different rapidities, $\Delta\eta > 1$  (long range), a famous recent example of which is the so-called ``ridge''~\cite{ATLAS:2012cix,CMS:2010ifv} in high-multiplicity minimum-bias events. 

Standard perturbation theory does not predict such effects; particle correlations in a perturbative jet are localised around $\Delta R = \sqrt{\Delta\eta^2 + \Delta\phi^2} \lesssim 1$. Due to transverse momentum conservation, there must also be a recoiling jet at $\Delta\phi\sim \pi$, which can be at any $\Delta\eta$ (the ``opposite-side ridge''). But the observed enhancement of ``near-side'' ($\Delta\phi \sim 0$) particles at large $\Delta\eta$ does not seem to have a perturbative origin. In heavy-ion contexts, the presence of this ``near-side ridge'' is explained as an effect of a collective flow of hadronic matter, thought to be caused by the spatial asymmetry of the initially almond-shaped drop of quark-gluon plasma (QGP), which expands faster along its minor axis than along its major axis. 

Among hadron-hadron event generators, \Epos and \Pythia implement dedicated non-perturbative models of flow-like effects. The \Epos model is based on the core-corona picture of~\cite{Werner:2023jps}, in which high string densities in central hadron-hadron collisions can precipitate the formation of small droplets of QGP, whose asymmetric shapes generate pressure-driven flow according to the same basic mechanism as in heavy-ion collisions. The \Pythia model is based on repulsion between strings, which induces collective flow among them, a process referred to as string shoving~\cite{Bierlich:2017vhg,Bierlich:2020naj}. Smaller flow-like effects can also be generated by its option for hadronic rescattering~\cite{Sjostrand:2020gyg,Bierlich:2021poz}, and -- as mentioned above -- by colour reconnections~\cite{OrtizVelasquez:2013ofg}.

\index{Strangeness enhancement}\paragraph{Strangeness enhancement:} strange quarks and hadrons have higher masses than their up and down counterparts. This leads to an expectation of a relative suppression of them. As discussed in \secRef{sec:MChad}, both string and cluster MC hadronisation models incorporate this baseline strange\-ness-suppression effect, with default parameters normally chosen to give as good agreement as possible with  measurements of strange-to-nonstrange hadron ratios (e.g.\ $K/\pi$, $K^*/\rho$, $\Lambda/p$, etc.) in hadronic $Z$ decays. 

Measurements at hadron colliders, however, tend to exhibit consistently higher strange-hadron fractions (relative to hadronic $Z$ decays), both on average~\cite{CMS:2011jlm} and especially in high-multiplicity minimum-bias events~\cite{ALICE:2016fzo,ALICE:2018pal,ALICE:2019avo,ALICE:2025aqz} and in the underlying  event~\cite{CMS:2013zgf,ATLAS:2024nbm}.\footnote{These trends are especially noticeable for multi-strange hadrons such as $\Xi$ and $\Omega$ baryons~\cite{CMS:2011jlm,ALICE:2016fzo} for which the relative production fractions can be up to an order of magnitude larger than in hadronic $Z$ decays.} The fact that strangeness production appears to be less suppressed in these environments (than in hadronic $Z$ decays) is referred to as strange\-ness enhancement.

Similarly to the case for collective flow, \Epos and \Pythia both implement dedicated nonperturbative models of strangeness enhancement. 

In \Epos, this follows from the same \index{Core-corona}core-corona picture as its modelling of collective flow: the small droplets of QGP that \Epos forms in the ``core'' of densely packed phase-space regions of hadron-hadron  collisions not only expand (producing collective flow), they are also ``hot''; this higher temperature drives a relatively higher production of strange hadrons. This is again quite  analogous to how \Epos models the (denser and much larger) environment of heavy-ion collisions. 

\Pythia currently offers two modelling options, \index{Colour ropes}colour ropes~\cite{Bierlich:2014xba,Bierlich:2022ned} and \index{String closepacking}string closepacking~\cite{Fischer:2016zzs,Altmann:2025afh}, which mainly differ in the degree of sophistication and detail they offer. 
Both models assume that a picture of hadronising strings remains valid, without QGP formation. 

The colour-rope model is the most elaborate of the two. It assumes that ``nearby'' strings can fuse together into a ``colour rope''~\cite{Biro:1984cf,Andersson:1991er}, with an effective tension that can be significantly higher than that of an ``ordinary'' string. 
The baseline assumption is that the effective tension scales like the total colour charge of the rope endpoints.\footnote{Since the effective colour charge of a given representation of SU(3) is given by the eigenvalue of the so-called quadratic Casimir operator, this baseline assumption is commonly referred to as Casimir scaling.} This is consistent with lattice results~\cite{Bali:2000un}.

The increased tension 
directly changes the strangeness suppression factor in \cref{eq:Schwinger}, leading to increased nonperturbative strangeness production in environments where rope formation is likely, i.e.\ high-density environments. 

The colour ropes are assumed to break down  ``strand by strand'', by sequential quark-antiquark (and diquark-antidiquark) pair formation with successively smaller strangeness enhancements at each step, until finally the last strand is just an ordinary string which breaks with the ordinary strangeness-suppression factors. Breaking the ropes strand by strand obviously implies that the relative time ordering of the breaks is physically significant. The rope model therefore augments the standard (momentum-space, acausal) formulation of string fragmentation by an explicit evolution of the strings/ropes in space-time.

The string closepacking model shares the same basic physical motivation of densely packed strings leading to increased string tensions, which result in lower strangeness suppression and hence increased strangeness fractions~\cite{Fischer:2016zzs,Altmann:2025afh}. It is somewhat more simple-minded than the rope model in that it retains the assumption of individual fundamental strings (as opposed to composite ones, i.e.\ ropes) undergoing acausal breaks. This circumvents the need for an explicit space-time evolution. 

Instead of rope formation, the closepacking model assumes that each string contributes to an effective average background field which any other strings nearby will experience. In dense string environments and/or on strings with very sharp kinks, this background field contributes to an increased string tension, the total magnitude of which is assumed to scale in the same way as it does in the rope model. However, instead of ropes breaking strand by strand, the closepacking model distributes the total enhancement roughly evenly among strings that are judged to overlap.\footnote{The current closepacking model of~\cite{Altmann:2025afh} defaults to the assumption that sets of strings with parallel flux orientations generate and receive twice as much enhancement as antiparallel ones do. This is in line with  lattice studies~\cite{Bali:2000un} and with the scaling in the rope model -- but options are provided to vary this assumption if desired.}  

Note that, in the current closepacking  implementation~\cite{Altmann:2025afh}, string overlaps are measured only as a function of rapidity along the beam axis, with no dependence on invariant time or on any other relevant axes (e.g.\ jets). The model therefore only applies to physical contexts for which the beam axis can indeed be regarded as the most relevant physical axis for measuring these overlaps. This includes minimum-bias and underlying-event studies but does not extend, e.g.\ to $e^+e^- \to W^+W^-\to \mathrm{hadrons}$ (for which the beam axis plays no role as far as soft QCD phenomena are concerned), nor does it apply to effects in or among high-\pT jets. Extending the closepacking model to these cases is a topic of active research.  



\subsection{Hadron decays}
\label{sec:decays}

\responsible{Andy B}

Having produced a set of primary hadrons\footnote{This terminology is very confusing: from an event-generator perspective, primary hadrons are the initial physical hadrons resulting from a hadronisation model; from an experimental perspective, primary particles (including hadrons) are those emerging from the beam-particle interaction before interacting with the detector (and producing \emph{secondary} particles. Experimental ``primary particles'' are frequently decay descendants -- on unresolvable length scales -- of the initially generated hadrons.} exclusive-state generators finally have to model particle decays -- and in particular hadron decays -- to obtain a realistic final-state which can be compared to data. While not the primary focus of activity in MC-generator physics studies or development, poor modelling of hadron decays can render all the hard-scattering technology unusable.

In general, accurate modelling of hadron decays can be complex, but in many cases sufficiently accurate results can be obtained with tabulated decay modes and branching ratios, with decay kinematics sampled (in the decaying particle's rest frame) from the appropriate dimensionality of phase-space configurations. Models with more specific matrix-element influences can be implemented as form factors.

More complexity enters when introducing the resonance widths of unstable hadrons, generally also tabulated and by default treated as Breit-Wigner profiles: there is a non-zero risk that intermediate resonances cannot be factorised since a low-mass fluctuation in a parent particle's production can accidentally close further decay channels. In some circumstances, the tabulated approach is not sufficient -- particularly when dealing with unobserved hadrons included in models for QCD consistency -- and the generator falls back to a partonic decay implemented in terms of the constituent quarks.

Generators can differ significantly in their implementation of such details, and in which particles are treated as receiving hadron-like vs electroweak-resonance-like decay treatments. This is worth being aware of for low-mass BSM particles injected into MC models, where the expected behaviour will not be encoded in the default tables and may require careful treatment to avoid wrong or qualitatively changing behaviours depending on e.g.~mass. Even for SM particles, the default capacity of a generator may not be sufficient for the required precision of hadron physics, the most notable situation being heavy-flavour hadron decays in CP-violation and rare-decay measurements such as by LHCb and the $b$-factory experiments. The \mc{EvtGen} ``afterburner'' package~\cite{Lange:2001uf,Curcio:2022gqk,Abudinen:2024wpc} was specifically implemented to provide state-of-the-art decay models for heavy-flavour physics, and is often used as standard to re-decay primary heavy-flavour hadrons according to its own specialist kinematic models and branching-ratio tables.

A final mention goes to \emph{non}-hadron decays: particularly those of the tau lepton, whose mass being above that of the pions is able to decay both leptonically and hadronically. An established analysis technique is use of the spin-analysing power of the virtual $W$-boson in tau decays to map the tau spin via its descendants' kinematics and hence distinguish between spin hypotheses of the tau's parent. Spin is typically ignored in hadronisation models -- a situation that is gradually changing with packages like \texttt{StringSpinner}~\cite{Kerbizi:2021pzn,Kerbizi:2023cde,Kerbizi:2024vpd} -- and therefore at this juncture careful spin treatments are typically reserved for colour-singlet particles, but all current major hadron-level generators implement spin-correlated tau decay models, in addition to which the legacy \mc{Tauola} afterburner~\cite{Jadach:1990mz,Chrzaszcz:2016fte,Antropov:2019ald} is still often used.\footnote{The use of afterburners in the handling of spin is limited by the absence of a standard spinor basis and technical interface for exchanging semi-contracted density matrices between tools, hence some guesswork and heuristics are involved, while internal treatments can use full information. This is an oft-discussed and somewhat evolving situation.} Similar internal and afterburner choices are available for inclusion of electroweak corrections, in particular photon radiation, in electroweak, hadron, and lepton decays.

\section{Model tuning}
\label{sec:tuning}

\responsible{Andy}

We have now finished our overview of the calculational methods and strategies within parton- and particle-level MC event simulations. But you cannot help but have noticed the recurring theme of free parameters in the models, particularly in the less fundamentally understood latter sections. Clever algorithms are one thing, but for practical emulation of real collider events, an MC simulation code is only as good as its configuration! So we must \emph{tune} the phenomenological parameters to achieve the best match to data.\footnote{In the ideal scenario, we would find that there is One True Tune that is able to describe all existing experimental data within expected statistical fluctuations. This has yet to occur!}

Model tuning is, like PDF fitting and the majority of experimental measurements and searches,  ultimately a high-dimensional optimisation problem: given a set of values for the set of $N_\mathrm{P}$ free parameters, a set of high-statistics MC runs will predict values (typically expected mean event rates or differential cross-sections) for every bin in a comprehensive set of analyses; these values are compared to the real data, the parameter-point is updated, and we recompute until a stable best-fit is found. Simple, but unlike many such problems, the naive computation of expected values for each parameter point would take several days -- such an approach would never converge in useful time. And so the approach largely taken, by e.g.~the \mc{Professor} and \mc{Apprentice} toolkits~\cite{Buckley:2009bj,Krishnamoorthy:2021nwv}, has been use of emulators or ``surrogate models'' -- fast approximations to the generator response to its parameters, such that numerical optimisation is tractable. Multiple nuances on this theme have been attempted, e.g.~Bayesian optimisation~\cite{AlKadhim:2025npf} and history-matching~\cite{Iskauskas:2026rxi}, in particular to attempt robust statistical interpretations with well-defined uncertainties via tune-variations. 

The most established uncertainty-encoding is \emph{eigentunes}, which similarly to PDF Hessian uncertainties diagonalise the parameter space and provide pairs of statistically equivalent variations along each principle axis.\footnote{A priority up to now has been compactness, as evaluating variations has meant an expensive re-run of the MC chain. $N_\mathrm{P} \sim \text{10--20}$, leading to an impractical 20--40 eigentune variations on the optimal configuration, but in practice the ``flatter'' principle axes -- corresponding to anticorrelated parameter combinations -- can be discarded, and uses of variation pairs made on case-by-case bases. This situation \emph{may} be changing, enabling e.g.~use of $\mathcal{O}(\text{100--1000})$ replica-type variations, due to the introduction of built-in MC reweighting methods in hadron-level modelling.} At the time of writing, there is still not a definitive approach to defining tune uncertainties, as the standard statistical formalisms for confidence intervals presume that the true model exists somewhere in the parameter space. This is almost certainly not the case for MC event generators, and so heuristic approaches remain common to balance the loss-function contributions of imperfectly modelled observables and to exclude unachieveable ones.

The best-known of such regularisers is multiplicative \emph{tuning weights} that effectively inverse-scale the error bars on the data being compared to the MC model; this biases the tuning loss/likelihood function to ensure that the imperfect model is not overfitted to the most precisely or frequently measured observables to the detriment of others important for the (perhaps novel) physics application. A complementary approach is an additive relative error, augmenting the experimental-data uncertainties with a nominal expected accuracy of the MC model beyond which the fit should not be strongly pulled toward perfect agreement. The arbitrariness or by-eye judgement involved in choosing both these classes of observable-specific factors has long been a source of concern, leading to attempts to empirically choose the weights such as in Ref.~\cite{Bellm:2019owc}, but there is as yet no consensus as to the ideal definition of the uncertainties on a model known not to be true!

\section{Detector simulation}
\label{sec:detector-sim}

\responsible{Chris G}

While this primer focuses primarily on event generation and analysis, \emph{detector simulation} is an essential part of the full Monte Carlo tool chain and forms the bridge between theoretical predictions and experimentally observable data. It is at this stage that abstract particle-level events are turned into realistic detector responses that can be processed by the same reconstruction software as real collision data.

\subsection{Full detector simulation and event reconstruction}

At its core, detector simulation models how particles propagate through matter and magnetic fields, and how they interact with detector components. Doing this this in detail, and on a particle-by-particle basis -- called ``full'' simulation -- includes charged-particle transport in inhomogeneous magnetic fields, electromagnetic and hadronic interactions with detector materials, energy loss through ionisation and radiation, and the production of secondary particles and showers. These processes span a wide range of length and energy scales and involve complex stochastic physics, making detector simulation one of the most computationally expensive parts of the overall workflow.

In current collider experiments, these tasks are almost universally handled by applications using the \texttt{Geant4}~\cite{GEANT4:2002zbu} framework. \texttt{Geant4} provides a comprehensive and highly configurable framework, interfaced as a C++ library, for modelling particle transport and interaction with matter, and underpins detector simulation at the LHC and many other experiments. Given a list of particles at the detector boundary -- typically provided by a generator in HepMC~\cite{Buckley:2019xhk} format -- it simulates their trajectories and interactions, producing a detailed record of energy deposits (``hits'') in sensitive detector elements.\footnote{The storage of these can be customised, for example to aggregate the millions of secondary particles and energy deposits in a typical calorimetric shower into a more manageable number of representative hits.} Expensive parts of this chain are increasingly replaced within experiments' simulation applications with faster approximations, for example simplified detector geometries or precomputed / ML-generated secondary-interaction showers; these are designed by each experiment to reduce computing load with minimal consequences for their targetted physics outputs.

In experimental workflows, detector simulation is followed by two conceptually distinct stages. The first is \emph{digitisation}, which emulates the response of detector readout electronics to the simulated energy deposits. This step models effects such as signal shaping, thresholds, electronic noise, and time sampling, converting idealised energy deposits into digitised signals that resemble those produced by real detector hardware. Digitisation is typically the simulation step at which pile-up in the same and preceding bunch crossings is introduced, by overlaying hits from different simulated events and computing the time-sensitive response to their sum.

The second stage is \emph{reconstruction}, which attempts to infer high-level physical objects -- tracks, vertices, calorimeter clusters, jets, or missing energy -- from these electronic signals.
A key point is that the reconstruction step is, by design, the same for simulated and real data. This symmetry is crucial: it ensures that comparisons between data and Monte Carlo predictions are meaningful and that detector effects are treated consistently. Reconstruction methods are typically developed and calibrated using a mixture of data-driven \emph{in situ} control measurements and corrections based on targetting reconstruction of a particular generator-level observable in detector-simulated MC event samples. As a consequence, the realism of simulated events depends on the \emph{combined fidelity} of event generation, particle transport, and digitisation. Any mismatch in these stages can lead to biases in reconstructed observables.

Within this chain, the greatest parametric freedom typically resides in the event generation stage, where physics models, perturbative calculations, and non-perturbative parameters are tuned to data. Detector simulation and digitisation, while complex, are more tightly constrained by detector design, calibration measurements, and test-beam data -- though substantial uncertainty can still exist in hadronic-interaction and multiple-scattering models, particularly the interactions of unstable hadrons with material which have little constraining data. This division of responsibility explains why much of the uncertainty modelling and tuning effort in collider physics focuses on event generation, even though detector simulation dominates the raw computing cost.

\subsection{Fast and parametrised detector simulation}

Because full detector simulation is computationally expensive, many workflows also rely on \emph{fast} or \emph{parametrised} detector models. These replace detailed particle transport, digitisation \emph{and} reconstruction with simplified descriptions of detector response, such as momentum- and energy-smearing, reconstruction efficiencies, and object-level approximations.

A commonly used standalone tool is \texttt{Delphes}~\cite{deFavereau:2013fsa}, which produces reconstructed objects from particle-level events using configurable parametrisations mixing simplified detector geometries and parametrised object efficiencies and resolutions. Since the reconstruction step by construction tries to reverse the biases induced by material interactions and digitisation, the reconstructed and true physics objects are often not so very different and can be mapped without recourse to explicit detector-geometry or reconstruction emulations, instead replacing the combination with a single ``smearing'' mapping. Both \texttt{Rivet}~\cite{Bierlich:2024vqo} and \texttt{MadAnalysis5}~\cite{Araz:2020lnp} provide built-in detector-smearing utilities, which use mapping functions to probabilistically create detector-level objects from particle-level equivalents. These are particularly useful for rapid studies, validation exercises, and reinterpretation workflows where running a full detector simulation would be impractical.

Such approaches (see also \cref{sec:acceleration}) can offer orders-of-magnitude speed-ups and are invaluable for exploratory studies, feasibility studies, and large parameter scans. They necessarily encode simplified assumptions about detector performance and typically bypass detailed digitisation and low-level reconstruction effects. As a result, they complement -- rather than replace --  full simulation with tools such as \texttt{Geant4}, which remains the reference for precision experimental analyses.

\section{Uncertainties}
\label{sec:uncertainties}

\responsible{Chris G}

A central goal of MC simulation is not only to provide a prediction for a given observable, but also to quantify how \emph{uncertainties in the physics modelling} propagate to the final analysis results. In practice, this means understanding how ``reasonable'' variations of the theoretical inputs and algorithmic choices affect distributions that are ultimately compared to data.

Such variations arise at many points in the simulation chain. They include uncertainties in PDFs, renormalisation and factorisation scales in perturbative calculations, matching and merging prescriptions between matrix elements and parton showers, and parameters governing non-perturbative effects such as hadronisation and the underlying event. Increasingly, these uncertainties are not treated as ad hoc parameter changes, but are instead grounded in statistical inference. In this approach, comparisons to experimental data are used to constrain the parameters of the model and to construct a likelihood (or posterior) distribution over them. Representative variations are then derived from this distribution, for example in the form of eigenvector directions (principal components) that probe independent sources of uncertainty, or ensemble of replica parameter sets that sample the full parameter space.

To turn these variations into uncertainty bands on observables, a number of standard recipes are employed. Depending on the source, uncertainties may be combined using Hessian methods, replica (Monte Carlo) ensembles, or envelope constructions. The details differ between PDFs (cf.~\cref{sec:pdfs}), scale variations, and generator-specific modelling uncertainties, but the underlying idea is the same: probe the sensitivity of predictions to plausible changes in the theoretical description, and propagate this sensitivity to the final results.

The most naive way to do this would be to rerun the event generator independently for each variation of interest. In modern LHC analyses, this approach is simply impractical. Single large-scale MC production campaigns already consume a substantial and barely sustainable fraction of the available computing resources. Repeating such campaigns dozens or hundreds of times to sample uncertainty variations would be prohibitively expensive, both in terms of CPU time and storage.

This reality has driven the widespread adoption of \emph{reweighting techniques}~\cite{Gainer:2014bta,Mrenna:2016sih,Mattelaer:2016gcx,Bothmann:2016nao,Bellm:2016voq}. In this approach, a single MC sample is generated, but each event carries multiple weights corresponding to different choices of scales, PDFs, or other perturbative inputs. These multi-weighted samples allow uncertainties to be evaluated a posteriori, without regenerating events, and at a computational cost that is often orders of magnitude lower than full reruns. For perturbative uncertainties -- such as QCD scale and PDF variations -- reweighting has become the standard approach and is deeply integrated into modern generator workflows. However, the statistical interpretation of these variations differ: while PDF replica or eigenvector sets are typically constructed to represent well-defined confidence intervals, scale variations are usually obtained by multiplying or dividing the renormalisation and factorisation scales by fixed factors, and should therefore be understood as heuristic estimates of missing higher-order corrections rather than statistically rigorous uncertainty intervals.

However, reweighting is not a universal solution. It relies on the assumption that variations can be expressed as smooth, event-by-event weight factors applied to an otherwise unchanged event structure. This assumption breaks down for many algorithmic and non-perturbative choices. For example, changes in hadronisation models -- such as switching between Lund string fragmentation and cluster hadronisation -- can lead to qualitatively different final states, including different particle species and multiplicities. In such cases, there is no meaningful way to reweight one event record into another: genuinely new events must be generated. Similar limitations arise for changes in shower ordering variables, recoil schemes, or other structural aspects of the simulation.

As a result, uncertainty estimation in practice involves a careful balance. Reweighting is exploited wherever it is theoretically and algorithmically justified, dramatically reducing computational cost. For uncertainties that cannot be reweighted, targeted alternative samples are produced, often at reduced statistics or for a limited set of benchmark variations. Designing MC workflows that make this balance explicit -- and that clearly document which uncertainties are covered by reweighting and which require dedicated samples -- is an essential part of sustainable simulation strategies for the HL-LHC era (see also \cref{sec:computing}).

\section{Analysis and reinterpretation}
\label{sec:analysis}

\responsible{Chris G}

Once simulated (and possibly reconstructed) events are available, they enter the analysis stage: raw event records are transformed into physics observables, compared to data, and ultimately interpreted in terms of theory parameters or constraints on models. This stage naturally splits into two closely related but conceptually distinct parts: \emph{event-level analysis} and \emph{statistical inference}.

\subsection{Event-level analysis}

Event-level analysis operates directly on individual events, whether at particle level (truth level) or detector level. Its purpose is to define and compute physically meaningful observables that can be constructed equally for experimental data and Monte Carlo predictions, allowing comparisons between them.

\subsubsection{Unfolding and fiducial analysis}

By default, the observables reconstructed via detectors and experiment software will \emph{not} be directly equivalent to those defined based on MC event records, as they contain biases from the experimental processing; this is where \emph{unfolding} comes in. Unfolding is at its root a mapping between reconstruction-level observables and MC-truth observables with strong enough correlations to make the mapping faithful and only minimally inflate the uncertainties. By performing such a transformation in the experimental analysis -- for example, converting the event populations of bins in a reconstruction variable to estimates of differential cross-sections in bins of the true variable, accounting for event migrations between bins -- an experiment can remove the need for MC comparisons to include a detailed (and computationally expensive) modelling of its detector effects and maximise the longevity and reusability of the measurement. Good design of the MC event-level observables used as the target of the unfolding is therefore crucial: not only must they be physically well-defined, but should also minimise extrapolations into unmeasured phase-space and maximise the faithfulness of the mapping: such a strategy, aligning MC observables with what the detector could actually see, is called \index{Fiducial analysis}{\emph{fiducial analysis}}.

\subsubsection{Jet finding}

A central concept at hadron colliders is \emph{jet clustering}: the reconstruction of collimated sprays of particles originating from high-energy quarks and gluons. Modern jet algorithms\footnote{Earlier cone-style jet algorithms were not always infrared safe, which complicated comparisons between theory and experiment. Modern sequential recombination algorithms were designed to avoid these problems; cf.~e.g.~\cite{Salam:2007xv}.} are designed to be \emph{infrared} safe, meaning that the addition of soft particles or the splitting of particles into collinear fragments does not change the resulting jets. This property is essential for meaningful comparisons between fixed-order calculations, parton showers, and experimental measurements.

In practice, jet finding is almost universally performed using \texttt{FastJet}~\cite{Cacciari:2011ma}, which provides efficient implementations of a broad family of sequential recombination algorithms, such as the $k_\mathrm{T}$~\cite{Ellis:1993tq}, Cambridge–Aachen~\cite{Dokshitzer:1997in,Wobisch:1998wt}, and anti-$k_\mathrm{T}$~\cite{Cacciari:2008gp} algorithms. The anti-$k_\mathrm{T}$ algorithm, in particular, has become the default choice at the LHC due to its robustness and the regular, cone-like jets it produces. Beyond basic jet clustering, \texttt{FastJet} and its extensions support a wide range of tools for jet grooming (e.g.\ trimming, pruning, soft drop), substructure analysis, and tagging, enabling detailed studies of boosted objects and complex final states. An accessible overview of jet substructure and boosted-object phenomenology is provided in~\cite{Marzani:2019hun}. 

Once objects such as jets, leptons, and missing transverse momentum have been defined, higher-level observables -- cross-sections, spectra, angular correlations -- can be constructed. Tools such as \texttt{Rivet}~\cite{Buckley:2010ar,Bierlich:2019rhm,Bierlich:2024vqo} and \texttt{MadAnalysis}~\cite{Conte:2012fm} play a key role at this stage. They provide frameworks for implementing analyses in a way that is independent of a specific experiment's internal software, facilitating reproducible comparisons between theoretical predictions and published measurements. By encoding analyses as standalone, version-controlled code operating on standard event formats, these tools make it possible to validate generators, compare different theoretical models, and preserve analyses for future reuse and reinterpretation.

\subsubsection{Truth-level object definitions, primary particles, and event history}
\label{sec:truth-level-defns}

A distinctive feature of truth-level (particle-level) analysis is that many object definitions are \emph{conventional rather than absolute}, and depend on choices made either at event generation or analysis time. One of the most fundamental of these concerns the definition of \emph{primary particles}: which particles in the event record are treated as stable, observable final-state objects, and which are regarded as intermediate or unstable.

In Monte Carlo generators, this is typically controlled via a lifetime or mean decay-length threshold. Particles with lifetimes longer than this threshold are left stable in the event record, while shorter-lived particles are decayed during generation. Although such thresholds are often chosen to roughly match detector scales, they are not universal and may not correspond precisely to a given experimental setup. This becomes particularly important in scenarios involving long-lived particles, where decay lengths can vary widely and different analyses may legitimately adopt different notions of what constitutes a ``final-state'' particle.

As a consequence, modern truth-level workflows increasingly rely on \emph{analysis-level reinterpretation} of the event record. Rather than taking generator-level decay choices at face value, analyses may dynamically redefine which particles are treated as inputs, for example by selectively decaying or reclassifying particles based on their lifetimes or ancestry. Event-record formats such as HepMC~\cite{Buckley:2019xhk}, together with analysis frameworks like \mc{Rivet}, are designed to support this flexibility, enabling transparent and reproducible object definitions that can be adapted to different experimental or phenomenological contexts.

Closely related to the notion of primary particles are the concepts of \emph{promptness} and \emph{directness}, which are often used interchangeably in informal discussion but encode distinct ideas. Promptness was introduced earlier in the context of hadronisation (see \cref{sec:hadrons}); here we revisit and extend that terminology for truth-level object definitions.

\begin{description}
    \item[Promptness] refers to \emph{where} a particle is produced in space–time. A prompt particle is produced sufficiently close to the primary interaction point that its origin is experimentally indistinguishable from the hard collision, whereas \emph{displaced} particles originate from decays that occur at measurable distances from the interaction point. Promptness is therefore tied to particle lifetimes, decay lengths, and experimental resolution, and applies to all types of final-state particle. 

    \item[Directness] refers instead to \emph{how} a particle is produced in the event history. In practice, directness cannot be unambiguously defined by tracing particles back to a unique hard scatter, especially in the presence of parton showers, hadronisation and multiple interactions. 
    Instead, directness is typically defined operationally, based on the physically meaningful particles stored in the event record after hadronisation. This reflects both a conservative approach to the practical limitation that final-state hadrons do not have unambiguous partonic ancestors, and the fact that the transition between perturbative and non-perturbative regimes corresponds to a genuine separation of physical time and energy scales. The effect is that \emph{hadrons cannot be direct}, as they are separated from the hard scatter by both practicalities and physical principles.

    A non-strongly interacting particle is then considered direct if it does not originate from a hadron decay. Ascertaining this from an event record involves recursion up decay-chain ancestry, only granting directness if hadronisation or partonic objects are encountered before any hadron. This convention is widely used, for example, when defining truth-level leptons or photons, where decay products of heavy-flavour hadrons are treated as background. There may be nuances to this -- whether the particles emerging from leptonic decay of a direct tau are themselves direct, for example -- but the general principle is that hadronisation introduces a key disconnection from the hard process.
\end{description}

More generally, promptness and directness should be regarded as orthogonal concepts. A particle can be prompt but not direct (such as a lepton from a promptly decaying heavy hadron), or direct but not prompt (as in certain long-lived new-physics scenarios). Making these distinctions explicit is essential for precision measurements, displaced-signature analyses, and robust reinterpretation.\footnote{This is an evolving area, as long-lived signatures receive more careful attention. Depending on the degree of displacement, certain hadrons may be directly identifiable as associated to a long-lived exotic state, with their hadronisation factorised from that of the prompt interaction. In such circumstances, particle-level analyses will need to develop fiducial definitions appropriately matched to both the experimental and theoretical issues.}

Finally, even for prompt Standard Model objects, additional conventions are required to obtain infrared-safe and experimentally meaningful definitions. A common example is \emph{photon dressing} of leptons, where photons within a cone around a charged lepton are recombined to account for QED final-state radiation. Similarly, \emph{democratic} jet definitions treat photons and hadrons on an equal footing during jet clustering, providing QED-aware observables that are robust against soft and collinear radiation.

Taken together, these choices highlight that event-level analysis is not a purely mechanical step, but one that encodes careful physics judgement at the interface between theory and experiment. Making truth-level object definitions explicit, configurable, and well documented is essential for reproducibility, meaningful data--MC comparisons, and long-term reuse of Monte Carlo simulations -- particularly as collider physics increasingly explores non-standard and long-lived signatures, but also future high-precision measurements at the HL-LHC and possible future colliders become sensitive to effects that were previously negligible. In such regimes, even seemingly minor details of object definitions -- for example the treatment of QED final-state radiation and subsequent photon splittings~\cite{Flower:2022iew} -- may require revisiting established conventions and best practices.

\subsection{Statistical inference and reinterpretation}

While event-level analysis produces histograms and summary observables\footnote{While most established collider analyses rely on inference based on summary statistics (e.g.\ binned likelihoods), there is a growing interest in \emph{simulation-based inference} (SBI) approaches that operate directly on high-dimensional event data using likelihood-free or density-estimation techniques. We refer to Ref.~\cite{sim-based-inference} for an overview of current directions.}, statistical inference operates on these aggregated quantities. Its goal is to extract quantitative information: constraints on model parameters, goodness-of-fit measures, exclusion limits, or tuned parameter values.

A number of specialised frameworks exist for this purpose, typically working on histograms produced by event-level analyses. \texttt{Contur}~\cite{Butterworth:2016sqg,CONTUR:2021qmv,CONTUR:2025yis} uses a wide range of preserved collider measurements to constrain new-physics models by comparing their predicted deviations from the Standard Model to data. \texttt{pyhf}~\cite{Feickert:2022lzh} provides a modern, pure-Python implementation of the \texttt{HistFactory}~\cite{Cranmer:2012sba} formalism, enabling likelihood-based inference and reinterpretation using binned data and uncertainties.

For generator tuning and model optimisation, tools such as \texttt{Professor}~\cite{Buckley:2009bj} and \texttt{Appren\-tice}~\cite{Krishnamoorthy:2021nwv} construct fast surrogate models of generator response, allowing parameter spaces to be explored and fitted efficiently using statistical techniques. These approaches are particularly powerful when direct generator evaluations are expensive, as is increasingly the case for high-precision simulations.

\section{Workflows, computing, and reproducibility}
\label{sec:computing}

\responsible{Chris G}

Modern MC simulation in HEP is inseparable from questions of workflow design, computing infrastructure, and long-term reproducibility. What once could be run on a local workstation has, for the LHC and especially the High-Luminosity LHC, evolved into a complex, distributed production chain spanning many software layers and computing centres worldwide.

At the front end of this chain sit event generators and parton-level calculations, which are increasingly expensive as higher-order corrections, sophisticated matching and merging schemes, and large parameter scans become standard. These are followed by detector simulation, reconstruction, and finally analysis. In large-scale production campaigns, each of these stages may be handled by different teams, use different software stacks, and be executed on heterogeneous resources ranging from grid sites and national HPC facilities to cloud infrastructure. Workflow orchestration therefore becomes a first-class concern: tasks must be split, scheduled, monitored, retried on failure, and validated in a way that scales to billions of events and months-long campaigns.

Computing constraints strongly shape how MC workflows are designed. Throughput, memory footprint, I/O performance, and increasingly energy efficiency all matter. The HL-LHC era amplifies these pressures: projected increases in luminosity translate directly into higher statistical demands on simulated samples, at a time when traditional CPU scaling is no longer keeping pace. This has driven interest in more performance-portable software, the use of accelerators such as GPUs (cf.~\cref{sec:acceleration}), and careful profiling to understand where computing time is actually spent. From a workflow perspective, it also encourages a clearer separation between well-defined stages, so that expensive steps can be reused, cached, or reweighted rather than regenerated from scratch.

A further and increasingly important consideration arises from the presence of \emph{negatively weighted events} (see \cref{sec:nlo-subtraction} for their theoretical origin). In calculations beyond simple leading-order descriptions, event weights can take both positive and negative values. While formally consistent, such samples reduce the \emph{effective statistical power} of a given number of generated events. In the simplest case where a fraction $f$ of events carry negative weights of comparable magnitude to the positive ones, the effective statistical power scales approximately with a factor $1-2f$, i.e.\ each negatively weighted event partially cancels a positively weighted one, so that as $f\to 50\,\%$, the effective sample size tends towards zero.

More generally, it is not only the fraction of negative weights but also the overall spread of the weight spectrum that determines the variance of observables: broad weight distributions dilute statistical precision for a fixed number of raw events. Since detector simulation and reconstruction costs scale with the total number of generated events rather than the effective statistical power, high negative-weight fractions -- or large weight variance -- can translate directly into substantial increases in computing requirements for a target precision.

A number of mitigation strategies are under active development. Some approaches aim to reduce the negative-weight fraction at generation time -- for example through modified matching schemes~\cite{Frederix:2020trv,Danziger:2021xvr,Sarmah:2024hdk} or folding techniques~\cite{Frixione:2007nw} -- while others attempt post-generation treatments, such as `resampling' strategies~\cite{Andersen:2020sjs,Nachman:2020fff,Andersen:2024mqh} that transform weighted samples into statistically equivalent positive-weight ensembles before detector simulation. Although no universally optimal solution has yet emerged, controlling weight distributions has become a central consideration in designing sustainable MC workflows at current and future colliders.

Reproducibility cuts across all of these concerns. In a distributed environment, it is no longer sufficient to record only the physics parameters of a simulation. One must also capture the exact software versions, compiler choices, external libraries, runtime configuration, and even operating system details that can influence numerical results. Container technologies (such as \texttt{Docker}~\cite{merkel2014docker} or \texttt{Apptainer}/\texttt{Singularity}~\cite{kurtzer_2018_1308868}) have therefore become central to MC production, allowing complete software environments to be packaged and deployed reproducibly across sites.

In practice, containerisation is complemented by shared software distribution infrastructures such as \texttt{CVMFS}~\cite{Blomer:2011tgq}, which provide a scalable, read-only filesystem for distributing experiment and community software stacks worldwide. Together, containers and \texttt{CVMFS} make it possible to decouple software deployment from local system administration, enabling consistent environments across laptops, grid worker nodes, and HPC systems, while still allowing centrally managed updates and long-term archival of software releases.

Alongside technical reproducibility, there is a growing emphasis on \emph{openness} and \emph{reuse} of both simulation outputs and experimental results~\cite{Bothmann:2026ogm}. Large MC production campaigns represent a substantial investment of public resources, and their scientific value extends well beyond the lifetime of a single analysis. In this context, infrastructures such as the CERN \texttt{OpenData} portal~\cite{cernopendata} and \texttt{HEPData}~\cite{Maguire:2017ypu} play complementary roles. While the \texttt{OpenData} portal provides access to event-level or derived datasets after appropriate embargo periods, \texttt{HEPData} serves as the curated repository for published measurements, uncertainties, and correlations in a machine-readable form.

The availability of results in \texttt{HEPData} is a key enabler for reproducible comparisons and reinterpretation. Tools such as \mc{Rivet}~\cite{Bierlich:2024vqo} rely on these preserved measurements to validate generators and to confront theoretical predictions with data in a consistent way. More broadly, the combination of preserved analyses, public measurements, and shared MC samples allows different communities to benchmark new methods, explore alternative models, and maximise the physics return of existing datasets -- an increasingly important consideration in the HL-LHC era.

\subsection{Emerging acceleration strategies: GPUs and machine learning}
\label{sec:acceleration}

In parallel with improvements to workflows and software engineering, significant effort is being invested in \emph{algorithmic} acceleration of the most computationally expensive parts of the MC chain. A prominent direction is the use of GPUs and other accelerators for parton-level event generation\footnote{Earlier work on parallelised matrix-element integration and GPU-accelerated event generation, as well as related developments based on VEGAS-inspired algorithms and flow-based integration frameworks, can be found in e.g.~\cite{Hagiwara:2010oca,Hagiwara:2010ujr,Giele:2010ks,Hagiwara:2013oka,Bothmann:2021nch} and related studies.}. Frameworks such as \mc{Pepper}~\cite{Bothmann:2023gew}, \mc{MadSpace}~\cite{Heimel:2026hgp}, and the \mc{MadGraph4GPU}~\cite{Hagebock:2025jyk} initiative aim to offload matrix-element evaluation and phase-space integration to massively parallel architectures, achieving substantial speed-ups for high-multi\-plicity processes that dominate future computing budgets. Exploratory studies have also begun to investigate GPU-accelerated implementations of parton showers and related components of the event-generation chain~\cite{Seymour:2025fpu}.

Complementing these hardware-oriented efforts, machine-learning-based approaches are increasingly being explored across the full event-generation pipeline~\cite{Butter:2022rso}, ranging from phase-space integration and unweighting to fast surrogates for expensive components of the hard-scattering calculation. 
A central goal of these developments is to improve sampling efficiency, reduce variance in Monte Carlo integration, and ultimately accelerate the generation of large, high-precision event samples.

In this context, \emph{normalising flows} and other neural density estimators are often trained to approximate the differential partonic cross-section and to construct adaptive, invertible mappings from simple base distributions to the physically relevant regions of phase space. In practice, these techniques act on top of physics-motivated phase-space mappings -- such as those encoding resonant propagators or soft and collinear limits -- and complement classical adaptive Monte Carlo algorithms like \texttt{VEGAS}~\cite{Lepage:1977sw,Lepage:1980dq,Lepage:2020tgj} rather than replacing them.

Concrete realisations of this idea are provided by neural importance-sampling frameworks such as \texttt{MadNIS}~\cite{Heimel:2022wyj,Heimel:2023ngj,Heimel:2024wph,DeCrescenzo:2026tsp} and related variants~\cite{Gao:2020vdv,Bothmann:2020ywa,Gao:2020zvv,Winterhalder:2021ngy,Deutschmann:2024lml,Janssen:2025zke,Bothmann:2025lwg,Bothmann:2026dar}, which combine traditional phase-space factorisations with trainable flow-based components to construct sampling densities that closely track the structure of the matrix element. Beyond improving sampling, ML models are also being investigated as fast \emph{surrogates} for expensive ingredients of the hard-scattering calculation itself. Neural-network-based amplitude surrogates~\cite{Bishara:2019iwh,Badger:2020uow,Aylett-Bullock:2021hmo,Maitre:2021uaa,Danziger:2021eeg,Badger:2022hwf,Janssen:2023ahv,Maitre:2023dqz,Bahl:2024gyt,Brehmer:2024yqw,Breso-Pla:2024pda,Herrmann:2025nnz,Bahl:2025xvx,Favaro:2025pgz,Villadamigo:2025our,Beccatini:2025tpk,Bahl:2026jvt,Bahl:2026qaf} aim to approximate fixed-order matrix elements -- or parts thereof, such as loop amplitudes or helicity components -- with high accuracy at a fraction of the computational cost. If successful, such approaches can enable rapid reweighting in theory-parameter spaces, efficient cross-section evaluation, and differentiable approximations of quantities that are traditionally obtained only through costly numerical codes.

Machine learning is also being actively explored -- and in some cases already employed -- in the context of \emph{detector simulation}, where detailed particle transport constitutes a major computational bottleneck. Surrogate models trained on high-fidelity simulations can learn effective representations of calorimeter response, shower development, or other detector effects, enabling fast approximations that retain essential physical features at a fraction of the computational cost. These approaches complement traditional fast-simulation techniques~\cite{ATLAS:2021pzo, Krause:2024avx, CMS-DP-2025-016}, but require careful validation, control of systematic uncertainties, and clearly defined domains of applicability, particularly when extrapolating beyond the training data. 

For readers interested in a broader overview of machine-learning applications in collider physics -- including detector simulation, event generation, reconstruction and analysis -- we refer to the \emph{HEP--ML Living Review}~\cite{Feickert:2021ajf,hepmllivingreview}, which provides a curated entry point to the rapidly evolving literature.

Taken together, these developments illustrate a broader trend: future MC workflows are likely to combine traditional physics-driven algorithms with accelerator hardware and ML methods in increasingly sophisticated ways. Integrating these advances into reproducible, maintainable, and well-validated toolchains remains an important challenge -- and opportunity -- for the community as it prepares for the HL-LHC era and beyond.

\section{Data formats and exchange}
\label{sec:io-formats}

\responsible{Chris G}

A defining feature of the high-energy physics Monte Carlo ecosystem is its strong reliance on well-defined data formats to connect otherwise independent pieces of software. Event generation, detector simulation, analysis, and reinterpretation are often developed by different teams, on different timescales, and in different computing environments. Standardised formats act as the ``glue'' that allows these components to interoperate, enabling flexible workflows and long-lived results.

At the same time, not all formats are mandatory in all workflows. Many are best thought of as optional interfaces: depending on the physics goal, scale of production, and available computing resources, using a particular format may or may not be advantageous.

\subsection{From theory to parton-level events}

At the most abstract end of the chain, physics models can be encoded in formats that describe particle content, parameters, and interactions in a generator-agnostic way. A prominent example is the UFO (Universal FeynRules Output) format~\cite{Degrande:2011ua,Darme:2023jdn}, which allows Lagrangian-level information to be exported from symbolic tools and ingested by a variety of matrix-element generators. Using such a format is optional: it is indispensable for Beyond-the-Standard-Model scans and rapid prototyping of new theories, but unnecessary for many Standard Model production campaigns where validated built-in implementations already exist.

Once a model is fixed, parton-level events are commonly exchanged using the Les Houches Event (LHE) format~\cite{Alwall:2006yp}. LHE files store information about incoming and outgoing partons, event weights, and generator settings in a structured but relatively lightweight way. They provide a natural boundary between matrix-element generation and subsequent steps such as parton showering, matching, and merging. More recently, the LHEH5~\cite{Bothmann:2023ozs} variant has been introduced, using HDF5 as a backend to address some of the scalability and I/O limitations of large XML-based LHE files. As with UFO, the use of LHE or LHEH5 is optional: tightly integrated generator workflows may bypass explicit files altogether, while large-scale campaigns or cross-tool workflows often benefit from the clear separation these formats provide.

\subsection{Particle-level events: a universal interface}

After parton showering, hadronisation, and decays, events reach the particle level. At this point, a single format has emerged as the universal exchange standard: HepMC~\cite{Buckley:2019xhk}. HepMC defines a common event record for particles, vertices, momenta, and metadata, and it is the format through which generators communicate their events to detector simulation and analysis software.

The importance of HepMC cannot be overstated. It decouples event generation from experiment-specific frameworks, allowing the same generator output to be used by different experiments, phenomenology studies, and preserved analyses. For experiments, HepMC acts as a stable contract: as long as a generator produces valid HepMC, it can be integrated into large production campaigns without bespoke interfaces. For theorists and tool developers, it provides a clear target that maximises downstream compatibility.

\subsection{Analysis-level data: histograms and observables}

Once events are analysed, the primary outputs are no longer individual particles but \emph{aggregated observables}: histograms, profiles, and summary statistics. At this stage, two C++-based  libraries dominate the HEP ecosystem, reflecting the different needs of truth-level generator studies and detector-level experimental analyses.

\texttt{YODA}~\cite{Buckley:2023xqh} is closely linked to \texttt{Rivet}~\cite{Bierlich:2024vqo} and is the standard data format for generator validation and truth-level analysis workflows. In this context, analyses are typically run directly on particle-level events, filling histograms incrementally inside tight event loops. \texttt{YODA} is designed specifically for this use case: it provides numerically robust, mergeable histogram objects with a simple and stable data model, and serialises them to a human-readable text format that is well suited to long-term preservation. As a result, \texttt{YODA} files are widely used for comparing generators, tuning models, validating new calculations, and preserving published analyses in a form that can be rerun on future Monte Carlo samples.

\texttt{ROOT}~\cite{ROOT_NIMA_1997}, by contrast, is deeply integrated into the software frameworks of the experimental collaborations. It underpins most detector-level workflows, from reconstruction and calibration through to final statistical inference. \texttt{ROOT} provides a rich ecosystem: high-per\-formance binary I/O, histogramming, fitting, visualisation, and tight coupling to experiment-specific data models. In addition to its role as an in-memory analysis toolkit, \texttt{ROOT} files are nowadays commonly used by experiments internally as an exchange format between processing stages. These files often contain so called \emph{ntuples}, a traditional HEP data format in which each recorded collision candidate (or simulated event) is represented by a fixed set of variables (for example particle momenta, energies, or detector-level observables). Conceptually, this resembles a tabular or columnar dataset in modern data science, although individual variables may themselves contain variable-length collections (for example lists of reconstructed particles) rather than single numerical values. Since the \texttt{ROOT} application itself is a large and dependency-heavy C++ code, analysis-level access to these data structures is nowadays often provided by the Python-ecosystem \texttt{uproot} interface~\cite{Pivarski:2020qcb}, enabling connection to wider data-science, visualisation and statistical inference workflows.

While Python-based analysis tools and plotting libraries are widely used for steering, post-processing, and visualisation, the bulk of event-loop-level analysis at the LHC -- both for Monte Carlo and real data -- is still performed in compiled, statically typed languages. The sheer scale of LHC datasets, often involving billions of events and complex per-event computations, places stringent demands on performance, memory usage, and parallel scalability. In this regime, compiled languages remain the practical choice, with higher-level languages layered on top for orchestration and user interaction rather than replacing the core analysis machinery.

\subsection{Standardisation and interoperability}

Underlying all these formats is a continuing effort to standardise semantics, not just syntax. For example, HepMC~\cite{Buckley:2019xhk} defines status codes to describe the role of particles (incoming, outgoing, intermediate, decayed), enabling analyses to interpret events consistently across generators. Similarly, there has been a concerted push to standardise the naming and interpretation of variation weights -- such as scale and PDF variations -- so that uncertainty estimates can be propagated reliably through complex workflows (see, for example, the community agreement documented in~\cite{Bothmann:2022pwf}).

Another cornerstone of interoperability is the particle identification scheme maintained by the Particle Data Group~\cite{ParticleDataGroup:2024cfk}. PDG ID numbers provide a universal labelling of particles and resonances, ensuring that a ``top quark'' or a ``neutral pion'' means the same thing across generators, analyses, and experiments. Without such shared conventions, meaningful exchange of events and results would be impossible.

\section{Sustainability and long-term support}
\label{sec:sustainability}

\responsible{Chris G}

The Monte Carlo tool ecosystem that underpins modern collider physics is the result of decades of cumulative development. Event generators, detector simulation toolkits, analysis frameworks, and data formats are not short-lived research prototypes: they form the backbone of scientific programmes with planned lifetimes of 30--50 years, such as the LHC and its upgrades, and potential future facilities like the Future Circular Collider (FCC). Ensuring that these tools remain \emph{correct}, \emph{performant}, and \emph{maintainable} over such timescales is therefore a central challenge for the field.

Historically, much of this software has been developed by physicists alongside their primary research activities, often with limited dedicated funding for long-term maintenance. While this model has produced remarkably successful tools, it is increasingly under strain. The demands placed on MC software are growing simultaneously in several dimensions: higher theoretical precision, vastly increased event volumes, more heterogeneous computing architectures, and stricter requirements on reproducibility and analysis preservation. Meeting these demands requires sustained effort in code refactoring, performance profiling, testing, documentation, and user support -- activities that are essential but do not map cleanly onto traditional academic reward structures.

This reality has led to a growing recognition of the importance of research-software engineering~(RSE)~\cite{8994167} expertise within high-energy physics. However, effective work in this area cannot be done by generic or externally embedded software engineers alone. The MC toolchain is highly specialised, with deep entanglement between physics assumptions, numerical algorithms, and software architecture. Meaningful progress therefore requires developers who are already fully trained particle physicists, with sufficient domain knowledge to make informed design and optimisation decisions. In many cases, there are no remaining ``low-hanging fruit'': performance, correctness, and maintainability are tightly coupled, and improvements must be made with full awareness of the underlying physics.

At present, this creates a structural problem. Experimentalists are funded to analyse data, theorists and phenomenologists are funded to develop new calculations and models, but there are currently no stable funding lines for the long-term software engineering work that connects and sustains these activities. Moreover, for early-career researchers, focusing primarily on RSE-type contributions -- despite their critical importance -- often comes at significant career risk, as such work is not consistently recognised within traditional academic evaluation frameworks.

A sustainable future for MC software therefore requires \emph{embedded} RSE career paths within experimental, theoretical, and phenomenological collaborations. These roles must allow domain experts to contribute long-term to software design, maintenance, and user support, while remaining scientifically integrated into their communities and recognised for their contributions. Such embedded expertise is essential not only for technical sustainability, but also for strategic planning, cross-tool coordination, and the effective support of increasingly large and diverse user bases.

The impact of this approach is already evident. Recent large-scale C++ profiling and optimisation efforts within the MC ecosystem (see, for instance, \cite{Bothmann:2022thx}) demonstrated that systematic, expert-led performance work can deliver substantial gains without compromising physics fidelity. These improvements were achieved through deep understanding of both the physics and the software, applying professional software-engineering practices to highly specialised scientific code. In an era where computing resources are a limiting factor for precision physics, such efforts directly expand the achievable physics programme.

Sustainable Monte Carlo software is therefore not just a technical challenge, but a sociotechnical one. Delivering the physics goals of the LHC, the HL-LHC, and future collider programmes requires long-term investment in people as well as code -- specifically, in career structures that enable expert physicist-developers to maintain, evolve, and support the MC ecosystem over decades.

\section{Summary and Outlook}
\label{sec:conclusions}

\responsible{Chris G}

Monte Carlo simulations sit at the heart of modern collider physics. From the formulation of theoretical models and precision calculations, through event generation and detector simulation, to analysis, reinterpretation, and statistical inference, MC tools provide the common language through which theory and experiment communicate. This primer has aimed to outline not only the individual components of this ecosystem, but also the principles that govern how they fit together and why they matter.

The modular structure of the MC ecosystem -- separating event generation, detector simulation, event-level analysis, and statistical inference -- reflects both historical development and the intrinsic complexity of collider phenomenology. While some tools aim to provide end-to-end functionality, many realistic workflows combine specialised components, each optimised for a particular task.
This modularity is a strength: it allows tools to evolve independently, encourages innovation, and makes it possible to reuse analyses and results across experiments and theoretical frameworks. At the same time, it places a premium on well-defined interfaces, common data formats, and shared conventions. Ensuring consistency, interoperability, and systematic improvement across this analysis and reinterpretation chain remains a central challenge -- and a central success -- for the high-energy physics Monte Carlo community.

Looking ahead, several themes recur across the full simulation and analysis chain. The scale of present and future collider programmes demands ever greater attention to computing performance, workflow design, and uncertainty estimation. At the same time, there is a growing emphasis on openness and reproducibility: containerised environments, preserved analyses, and public data releases are increasingly recognised as essential for maximising the long-term scientific value of large-scale production campaigns. These practices not only support validation and reinterpretation, but also lower the barrier for new users and neighbouring communities to engage with collider data.

Equally important is sustainability. The tools discussed in this primer are long-lived scientific infrastructure, underpinning research programmes that will span decades. Maintaining and evolving them requires sustained investment in software maintenance, documentation, user support, and performance engineering, as well as viable career paths for the people who do this work. Models that recognise software development and maintenance as first-class research contributions -- such as the Research Software Engineering model -- are therefore not optional extras, but a prerequisite for the future success of collider physics.

For new PhD students, especially those entering the field from either a primarily theoretical or experimental background, the breadth of the MC ecosystem can initially feel daunting. One of the aims of this primer is to demystify that landscape and to highlight the common concepts and interfaces that link its components together. Training initiatives such as the MCnet Summer Schools play a crucial role here, bringing together students and experts from across theory, phenomenology, experiment, and computing to provide hands-on experience with modern MC tools and workflows, and to foster a shared culture of best practice.

In closing, the Monte Carlo ecosystem is both technically sophisticated and deeply collaborative. Its continued success relies on openness, reproducibility, and sustained community investment, as well as on the next generation of researchers who will use, develop, and ultimately redefine these tools. We hope that this primer serves as a useful entry point -- and an invitation to engage -- for those beginning their journey in collider physics.

\section*{Acknowledgements}

The authors would like to thank the MCnet community for ongoing discussions and contributions to the development of Monte Carlo tools and workflows. We are particularly grateful to the students participating in MCnet schools and training events, whose questions and perspectives have helped shape the motivation and scope of this primer.
MvB acknowledges support from the Dutch Research Council (NWO) under project number
VI.Veni.232.190. 
AB acknowledges funding from the UK STFC grant ST/K001205/1 and CHIST-ERA OpenMAPP project EP/Y036360/1.
CG acknowledges support from the STFC SWIFT-HEP project (grant ST/V002627/1) and the Software Sustainability Institute (funded through the UKRI Digital Research Infrastructure Programme through grant number AH/Z000114/1). 
PS is supported by ARC grant DP230103014.

\clearpage
\bibliography{refs}
\end{document}

%% file: incl_settings.tex
\usepackage[utf8]{inputenc} 
\usepackage[T1]{fontenc} 	
\usepackage[english]{babel} 

\usepackage[bitstream-charter]{mathdesign}
\usepackage{geometry} 		
\usepackage{amsmath} 		
\usepackage{mathtools} 		
\usepackage{float} 			
\usepackage{graphicx} 		
\usepackage{tabularx} 		
\usepackage{booktabs} 		
\usepackage{color} 	
\usepackage{pdfpages} 		
\usepackage{extarrows} 		
\usepackage{multirow} 		
\usepackage{multicol} 		
\usepackage{enumitem} 		
\usepackage{xspace} 		
\usepackage{stackrel} 		
\usepackage[rgb,table]{xcolor}
\usepackage{tikz} 			
\usetikzlibrary{patterns}
\usetikzlibrary{arrows.meta}
\usetikzlibrary{decorations.markings}
\usepackage[compat=1.1.0]{tikz-feynman}
\usepackage{braket} 		
\usepackage{bm} 			
\usepackage{tensor} 		
\usepackage{slashed} 		
\usepackage{siunitx} 		
\usepackage{lastpage} 		
\usepackage{cite} 			
\usepackage[normalem]{ulem} 
\usepackage{fontawesome} 	
\usepackage{tocloft} 		
\usepackage{titlesec} 		
\usepackage{doi} 			
\usepackage{hyperref} 		
\usepackage[most]{tcolorbox} 					
\usepackage[nameinlink, capitalize]{cleveref} 	
\usepackage[nottoc, notlot, notlof]{tocbibind} 	
\usepackage[ruled, vlined]{algorithm2e} 		
\usepackage{makecell}
\usepackage[makeroom]{cancel}
\usepackage{feynmf}
\usepackage{boxedminipage}
\usepackage{subfig}

\makeatletter
\def\BState{\State\hskip-\ALG@thistlm}
\makeatother

\makeatletter
\@ifundefined{pdfoutput}{}{\DeclareGraphicsRule{*}{mps}{*}{}}
\makeatother

\makeatletter
\DeclareRobustCommand*{\bfseries}{%
   \not@math@alphabet\bfseries\mathbf
   \fontseries\bfdefault\selectfont
   \boldmath
}
\makeatother

\hypersetup{
	pdftitle={The HEP Monte Carlo Ecosystem - A Primer},
	pdfauthor={van Beekveld et al.},
    colorlinks=true,
    linkcolor={red!50!black},
    citecolor={blue!50!black},
    urlcolor={blue!80!black}
} 

\DeclareSymbolFont{usualmathcal}{OMS}{cmsy}{m}{n}
\DeclareSymbolFontAlphabet{\mathcal}{usualmathcal}

\SetArgSty{textnormal}
\SetKwComment{Comment}{{\small\#}~}{}
\SetCommentSty{mycommfont}

\setitemize{itemsep=2pt,topsep=2pt,parsep=0pt,partopsep=0pt,leftmargin=*}
\setenumerate{itemsep=0pt,topsep=2pt,parsep=0pt,partopsep=0pt,labelindent=3pt,leftmargin=*}
\newlist{todolist}{itemize}{2}
\setlist[todolist]{label=$\square$}
\usepackage{pifont}

\usepackage{amsmath}
 
\usepackage{amsthm} 		
\theoremstyle{definition}

\usepackage{dashrule}

\usepackage{makeidx}
\makeindex

\definecolor{softgreen}{RGB}{120,150,90}
\definecolor{midteal}{RGB}{0,155,155}
\definecolor{myteal}{RGB}{0,155,155}

%% file: incl_shortcuts.tex
\definecolor{red_cb}{HTML}{e41a1c}
\definecolor{blue_cb}{HTML}{377eb8}
\definecolor{green_cb}{HTML}{4daf4a}
\definecolor{purple_cb}{HTML}{984ea3}
\definecolor{orange_cb}{HTML}{ff7f00}

\newcommand{\Eg}{\text{E.g.}\xspace}

\newcommand{\mwith}{\text{with}}
\newcommand{\mand}{\text{and}}

\def\d{\mathrm{d}}

\newcommand\one{\leavevmode\hbox{\small1\normalsize\kern-.33em1}}

\newcommand{\pT}{\ensuremath{p_{\mathrm{T}}}\xspace} 	
\newcommand{\pTi}[1][i]{\ensuremath{p_{\mathrm{T},#1}}\xspace} 

\newcommand{\HT}{\ensuremath{H_{\mathrm{T}}}\xspace} 	

\newcommand{\Nc}{\ensuremath{N_{c}}\xspace} 

\newcommand{\arXiv}[2][]{%
	\ifthenelse{\equal{#1}{}}%
	{\href{http://arxiv.org/abs/#2}{arXiv:#2}}%
	{\href{http://arxiv.org/abs/#2}{arXiv:#2~[#1]}}}

\newcommand{\fm}{\text{fm}}

\newcommand{\mev}{\text{Me\kern-0.1ex V}}
\newcommand{\gev}{\text{Ge\kern-0.1ex V}}
\newcommand{\tev}{\text{Te\kern-0.1ex V}}

\def\slashchar#1{\setbox0=\hbox{$#1$}           
   \dimen0=\wd0                                 
   \setbox1=\hbox{/} \dimen1=\wd1               
   \ifdim\dimen0>\dimen1                        
      \rlap{\hbox to \dimen0{\hfil/\hfil}}      
      #1                                        
   \else                                        
      \rlap{\hbox to \dimen1{\hfil$#1$\hfil}}   
      /                                         
   \fi}

\newcommand{\tikznode}[2]{%
\ifmmode%
\tikz[remember picture,baseline=(#1.base),inner sep=0pt] \node (#1) {$#2$};%
\else
\tikz[remember picture,baseline=(#1.base),inner sep=0pt] \node (#1) {#2};%
\fi}

\def\mathswitchr#1{\relax\ifmmode{\mathrm{#1}}\else$\mathrm{#1}$\xspace\fi}
\def\mathswitch#1{\relax\ifmmode#1\else$#1$\xspace\fi}

\newcommand{\figRef}[1]{\cref{#1}}

\newcommand{\secRef}[1]{\cref{#1}}

\newcommand{\secsRef}[1]{\cref{#1}}

\newcommand{\muF}{\ensuremath{\mu_\mathrm{F}}\xspace}
\newcommand{\muR}{\ensuremath{\mu_\mathrm{R}}\xspace}
\newcommand{\alphaS}{\ensuremath{\alpha_\mathrm{s}}\xspace}
\newcommand{\Qcut}{\ensuremath{Q_\mathrm{cut}}\xspace}
\newcommand{\hdamp}{\ensuremath{h_\mathrm{damp}}\xspace}
\newcommand{\MeV}{\ensuremath{\mev}\xspace}
\newcommand{\GeV}{\ensuremath{\gev}\xspace}
\newcommand{\TeV}{\ensuremath{\tev}\xspace}

\NewDocumentCommand {\mc} { o m } {%
  \IfNoValueTF {#1} {%
    \texttt{#2}\xspace%
  }{%
    \texttt{#2}\,#1\xspace%
  }%
}
\NewDocumentCommand {\Pythia} { o } {\mc[#1]{Pythia}}
\NewDocumentCommand {\Herwig} { o } {\mc[#1]{Herwig}}
\NewDocumentCommand {\Sherpa} { o } {\mc[#1]{Sherpa}}
\NewDocumentCommand {\Panscales} { o } {\mc[#1]{PanScales}}
\NewDocumentCommand {\Madgraph} { o } {\mc[#1]{MadGraph}}
\NewDocumentCommand {\Rivet} { o } {\mc[#1]{Rivet}}
\NewDocumentCommand {\Epos} { o } {\mc[#1]{Epos}}

\newcommand{\pqcd}[2][]{\ensuremath{\d\sigma^{(#1)}_{#2}}}
\newlength{\tabcolsepsave}
\newlength{\boxlen}
\newlength{\skipback}
\newlength{\skipup}
\newenvironment{loopsnlegsFO}[1][t]{
\setlength{\arrayrulewidth}{1.1mm} 
\setlength{\tabcolsepsave}{\tabcolsep}
\setlength{\tabcolsep}{0pt}
\begin{tabular}{ccc}\parbox[c]{0.5cm}{\rotatebox{90}{~~\# of loops}}&%
\begin{tabular}[c]{|c|c|c|cc}}{
\end{tabular}&%
\parbox[c]{0.5cm}{\vspace*{0.1cm}\hspace*{-\skipback}\tikz{\draw [line width=1.5pt, -{Stealth}] (0,0cm) -- (0,3.7cm);
}}\\[-\skipup]
\multicolumn{3}{c}{\hspace*{-0.5cm}
\tikz{\draw [line width=1.5pt, -{Stealth}] (0,0) -- (7.5cm,0);
}}\\[-1.5mm]
& \# of final-state particles
\end{tabular}%
\setlength{\tabcolsep}{\tabcolsepsave}
}

\definecolor{myExact}{HTML}{5ECFA3}   
\definecolor{myApprox}{HTML}{ffd840}
\definecolor{myBad}{HTML}{ffa600}
\definecolor{myFalse}{HTML}{E77C7C}  

\newlength{\boxht}
\newcommand{\emptyBox}[2][]{%
\begin{minipage}[c]{\boxlen}%
\parbox[c]{\boxlen}{
\begin{tikzpicture}
  \fill[white] (0,0) rectangle (\boxlen,1);
\end{tikzpicture}}
\end{minipage}%
\hspace*{-\boxlen}%
\begin{minipage}[c]{\boxlen}%
\centering
\pqcd[#1]{#2}\end{minipage}}

\newcommand{\exactBox}[2][]{%
\begin{minipage}[c]{\boxlen}%
\parbox[c]{\boxlen}{
\begin{tikzpicture}
  \fill[myExact] (0,0) rectangle (\boxlen,1cm);
\end{tikzpicture}}
\end{minipage}%
\hspace*{-\boxlen}%
\begin{minipage}[c]{\boxlen}%
\centering
\pqcd[#1]{#2}%
\end{minipage}}

\newcommand{\addCut}[1]{%
\hspace*{-\boxlen}%
\begin{minipage}[c]{\boxlen}%
\parbox[c]{\boxlen}{%
\begin{tikzpicture}
  \fill[fill=white!60!myExact] (0,0) rectangle (#1,1cm);
  \fill[pattern=crosshatch, pattern color=white!70!black] (0,0) rectangle (#1,1cm);
  \draw[white!70!black] (#1,0) -- (#1,1cm);
\end{tikzpicture}}
\end{minipage}}